\documentclass[a4paper,11pt]{article}
\pdfoutput=1 

\usepackage[T1]{fontenc} 
\usepackage{graphicx}
\usepackage{epsfig}
\usepackage{amssymb}
\usepackage{amsfonts}
\usepackage{dsfont}
\usepackage{psfrag}
\usepackage{amsmath,euscript,array,mathrsfs,nccmath}
\usepackage{multirow}
\usepackage{booktabs} 
\usepackage{bbold,bm,bbm}
\usepackage{epsf}
\usepackage{slashed}
\usepackage{comment}
\usepackage{colortbl}
\usepackage{lmodern}
\usepackage[utf8]{inputenc}
\usepackage[noadjust]{cite}
\usepackage{jheppub}
\usepackage[a4paper,left=5cm,right=0cm,top=5cm,bottom=0.5cm]{geometry}
\usepackage{caption,subcaption,wrapfig}
\usepackage[colorlinks=true,linktocpage=true,linkcolor=blue,citecolor=blue]{hyperref}

\usepackage{tikz}
\usetikzlibrary{decorations.markings,decorations.pathmorphing}

\newcommand{\Be}{\mathds{B}_8}

\newcommand{\Bconf}{\mathds{B}_8^{\rm{conf}}}
\newcommand{\Binf}{\mathds{B}_8^{\infty}}
\newcommand{\BOP}{\mathds{B}_8^{\rm{OP}}}

\newcommand{\dd}{\text{d}}

\newcommand{\WH}{\Sigma_{\text{\tiny WH}}}
\newcommand{\vol}{V_{\mathds{R}^2}}
\newcommand{\volcp}{V_{\mathds{C}\text{P}^3}}
\newcommand{\W}{W}
\newcommand{\rH}{r_{\text{\tiny H}}}
\newcommand{\growth}{\mathcal{G}}

\newcommand{\parwh}{s}
\newcommand{\CCV}{\mathcal{C}_{\text{\tiny CV}}}
\newcommand{\CK}{\mathcal{C}_{\text{\tiny K}}}
\newcommand{\rUV}{{r_i}}
\newcommand{\rsat}{r_m^{\rm{sat}}}
\newcommand{\uH}{u_{\text{\tiny H}}}
\newcommand{\CP}{\mathds{C}\text{P}}
\newcommand{\ene}{\mathsf{E}}
\newcommand{\free}{\mathsf{F}}
\newcommand{\temp}{\mathsf{T}}
\newcommand{\us}{u_s}

\newcommand{\Tmin}{T_{\text{\tiny m}}}
\newcommand{\rs}{r_s}
\newcommand{\TD}{T_{\text{\tiny D0}}}
\newcommand{\rmin}{r_{\text{\tiny m}}}
\newcommand{\lcWL}{ l_c^{\text{\tiny(WL)}}}
\newcommand{\lcEE}{ l_c^{\text{\tiny(EE)}}}
\newcommand{\Tuv}{T_{\text{\tiny UV}}}
\newcommand{\Tir}{T_{\text{\tiny IR}}}
\newcommand{\hH}{h_{\text{\tiny H}}}
\newcommand{\fH}{f_{\text{\tiny H}}}
\newcommand{\gH}{g_{\text{\tiny H}}}
\newcommand{\lH}{\lambda_{\text{\tiny H}}}
\newcommand{\bH}{\mathsf{b}_{\text{\tiny H}}}
\newcommand{\uS}{u_s}
\newcommand{\fs}{f_s}
\newcommand{\hir}{h_{\text{\tiny IR}}}
\newcommand{\rz}{\rho_0}
\newcommand{\xih}{\xi_{\text{\tiny H}}}
\newcommand{\bconf}{b_0^{\text{\tiny conf}}}
\newcommand{\Lt}{\Lambda_c}
\newcommand{\be}{\begin{equation}}
\newcommand{\bea}{\begin{eqnarray}}
\newcommand{\lm}{\ell_m}

\newcommand{\ee}{\end{equation}}
\newcommand{\eea}{\end{eqnarray}}

\definecolor{c1}{RGB}{189,64,8}
\definecolor{c2}{RGB}{247,128,38}
\definecolor{c3}{RGB}{253,190,17}
\definecolor{c4}{RGB}{68,199,33}
\definecolor{c5}{RGB}{28,173,133}
\definecolor{c6}{RGB}{39,146,182}
\definecolor{c7}{RGB}{2,83,148}
\definecolor{c8}{RGB}{29,53,133}
\definecolor{c9}{rgb}{.5,0,.5}
\definecolor{gray2}{rgb}{.42,.42,.42}

\definecolor{mycolor}{RGB}{255,238,140}
\usepackage{soul}
\usepackage{cancel}

\preprint{\texttt{IFT-UAM/CSIC-26-105}}

\title{\boldmath Complexity measures in holographic cascading theories with multiscale dynamics
}
\author[a]{Carlos Nunez,}
\author[b]{Juan F. Pedraza}
\author[c]{and Javier G. Subils}

\affiliation[a]{Centre for Quantum Fields and Gravity, Department of Physics, Swansea University, Swansea SA2 8PP,
United Kingdom}
\affiliation[b]{
Instituto de F\'isica Te\'orica UAM/CSIC, Calle Nicol\'as Cabrera 13-15, Madrid, E-28049, Spain}
\affiliation[c]{Institute for Theoretical Physics, Utrecht University, 3584 CC Utrecht, The Netherlands}

\emailAdd{c.nunez@swansea.ac.uk}
\emailAdd{j.pedraza@csic.es}
\emailAdd{j.gomezsubils@uu.nl}

\abstract{
Using their gravitational duals, we perform a systematic study of two notions of complexity in a family of three-dimensional gauge theories with rich infrared structure. For thermofield double states, we employ the complexity=volume prescription, which relates computational complexity to the volume of the dual Einstein–Rosen bridge. For one-particle states created by the insertion of a local operator on the vacuum, we study their spreading in Krylov space, encoded holographically by the radial momentum of bulk excitations. We investigate these two notions of complexity across the parameter space of the theories, focusing on the two limiting values of a tunable parameter. Near the limit where the theories flow close to an intermediate conformal fixed point, both notions reveal distinct manifestations of ``walking'' dynamics. Near the opposite limit, computational complexity is largely insensitive to the confining nature of the ground state, whereas the frequency of oscillations in the Krylov spread complexity ---set by the emerging infrared scale--- is sensitive to the presence of confinement.
}

\begin{document} 
\maketitle
\flushbottom

\section{Introduction}

Quantum-information-theoretic quantities provide valuable probes of strongly coupled quantum field theories, particularly when conventional local observables are insensitive to important aspects of the underlying dynamics. Among them, different notions of quantum complexity characterize complementary aspects of how difficult quantum states are to prepare and how they spread under unitary time evolution. In holographic theories, several proposals associate these measures with geometric quantities in the bulk, allowing them to be studied through the structure and dynamics of the corresponding gravitational duals.

In this work, we apply this perspective to the $\mathds{B}_8$ family of three-dimensional $\mathcal{N}=1$ quiver gauge theories, whose gravitational duals were originally constructed in Refs.~\cite{Cvetic:2001bw,Herzog:2002ss,Cvetic:2001pga} and reviewed in Refs.~\cite{Faedo:2017fbv,GomezSubils:2021dzb}. These theories exhibit multiscale Renormalization Group (RG) flows: their ultraviolet (UV) dynamics are governed by three-dimensional super-Yang--Mills theory, while their infrared (IR) behavior is controlled by a continuous parameter $b_0\in[0,1]$. At $b_0=0$, the flow terminates at the Ooguri--Park (OP) conformal fixed point. For small but nonzero $b_0$, the RG trajectories remain close to this fixed point over a parametrically large range of scales, giving rise to walking, or approximately scale-invariant, dynamics. At the opposite endpoint, $b_0=1$, reached through a limit in which the Chern-Simons (CS) interactions are switched off, the theory exhibits Wilson-loop confinement: the potential between external fundamental charges grows linearly at large separation.

For generic $0<b_0<1$, the eleven-dimensional infrared geometry caps off smoothly, giving rise to a discrete tower of massive excitations \cite{Elander:2018gte}. This feature alone, however, does not imply confinement. Because the Chern--Simons interactions remain nonzero, external fundamental charges are screened, and large Wilson loops do not exhibit an area law. Moreover, despite the presence of a discrete massive sector, the theory is not strictly gapped owing to an exact Goldstone mode \cite{Faedo:2025nqc}.\footnote{The coexistence of a discrete massive spectrum with an exact Goldstone mode is not unique to the $\mathds{B}_8$ family. On the baryonic branch of Klebanov--Strassler, the baryonic U$(1)$ symmetry is spontaneously broken \cite{Gubser:2004qj,Butti:2004pk}, producing an axionic Goldstone mode of the type discussed in Ref.~\cite{Dymarsky:2005xt}, while the theory remains confining. The $\mathds{B}_8$ family realizes a complementary situation: for $Q_k\neq0$, the same symmetry-breaking pattern is present, but the nonvanishing Chern--Simons coupling screens the confining string while leaving the Goldstone mode gapless.} The $\mathds{B}_8$ family therefore provides a useful setting in which approximate conformality, a discrete massive infrared spectrum, charge screening, and genuine Wilson-loop confinement can be cleanly disentangled. Previous studies have shown that the associated rich infrared structure leaves characteristic signatures in thermodynamic quantities~\cite{Elander:2020rgv} and quantum-information-theoretic observables~\cite{Jokela:2020wgs,Jokela:2023lvr}. Our aim is to determine how this infrared dynamics is encoded in two complementary notions of complexity, each associated with a different class of states and a distinct bulk observable. For finite-temperature thermofield-double states, we probe this structure through the growth of the dual Einstein--Rosen bridge, whereas for locally excited states above the vacuum, we examine their dynamical spreading in Krylov space through the radial motion of the corresponding bulk probes. Comparing these two constructions allows us to assess which features of the complexity dynamics are associated with approximate conformality and the discrete massive infrared sector, and which, are specifically sensitive to genuine Wilson-loop confinement.

\paragraph{Thermofield-double states and wormhole growth.}
A standard prescription for purifying a thermal density matrix is to introduce two identical copies of the
theory and construct the so-called thermofield-double (TFD) state
\begin{equation}
 |\mathrm{TFD}(\beta)\rangle
 =
 \frac{1}{\sqrt{Z(\beta)}}
 \sum_n e^{-\beta E_n/2}
 |n\rangle_L |n\rangle_R .
\end{equation}
In holographic theories this state is dual to a
two-sided eternal black hole or black brane, whose two asymptotic regions are
connected by an Einstein--Rosen bridge \cite{Maldacena:2001kr}. Although the reduced density matrix on
either side is time independent, the geometry behind the horizons is dynamical and continues to grow
for parametrically long times. Crucially, this growth persists well beyond  thermalization and scrambling, i.e., after simple correlation functions have relaxed to their thermal values and the entanglement entropies of finite subsystems have saturated. The continued growth of the wormhole therefore probes a regime that is largely invisible to conventional observables, motivating the search for a boundary quantity capable of capturing its late-time behavior. This led to the proposal that the growth of the Einstein--Rosen bridge should be interpreted as the growth of the computational complexity of the TFD state \cite{Susskind:2014rva,Susskind:2014moa}.

A natural notion of computational complexity is circuit complexity, defined as the minimum cost of a quantum circuit that prepares a target state
$|\psi_T\rangle$ from a chosen reference state $|\psi_R\rangle$. In the simplest setting, this cost is given by the number of elementary operations, or gates, in the circuit. More generally, different gates may be assigned different weights, leading to
\begin{equation}
 \mathcal{C}(|\psi_T\rangle,|\psi_R\rangle)
 =
 \min_{U(1)|\psi_R\rangle=|\psi_T\rangle}
 \operatorname{Cost}[U] .
\end{equation}
In Nielsen's geometric formulation
\cite{Nielsen:2005mkt,Nielsen:2006cea}, a continuous circuit is described by
\begin{equation}
 \dot U(s)U^{-1}(s)
 =
 -i\sum_I Y^I(s)\mathcal{O}_I,
 \qquad
 U(0)=\mathbf{1},
\end{equation}
and its cost is obtained by minimizing a functional $\int_0^1 \dd s\,F(Y^I)$ over all paths connecting the reference and target states. In quantum field theory, this construction requires specifying the allowed gates, penalty factors, ultraviolet regulator, and reference state. Early systematic implementations considered Gaussian states in free scalar theories, using either Nielsen geometry~\cite{Jefferson:2017sdb} or a Fubini-Study metric on the manifold of Gaussian states \cite{Chapman:2017rqy}.

In holographic theories, the complexity=volume (CV) proposal provides a natural geometric quantity that has long been conjectured to capture the computational complexity of holographic states \cite{Stanford:2014jda}. For a two-sided eternal black hole and up to an order-one, convention-dependent normalization\footnote{The Lorentzian-flow reformulation of CV recasts the proposal in computational terms. Upon discretization, Lorentzian threads can be viewed as elementary operations in an optimized tensor network, with the maximal volume providing a coarse-grained count of gates required to prepare the associated state \cite{Pedraza:2021mkh,Pedraza:2021fgp,Caceres:2025ypk}. Moreover, in perturbative settings, the holographic first law of CV can be mapped to the linearized Einstein equations \cite{Pedraza:2022dqi,Carrasco:2023fcj}, supporting a picture in which spacetime dynamics may arise from principles of optimized quantum computation.}, it assigns
\begin{equation}\label{eq:definition_complexity}
 \CCV(t_L,t_R)
 =
 \frac{1}{G_N L}
 \max_{\partial\Sigma=t_L\cup t_R}
 \operatorname{Vol}(\Sigma)\,.
\end{equation}
Here $G_N$ is gravitational Newton's constant, $L$ is an appropriate bulk length scale, and $\Sigma$ is a maximal codimension-one spatial hypersurface anchored at the boundary times $(t_L,t_R)$ and extending through the black-hole interior. As the boundary times increase, the hypersurface probes an increasingly large portion of the Einstein–Rosen bridge. At late times, after the system has locally equilibrated, its volume keeps growing linearly with time, providing a geometric realization of the persistent growth of the complexity of the thermofield-double state even after conventional observables have reached thermal equilibrium.

Other holographic prescriptions have also been proposed. Most notably, the
complexity=action (CA) conjecture identifies the computational complexity of the boundary state
with the on-shell gravitational action evaluated on the Wheeler--DeWitt patch
\cite{Brown:2015bva,Brown:2015lvg}. More generally, the
``complexity=anything'' framework encompasses broad classes of
diffeomorphism-invariant observables supported on codimension-one or
codimension-zero bulk regions, with CV and CA arising as particular choices or
limiting cases \cite{Belin:2021bga,Myers:2024vve,Caceres:2025myu}. The resulting multiplicity of
bulk candidates may be viewed as reflecting the ambiguities inherent in defining
complexity in quantum field theory, where the answer depends on the choice of
reference state, gate set, cost functions, penalty factors, and ultraviolet
regularization. Rather than selecting a unique bulk observable, the
complexity=anything proposal identifies a family of candidates that share the
characteristic features expected of complexity, including linear growth at late
times and the switchback effect in shock-wave geometries. These holographic prescriptions have been extensively used to investigate ultraviolet divergences, evolution following quantum quenches, complexity of formation and aspects of black hole interiors and their singularities
\cite{Carmi:2016wjl,Carmi:2017jqz,Chapman:2016hwi,
Reynolds:2016rvl,Chapman:2018dem,Jorstad:2023kmq,Arean:2024pzo}. In the present work, for concreteness, we employ
the CV prescription and focus on the late-time growth of the complexity of
TFD states across the different infrared regimes of the
$\mathds{B}_8$ family.

\paragraph{Locally excited states and radial momentum.}

The second class of states considered here consists of one-particle excitations
created by local operator insertions in the vacuum. Schematically, these states
take the form
\begin{equation}
 |\psi_0\rangle
 \propto
 \mathcal{O}(\mathbf{x})|0\rangle ,
\end{equation}
with an appropriate regulator understood when necessary. Such states are generally
not energy eigenstates and therefore undergo nontrivial Schrödinger evolution,
\begin{equation}
 |\psi(t)\rangle=e^{-iHt}|\psi_0\rangle.
\end{equation}
In holography, when the excitation admits a semiclassical description,
such a state is represented by a massive particle or brane propagating through the
dual ground-state geometry
\cite{Nozaki:2013wia,Caputa:2014eta,Agon:2020fqs}.

We characterize the real-time evolution of these states using Krylov state
complexity, often referred to as spread complexity
\cite{Balasubramanian:2022tpr}.\footnote{The original Krylov construction concerns 
operators evolving in the Heisenberg picture \cite{Parker:2018yvk}.
Although state and operator Krylov complexities are conceptually distinct
from circuit and Nielsen complexity, several connections have been
established for particular choices of gates, cost functions, and complexity
metrics~\cite{Chattopadhyay:2023spread,Lv:2023jbv,Craps:2023ivc,
Beetar:2025spread,Craps:2025explicit}.} A priori, the spread of a state depends on the basis in which it is expanded.
To make this ambiguity explicit, consider a complete ordered orthonormal basis
$\mathcal{B}=\{|B_n\rangle\}_{n\geq0}$, with
$|B_0\rangle=|\psi_0\rangle$, and define the cost
\begin{equation}
 \mathcal{C}_{\mathcal{B}}(t)
 =
 \sum_{n\geq0}c_n
 \left|\langle B_n|\psi(t)\rangle\right|^2,
 \qquad
 0=c_0<c_1<c_2<\cdots .
\end{equation}
The optimal-basis theorem of Ref.~\cite{Balasubramanian:2022tpr} singles out
the Krylov basis generated from the sequence
\begin{equation}
 |\psi_0\rangle,\qquad
 H|\psi_0\rangle,\qquad
 H^2|\psi_0\rangle,\qquad\ldots .
\end{equation}
For continuous Hamiltonian evolution, this basis minimizes the short-time
growth of the cost: whenever another ordered basis first differs from the
Krylov basis, the first coefficient in the short-time expansion at which the
two costs differ is larger for the alternative basis. Thus, among all
orthonormal bases containing the initial state, the Krylov basis delays the
spreading of the wavefunction into increasingly costly directions as
efficiently as possible.

Concretely, one constructs the orthonormal Krylov chain
$\{|K_n\rangle\}$ by applying the Lanczos algorithm to the sequence above,
with $|K_0\rangle=|\psi_0\rangle$. The Hamiltonian is tridiagonal in this
basis, so that the dynamics reduce to a nearest-neighbor hopping problem
along a one-dimensional chain. Expanding
\begin{equation}
 |\psi(t)\rangle
 =
 \sum_{n\geq0}\phi_n(t)|K_n\rangle ,
 \qquad
 |K_0\rangle=|\psi_0\rangle ,
\end{equation}
and choosing the standard linear weights $c_n=n$, one obtains the spread
complexity\footnote{In specific 2D holographic dualities, notably in DSSYK/JT gravity, Krylov complexity of TFD states admits a precise bulk interpretation in terms of wormhole growth \cite{Rabinovici:2023yex,Heller:2024ldz,Balasubramanian:2024lqk}. More general choices of weights, defining higher-order Krylov complexities, probe connected bulk contributions associated with replica wormholes \cite{Fu:2025kkh}.}
\begin{equation}
 \CK(t)
 =
 \sum_{n\geq0}n\,|\phi_n(t)|^2 .
\end{equation}
It measures the average position of the state along the Krylov
chain. Because this chain is determined canonically by the pair
$(H,|\psi_0\rangle)$ and spans the smallest Hamiltonian-invariant subspace
accessible from the initial state, $\CK(t)$ quantifies the extent of the dynamical spreading required to represent the evolution. A state localized near the beginning of the chain is well approximated by a few Krylov vectors, whereas substantial support at large $n$ signals that the evolution explores progressively more complex states generated by repeated action of the Hamiltonian.

Notably, the Lanczos coefficients and Krylov amplitudes can equivalently be reconstructed directly from the survival amplitude
\begin{equation}
 S(t)=\langle\psi_0|e^{-iHt}|\psi_0\rangle
\end{equation}
and its moments. This relation makes spread complexity a dynamically adapted probe of spectral correlations and, through them, of quantum-chaotic behavior across a broad class of systems~\cite{Erdmenger:2023wjg,Balasubramanian:2023kwd,Caputa:2024vrn,Camargo:2024deu,Baggioli:2024wbz,Huh:2024ytz,Baggioli:2025knt,Erdmenger:2026iga}. Recent reviews
include Refs.~\cite{Nandy:2024evd,Baiguera:2025dkc,Rabinovici:2025otw,Jeong:2026toappear}.

For states created by local operators in conformal field theories, a simple holographic interpretation of spread complexity has recently been proposed. In this picture, the growth rate of spread complexity is related to the proper radial momentum of the dual massive excitation propagating through the emergent bulk geometry \cite{Caputa:2024sux}; refinements and extensions were subsequently developed in Refs.~\cite{He:2024pox,Fan:2024iop,Li:2025fqz}. This motivates the prescription
\begin{equation}
 \dot{\CK}(t)\simeq - P_{\bar y}(t),\label{lionelm}
\end{equation}
where $P_{\bar y}$ is the momentum conjugate to the proper radial coordinate. 
The present work forms part of a broader program developing and testing this proposal in increasingly general holographic settings. It was first examined in a smooth confining geometry, where the finite radial extent of the bulk induces oscillations in spread complexity \cite{Fatemiabhari:2025usn}, and was later extended to conformal quiver theories, in which motion along an internal quiver direction probes colour and flavour data \cite{Fatemiabhari:2025poq}. A systematic analysis of confining backgrounds subsequently identified the confinement scale as the characteristic frequency controlling these oscillations \cite{Fatemiabhari:2026goj}. Further developments incorporated higher-dimensional conformal theories, quiver dynamics, and \(R\)-symmetry charge \cite{Fatemiabhari:2026rob}; charged, composite, and extended probes \cite{Nastase:2026lhz}; and genuine string and D-brane excitations, together with the appropriate fixed-charge Routhian prescription \cite{Chatzis:2026ekd}. Connections between complexity and holographic \(c\)-functions were explored in \cite{Nunez:2026kwr}.
Taken together, these results indicate that the proper-momentum prescription is sensitive not only to radial motion and energy scales, but also to confinement, internal symmetries, quiver structure, and the composite or extended nature of the boundary excitation. The subject is developing rapidly, with several recent works exploring further aspects of the relation between spread complexity and bulk dynamics in holography \cite{Roychowdhury:2026eds,Zoakos:2026obl,Roychowdhury:2026sgg,Roychowdhury:2026vzq,Roychowdhury:2026igc,Alfinito:2026vah,BitaghsirFadafan:2026lek,Graef:2026pzv,Roychowdhury:2026mpd,Qu:2025lgo,Alfinito:2026yex,Baume:2026jyt,Muck:2026top,Li:2026pdh,Pedraza:2026zji,Qu:2026dmv,Bhattacharya:2026ybq,Bhattacharyya:2026zpu}.\footnote{Deviations from this prescription were recently observed in DSSYK/sine-dilaton gravity \cite{Fu:2025kkh} beyond the Schwarzian low-energy regime, plausibly reflecting UV-sensitive, nonperturbative quantum-gravitational effects.}

\paragraph{General idea and organization of this work.} In this work, we investigate these two complementary notions of complexity in the $\mathds{B}_8$ family. Let us emphasize that these are not competing prescriptions for the same state. CV probes the computational complexity of a finite-temperature TFD state through the growth of a maximal hypersurface behind a black-brane horizon, whereas spread complexity probes the unitary evolution of a localized excitation above the vacuum through the radial motion of a bulk probe in the ground-state geometry. They nevertheless provide complementary diagnostics of the same RG dynamics: the former through wormhole growth, while the latter through the spreading and recurrence of local excitations.

Our results reveal two distinct manifestations of walking behavior. For TFD states, we focus on the late-time CV growth rate normalized by the energy of the thermal state. At the Ooguri–Park fixed point, this ratio is constant. For $0<b_0\ll1$, the theories flow close to the OP fixed point, and the ratio remains approximately equal to its conformal-point value over an extended range of energies. This provides a direct complexity-based characterization of the walking regime. As $b_0$ approaches the opposite endpoint, $b_0\sim1$, the CV growth smoothly approaches the confining result without developing a qualitatively new feature precisely at the onset of confinement.

For locally excited states, we apply the proper-momentum prescription to point particles and D0-branes, which are natural in the ten-dimensional description, since they correspond to physical excitations of type IIA string theory. Because our ultraviolet geometries are D2-brane-like rather than asymptotically AdS, this prescription is being extrapolated beyond the setting in which it was derived. Nevertheless, it provides a geometric diagnostic of how the multiscale RG flow affects the spreading of locally excited states.

The smoothly capped infrared geometries lead to bounded radial motion and an oscillatory rate of spread complexity. The corresponding oscillation periods are controlled by the infrared scale and become parametrically long as the flow approaches the Ooguri--Park fixed point. The oscillations persist for generic $0<b_0<1$, where external fundamental charges remain screened, showing that they diagnose the presence of a smooth cap and a discrete spectrum more generally, rather than Wilson-loop confinement itself. Genuine confinement is reached only in the correlated $b_0\to1$ limit, in which the Chern--Simons interactions are switched off. In this limit, the oscillations persist, but with qualitative differences, suggesting that the spread complexity of locally excited states is more sensitive to the confining nature of the ground state.

Hence, we will see that these two notions studied here yield an intriguing picture of how complexity measures encode infrared physics. Notably, both the CV growth-to-energy ratio and the period of the Krylov oscillations respond sharply to approximate conformality. By contrast, while the former is rather insensitive to the confining nature of the IR, the latter is not. This is precisely the distinction that the $\mathds{B}_8$ family is uniquely suited to expose.

The remainder of the paper is organized as follows. In Section~\ref{sec:B8_family}, we introduce the $\mathds{B}_8$ family of backgrounds, review their renormalization-group structure, and summarize the relevant zero- and finite-temperature geometries. In Section~\ref{sec:CV}, we study the complexity of thermofield-double states using the CV prescription, with particular emphasis on its late-time growth, walking behavior, and approach to the confining limit. In Section~\ref{sec:Krylov}, we investigate the Krylov spreading of locally excited states through the radial motion of point particles and D0-branes in the ground-state geometries. We summarize and discuss our results in Section~\ref{sec:discussion}. Technical details concerning the geometries, their eleven-dimensional uplift, and the confining limit are collected in Appendices~\ref{app:geometry}, \ref{app11d}, and \ref{app:Bconf}.

\section{The $\Be$ family}
\label{sec:B8_family}

\begin{figure}[t!]
    \begin{center}
	\begin{tikzpicture}[scale=3.5,very thick,decoration={markings,mark=at position .5 with {\arrow{stealth}}}]
	\node[above] at (0,0) {SYM--CSM $|$ D2};
	\node[below] at (0,-2.2) {Mass gap};
	\node[below] at (0,-2.35) {$\mathds{R}^7\times {\rm S}^4$};
	\node[left,red] at (-1,-.85) {OP CFT~~~};
	\node[left,red] at (-1,-1) {AdS$_4\ \times \ $S$^7/\mathds{Z}_k$~~};
	\node[right] at (1.6,-2.2) {Confinement};
	\node[right] at (1.6,-2.35) {$\mathds{R}^6\times{\rm S}^1\times {\rm S}^4$};
	\node at (-.6,-1) {$\Be^+$};
	\node at (.63,-1) {$\Be^-$};
	\draw[postaction={decorate},ultra thick,c4] (0,0) --  (0,-2) node[left,midway]{$\mathds{B}_8$};
	\draw[postaction={decorate},ultra thick, gray2] (-1,-1) -- (-1,-2) node[left,midway]{$\BOP$\,\,};
	\draw[postaction={decorate},ultra thick, c2] (0,0) .. controls (-.9,-.9) and (-1,-1) .. (-0.98,-2);
	\draw[postaction={decorate},ultra thick, c3] (0,0) .. controls (-.5,-.5) and (-.5,-1.5) .. (-.5,-2);
	\draw[postaction={decorate},ultra thick, c5] (0,0) .. controls (.5,-.5) and (.5,-1.5) .. (.5,-2);
	\draw[postaction={decorate},ultra thick, c6] (0,0) .. controls (.9,-.9) and (.95,-1.05) .. (1,-2);
	\draw[postaction={decorate},ultra thick, c7] (0,0) .. controls (.9,-.9) and (1.5,-1.6) .. (1.5,-2);
	\draw[postaction={decorate},ultra thick, c8] (0,0) .. controls (.95,-.95) and (1.8,-1.8) .. (1.9,-2);
	\draw[|-|] (-1,-2) -- (0,-2);
	\draw[-stealth] (0,-2) -- (2,-2);
	\node[left=5] at (-1,-2) {$b_0$};
	\node[below=5] at (-1,-2) {$0$};
	\node[below=5] at (0,-2) {$2/5$};
	\node[below=5] at (2,-1.95) {$1$};
	\draw[postaction={decorate},ultra thick,c1] (0,0) -- (-1,-1) node[left,midway]{$\Binf$\,};
	\draw[postaction={decorate},ultra thick,c9] (0,0) -- (2,-2) node[right,midway]{\, $\Bconf$};
	\draw [red, ultra thick,fill=red] (-1,-1) circle [radius=0.03]; 
	\draw [black, fill=black, ultra thick] (0,0) circle [radius=0.03];
	\end{tikzpicture}
    \end{center}
	\caption{\small Summary of the $\Be$ family. The asymptotic UV geometry is that of a D2-brane, featuring the 3D super Yang--Mills microscopic theory. The arrows represent various RG flows, from high to low energies. The different IR regimes are indicated explicitly, and their properties follow from the behavior of the eleven-dimensional metric in each case.
    At the endpoint value $b_0=1$ the theory is confining. 
	For $b_0=0$ the IR becomes the Ooguri-Park (OP) CFT, dual to AdS$_4$ times a squashed orbifolded seven-sphere. The hue or the warmth of the curves indicates the value of $b_0$. We use this color coding in the rest of the paper. Figure taken from Ref.~\cite{Jokela:2021knd}.
	}\label{fig:triangle}
\end{figure}
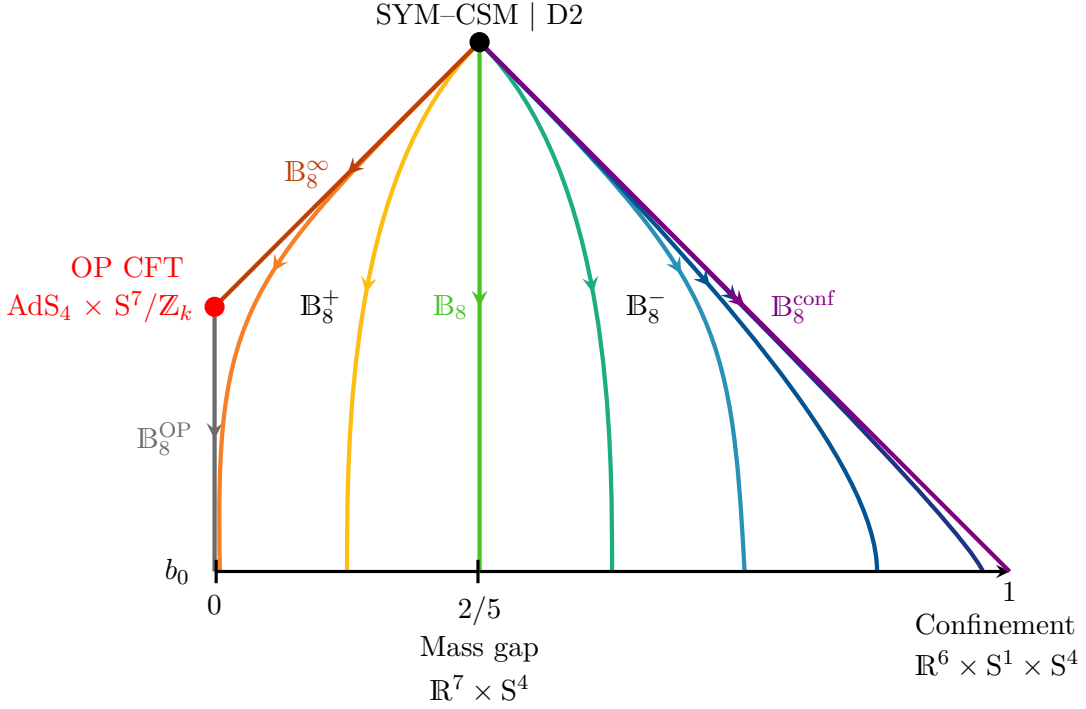

In the rest of the paper, we focus on  the family of $\mathcal{N}=1$ super Yang-Mills quiver gauge theories with group U$(N)_{k}\times$U$(N+M)_{-k}$ and Chern-Simons (CS) interactions at level $k<0$. These theories are studied through their gravity duals. Even though we lack a full field theory description of them, valuable insight is available due to their similarity to ABJM~\cite{Aharony:2008ug}. For instance, the quiver structure of the gauge group arises as a result of the non-trivial geometric structure of the internal space~\cite{Loewy:2002hu}, while the relative shift is a consequence of the presence of ``fractional branes''~\cite{Aharony:2008gk}.

The model is often referred to as the $\Be$ family, and its Renormalization Group (RG) structure ---understood through the eleven-dimensional supergravity holographic dual--- is depicted in Fig.~\ref{fig:triangle}. Even though the geometries dual to the ground states of the theories are regular only in eleven dimensions in general, some of their features are better understood in terms of their reduction to ten-dimensional type IIA supergravity. An ansatz for the metric is constructed by viewing complex projective three-space, $\CP^3$, as a two-sphere fibration over a four-sphere. More precisely, the (string-frame) metric and dilaton ansatz read
\begin{equation}
\label{10Dansatz}
\begin{aligned}
\dd s_{\rm st}^2 &=h^{-\frac12}\left(-\mathsf{b}\dd t^2 + \dd x_1^2 + \dd x_ 2^2\right)+h^{\frac12} \left(\frac{\dd r^2}{\mathsf{b}}+e^{2f}\dd\Omega_4^2+e^{2g}\left[\left(E^1\right)^2+\left(E^2\right)^2\right] \right)\,,\\[2mm]
e^\Phi&=h^{\frac14} \, e^\Lambda \,,
\end{aligned}
\end{equation}
where $E_1$, $E_2$ are vielbeins on the two-sphere and  $\dd\Omega_4^2$ is the line element of the four-sphere, see Appendix~\ref{app:geometry} for details. 
Using the left-invariant forms $J_2$, $X_2$, $J_3$ and the volume form $\omega_6$ on $\CP^3$, defined explicitly in Eqs.~\eqref{eq:X2J2}, \eqref{eq:X3J3} and~\eqref{eq:omega6}, the type IIA Ramond–Ramond and Neveu-Schwarz fluxes are 
\begin{equation}\label{eqfluxesansatzFT}
\begin{array}{rclcrcl}
F_4 &=&  \mathbf{f}_4 *\omega_6 + G_4 + B_2\wedge F_2 \,,&\qquad& F_2& =& Q_k (X_2 - J_2)\,,\\[2mm]
G_4 &=& \dd(a_J J_3) + q_c\left(J_2\wedge J_2 - X_2\wedge J_2\right)\,,&\qquad& B_2& =& b_X X_2 + b_J J_2\,,
\end{array}
\end{equation}
with\footnote{In previous works, an additional constant $Q_c$ was considered here. It turns out that regularity ---in this case, absence of sources in the IR of the geometry describing the ground-state--- fixes this constant to zero.} $\mathbf{f}_4= Q_k b_J^2 +2(q_c+2a_J)b_X - 2 b_J(q_c-2a_J+Q_kb_X)$.  In these expressions, $f,\, g,\, \Lambda,\, h,\, {\mathsf{b}},\, b_X, \, b_J, $ and $a_J$ are functions of the radial coordinate $r$;  while the constants $Q_k$ and $q_c$ are related to the Chern-Simons level and the shift in the rank of the groups through \cite{Hashimoto:2010bq}
\begin{equation}\label{eq:constants_Qk_qc}
    Q_k = \frac{\ell_s g_s}{2}k\,, \qquad \, q_c = \frac{3\pi \ell_s^3 g_s }{4}\left(M-\frac{k}{2}\right)\,,
\end{equation}
with $\ell_s$ and $g_s$ denote the string length and string coupling respectively. Since the microscopic theory is three-dimensional, its ’t~Hooft coupling 
\begin{equation}\label{eq:tHooft}
    \lambda=\ell_s^{-1} g_s N
\end{equation} has dimensions of energy. All the observables that we  compute below depend on the combinations
\begin{equation}\label{eq:def_Lt_alpha}
    \Lt = \frac{k^2\lambda}{6\pi N(M-k/2)}\,,\qquad \alpha =\frac{9}{256|k|\pi}\left(M-\frac{k}{2}\right)^3\,.
\end{equation}
Note that $\Lt$ has dimensions of energy, while $\alpha$ is dimensionless.

Each representative in the family is characterized by a parameter $b_0\in[0,1]$, which provides the asymptotic value of the $B_2$ field in the way specified below. It is therefore expected to capture the difference in the gauge couplings of the two gauge groups in the quiver,
\begin{equation}
     b_0 \sim \frac{1}{g_1^2} - \frac{1}{g_2^2}\,.
\end{equation}
In addition, $B_2$ is not gauge invariant, and we can add to it a closed but non-exact piece, $B_2\mapsto B_2+\mathcal{B}(X_2-J_2), $ with $\mathcal{B}$ constant. This shifts the Page charge and, in the dual quiver description, changes the ranks of the gauge groups, giving rise to a Seiberg-like duality cascade~\cite{Hashimoto:2010bq,Aharony:2009fc,Faedo:2025nqc, Benini:2007gx} similar to that of Klebanov--Strassler. In the present case, however, the cascade terminates after a finite number of steps\footnote{See Ref.~\cite{Aramini:2025twg, Aramini:2026stj} for a $(3+1)$-dimensional model with a duality cascade involving a finite number of steps.}.

For $b_0\in[0,1)$, in the dimensionless coordinate 
\begin{equation}\label{eq:u_coord}
    u =\frac{|Q_k|}{r}
\end{equation}
the near boundary expansions (at $r\to\infty$) read
\begin{equation}\label{eq:UV_expansions}
\begin{aligned}
e^{2f}
&=
\frac{|Q_k|^2}{2u^2}
\left[
1-2u-4u^2-8u^3
+\left(2f_4+\frac{77}{4}\right)u^4
+\left(-2f_4+2f_5+\frac{65}{2}\right)u^5
+\cdots
\right],
\\[4pt]
e^{2g}
&=
\frac{|Q_k|^2}{4u^2}
\left[
1-4u-4u^2
-\left(4f_4+\frac{109}{2}\right)u^4
+\left(14f_4+2f_5+\frac{821}{2}\right)u^5
+\cdots
\right],
\\[4pt]
e^{{\Lambda}}
&=
1-4u^2+\cdots,
\qquad
\mathsf{b}
=
1+\mathrm{b}_5 u^5+\cdots,
\qquad
h
=
\frac{4q_c^2}{|Q_k|^6}
\frac{16}{15}
\left(1-{b}_0^2\right)u^5
+\cdots .\\[4pt]
b_X
&=
\frac{2q_c}{3|Q_k|}
\left(
1-b_0-4b_0u-12b_0u^2+\cdots
\right),
\qquad
b_J
=
\frac{2q_c}{3|Q_k|}
\left(
-1+b_0+4b_0u+8b_0u^2+\cdots
\right),
\\[4pt]
a_J
&=
\frac{q_c}{6}
\left(
-1+2b_0+16b_0u+96b_0u^2+\cdots
\right).
\end{aligned}
\end{equation}
Note the appearance of the parameter $b_0$ in the leading orders of the components of $B_2$, namely $b_X$ and $b_J$. The remaining undetermined parameters in these expansions ---from which we are only showing $f_4$, $f_5$ and $\mathsf{b}_5$--- are fixed by a regularity condition in the bulk. The two regularity conditions we consider are
\begin{itemize}
    \item \textbf{There is a black brane horizon at} $u=\uH$, about which the functions appearing in the metric and the dilaton read
    \begin{equation}\label{eqBH_expansions}
\begin{array}{lll}
     e^{2f} = |Q_k|^2 \fH+\cdots\,, \quad
     & e^{2g} = |Q_k|^2 \gH+\cdots \,, \quad
     &e^{\Lambda} =  \lH + \cdots \,, 
     \\[2mm]
     \mathsf{b} = \bH (u-\uH)+\cdots \,,\quad 
      &\displaystyle h 
=
\frac{4q_c^2}{|Q_k|^6}
\frac{16}{15}
\left(1-{b}_0^2\right) \hH +\cdots \,. \\
\end{array}
    \end{equation}
    At the horizon, the fluxes $b_J$, $b_X$, $a_J$ reach finite values which will not play any role here. The full solutions are constructed numerically, by matching the two expansions~\eqref{eq:UV_expansions} and~\eqref{eqBH_expansions}, for example, by means of a \textit{shooting} method, as done in Ref.~\cite{Elander:2020rgv}. 
    \item \textbf{The eleven-dimensional uplift is regular at $u = \uS$}, and has $\mathsf{b} = 1$ (and thus, no horizon). In this case supersymmetry provides a system of first-order BPS equations, and some of the functions can be given in terms of hypergeometric functions, as summarized in Refs.~\cite{Faedo:2017fbv,GomezSubils:2021dzb}. We nonetheless decided to also construct them in a fully numerical way.

    The cases $b_0=0$ and $b_0=1$ are discussed separately. For $b_0\in (0,1)$, near the end of the space the metric functions read 
    \begin{equation}\label{eq:parameters_IR_B8}
\begin{array}{lll}
     \displaystyle e^{2f} = |Q_k|^2 \frac{\fs^2}{\uS^2}(\uS-u)+\cdots,\quad 
     & \displaystyle e^{2g} = \frac{|Q_k|^2}{\uS^4}(\us-u)^2+\cdots ,\quad
     &  \displaystyle e^{\Lambda} = \frac{1}{\uS^2}(\us-u)+\cdots ,\quad
     \\[4mm]
    \displaystyle h = \frac{4q_c^2}{|Q_k|^6}
 \frac{\hir}{\us-u}+\cdots\,,\quad&
       \displaystyle e^{2\Phi}= \frac{2q_c}{|Q_k|^3}\frac{\hir^{{1}/{2}}}{\uS^4}(\us-u)^{3/2}+\cdots\,. \\
\end{array}
    \end{equation}
    Note that the ten-dimensional warp-factor $h$ diverges at $u=\uS$ (or, equivalently, at $r=\rs=|Q_k|/\us$), inducing a singularity in the type~IIA geometries. This is regularized by resolving the M-theory circle in eleven dimensions (see Appendix~\ref{app11d}).
    
    More precisely, part of the internal geometry (containing the M-theory circle) shrinks smoothly in the IR, leading to a discrete spectrum in the dual gauge theories~\cite{Elander:2018gte}. Nonetheless, the non-trivial fibration and shrinking of the M-theory circle spoil confinement---in the sense of a linear growth of the quark-antiquark potential for large separations~\cite{Faedo:2017fbv}. This can be attributed to the presence of CS interactions, as we will explain later. 

    Even though there is a massless Goldstone mode due to a broken global symmetry \cite{Faedo:2025nqc}, we will refer to this phase as \textit{gapped}, due to its discrete spectrum. 
\end{itemize}

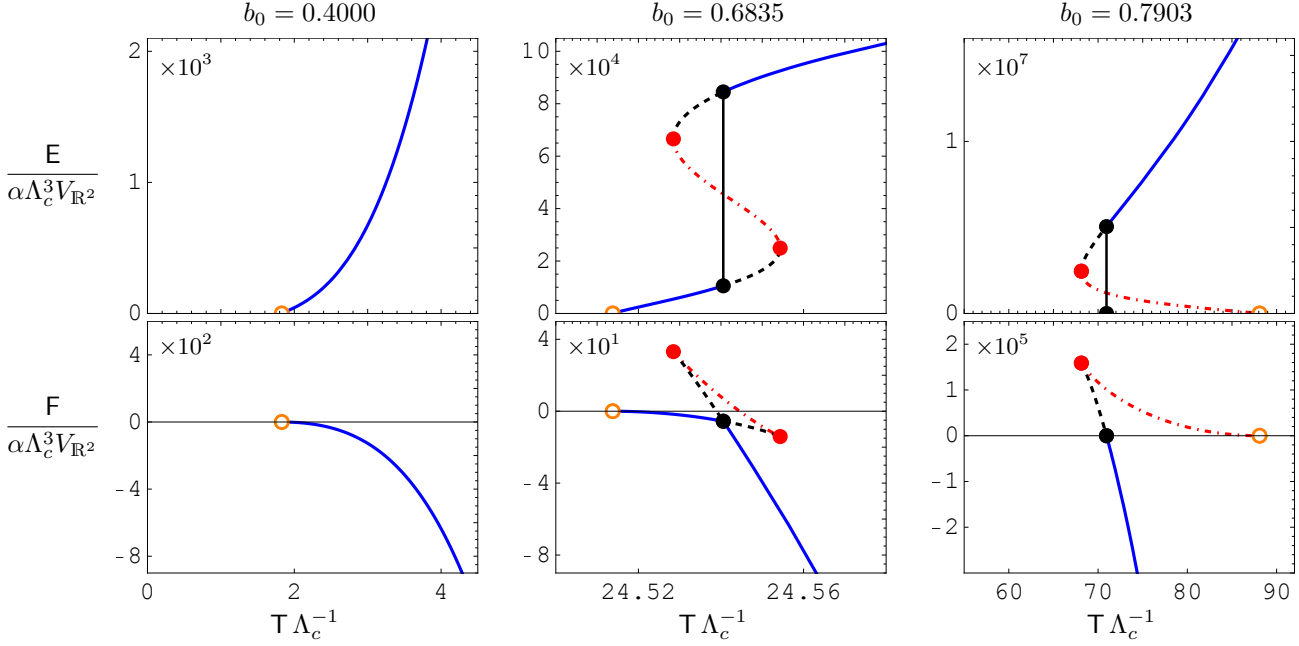
\begin{figure}[t]
	\begin{center}\noindent
    \includegraphics[width=0.325\textwidth]{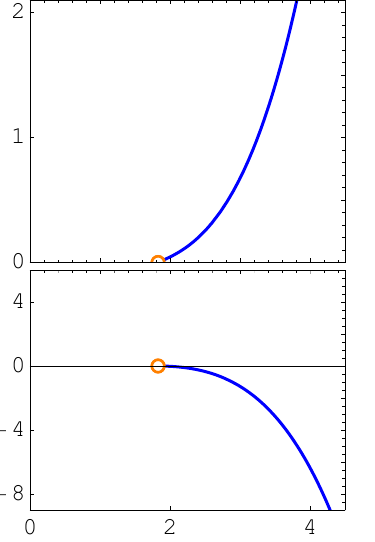} 
    \put(-190,160){\small $ \displaystyle \frac{\ene}{\alpha\Lt^3\vol}$}
    \put(-190,65){\small $ \displaystyle \frac{\free}{\alpha\Lt^3\vol}$}
    \put(-90,-10){\small $ \displaystyle \temp\, \Lt^{-1}$}
    \put(-100,220){\small $ \displaystyle b_0 = 0.4000$}
    \put(-132,200){\footnotesize $\times 10^{3}$}
    \put(-132,95){\footnotesize $\times 10^{2}$}
    \hfill
    \includegraphics[width=0.325\textwidth]{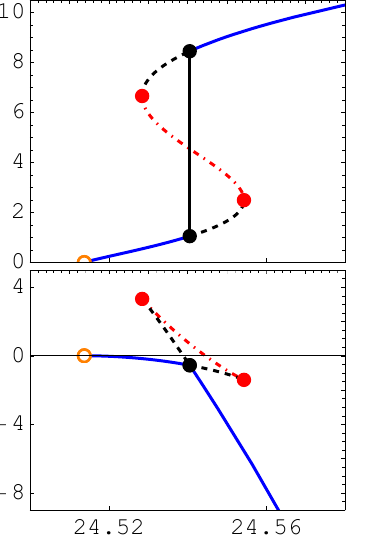}
    \put(-132,200){\footnotesize $\times 10^{4}$}
    \put(-132,95){\footnotesize $\times 10^{1}$}
    \put(-90,-10){\small $ \displaystyle \temp\, \Lt^{-1}$}
    \put(-100,220){\small $ \displaystyle b_0 = 0.6835$}
    \hfill
    \includegraphics[width=0.325\textwidth]{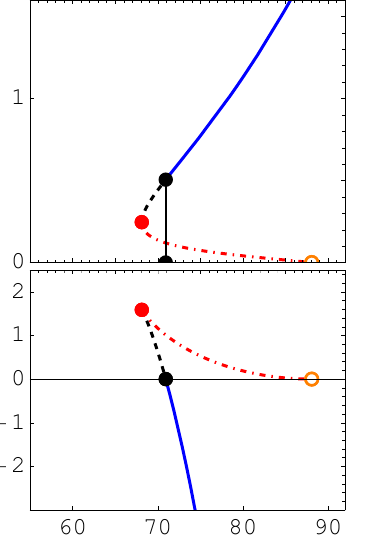} 
    \put(-132,200){\footnotesize $\times 10^{7}$}
    \put(-132,95){\footnotesize $\times 10^{5}$}
    \put(-90,-10){\small $ \displaystyle \temp\, \Lt^{-1}$}
    \put(-100,220){\small $ \displaystyle b_0 = 0.7903$}
	 \caption{\small Energy (top) and free energy (bottom) as a function of temperature for states in the theories with three representative values of $b_0$. Blue solid curves correspond to stable plasma phases. The horizontal $\ene=\free=0$ axis corresponds to the gapped phase. Phase transitions to this phase  happening at finite temperature and zero energy are depicted using hollowed orange circles. First-order PT between plasma phases, or between a plasma phase and a gapped phase, are represented by vertical solid black lines connecting the two available states at the critical temperature.
     The black dashed and red dot-dashed curves indicate metastable and dynamically unstable branches, respectively. Note the additional factors of ten for the vertical axes, indicated inside each plot. See Ref.~\cite{Elander:2020rgv} for details.}\label{fig.DifferentCasesThermalPT}
	\end{center}
\end{figure}
The parameters appearing in the asymptotic expansions~\eqref{eq:UV_expansions} contain information about the thermodynamic properties of each state. For instance, the total energy $\ene$ and free energy $\free$ read \cite{Elander:2020rgv}
\begin{equation}
\begin{aligned}
    \label{eq:energy_formula}
    \frac{\ene}{\vol
    }
    &=
    \alpha \Lt^3 \left(-\frac{411}{2} - 6 f_4 - 2 f_5 - \frac{7}{2} b_5\right)\,, \\[2mm] 
    \frac{\free}{\vol
    }&=
     \alpha \Lt^3\left(-\frac{411}{2} - 6 f_4 - 2 f_5 + \frac{3}{2} b_5\right)
    \,,
\end{aligned}
\end{equation}
with $\vol$ the volume of flat space along the gauge theory directions. Moreover, the temperature of each black brane solution is fixed to avoid a conical singularity in terms of horizon data,
\begin{equation}
    \temp = \Lt\left(-\frac{1}{4\pi}\frac{\bH \uH^2}{\sqrt{\hH}}\right)\,.
\end{equation}
In contrast, no regularity condition fixes the temperature of the ground-state solutions, which can therefore be considered to exist at any temperature.

The resulting finite temperature phase diagram is very rich. 
For non-vanishing but sufficiently small values of $b_0$, the temperature of the black brane solutions saturates to a finite value as the \textit{extremal} limit---meaning, zero energy limit--- is taken, as in Fig.~\ref{fig.DifferentCasesThermalPT}~(left). Below this temperature the theories transition to the ground state, which has $\ene=\free=0$ in our conventions. For an intermediate range of values of $b_0$, a phase transition between black brane solutions is found before the extremal limit is reached, see Fig.~\ref{fig.DifferentCasesThermalPT}~(middle). Finally, for sufficiently large values of $b_0$, there is a first-order phase transition between black brane solutions and the ground state geometries, as in Fig.~\ref{fig.DifferentCasesThermalPT}~(right). Further details can be found in Ref.~\cite{Elander:2020rgv}.

For later convenience, let us emphasize that the leading order of Eq.~\eqref{eq:UV_expansions},
\begin{equation}\label{eq:D2metric}
\begin{aligned}
    e^{2f} &= 2 e^ {2g} = \frac{|Q_k|^2}{2u^2}\,,\qquad e^\Lambda = 1\,, \qquad h = \frac{4q_c^2}{|Q_k|^6}
\frac{16}{15}
\left(1-{b}_0^2\right)u^5\,,\qquad \mathsf{b}= 1+\mathsf{b}_5{u^5}\,, \\[2mm]
    b_X&= -b_J = \frac{2q_c}{3|Q_k|}(1-b_0)\,,\qquad a_J = \frac{q_c}{6}(-1+2b_0)\,.
\end{aligned}
\end{equation}
is nothing but the metric of a black D2-brane (we kept the next to-leading order term in the function $\mathsf{b}$ to account for the temperature). Note, however, that it is not an \textit{exact solution} to the equations of motion, but an approximate one at high energies.

We conclude this section with a comment on the two special limiting values $b_0=0$ and $1$. In the limit $b_0=1$, the theory becomes confining, in the sense that the potential between probe charges grows linearly at large separations (this is calculated using \cite{Maldacena:1998im, Sonnenschein:1999if, Nunez_2010}). Its UV expansion, shown in Eq.~\eqref{eq:UV_expansions_conf}, is analogous to Eq.~\eqref{eq:UV_expansions}. A crucial difference, however, is that this solution has $Q_k=0$, so the dual theory lacks CS interactions. Hence, the $\Be$ family realizes a natural field-theory expectation: CS interactions give gauge bosons an effective mass, which screens color charges \cite{Dunne:1998qy,Karabali:1999ef}; when these interactions are absent, the theory becomes confining instead. To recover this geometry in the $b_0\to1$ limit, $Q_k$ must be rescaled appropriately; see Eq.~\eqref{eq:rescaling_Qk}.

In the opposite limit $b_0 = 0$, we recover the $\Binf$ geometry, which describes a theory flowing to the Ooguri-Park (OP) CFT \cite{Ooguri:2008dk} in the IR. This fixed point, in turn, corresponds to the UV of a theory that flows to the gapped IR and whose dual geometry is dubbed $\BOP$. Theories with $b_0\gtrsim0$ develop ``walking dynamics'':
they resemble $\Binf$ until some intermediate energies in which a quasi-conformal regime emerges. Eventually, they depart from this regime, mimicking the $\BOP$ theory.

Exactly at the fixed point we have  the OP CFT. The ground state is dual to  AdS$_4$ times a \textit{squashed} seven-sphere
(orbifolded by $\mathds{Z}_k$)---in contrast to ABJM, where the sphere is round. The OP background preserves two supercharges (plus the usual enhancement at the fixed point). In general, the backgrounds in Eqs.~\eqref{10Dansatz}-\eqref{eq:UV_expansions}
preserve two supercharges or ${\cal N}=1$ SUSY in $(2+1)$-dimensions.

Finite temperature states at this fixed point are realized holographically by the AdS-Schwarzschild black branes
\begin{equation}\label{eq:OPmetric}
\begin{aligned}
    e^{2f} &= 5 e^ {2g} = \frac{9}{5}{r^2}\,,\qquad e^\Lambda = \frac{3r}{5|Q_k|}\,, \qquad h = \frac{500\, q_c^2 }{2187 |Q_k|^ 2}\, \frac{1}{r^4}\,,\qquad \mathsf{b}= 1-\frac{\rH^3}{r^3}\,, \\[2mm]
    b_X&= -b_J = \frac{2q_c}{3|Q_k|}\,,\qquad a_J = -\frac{q_c}{6}\,.
\end{aligned}
\end{equation}
We will use these expressions to show analytic results in this limit.

\section{Complexity of thermal states}

\label{sec:CV}

\subsection{General formulae}

In this section we exploit the ``complexity =  volume'' (CV) proposal, which is expected to capture the quantum circuit complexity of a boundary state. Let us review how the computation goes. We begin by considering the black brane solutions dual to plasma states of the dual quantum field theory. In Einstein frame, the black brane ten-dimensional ansatz~\eqref{10Dansatz} reads
\begin{equation}
\label{eq:10Dansatz_Einstein}\begin{aligned}
\dd s_{\text{\tiny E}}^2 &=e^{-\Phi/2}\left[h^{-\frac12}\left(-\mathsf{b}\dd t^2 + \dd x_1^2 + \dd x_ 2^2\right)+h^{\frac12} \left(\frac{\dd r^2}{\mathsf{b}}+e^{2f}\dd\Omega_4^2+e^{2g}\left[\left(E^1\right)^2+\left(E^2\right)^2\right] \right)\right]\,.
\end{aligned}
\end{equation}
For the CV calculation, we will look for codimension-one extremal wormholes anchored at the two asymptotic boundaries at equal times. The derivation is analogous to that of Refs.~\cite{Carmi:2017jqz, Stanford:2014jda}, but now with the appropriate metric \eqref{eq:10Dansatz_Einstein}.

To handle the relevant wormhole slices we first need to perform a change of coordinates involving  $(t,r)$, in such a way that the black brane interior can be accessed. Demanding that the  $(t,r)$ part of the geometry 
\begin{equation}\label{eq:t_t_part_1}
\dd s_{(t,r)}^2
=
e^{-\frac\Phi2}\left(
-h^{-\frac12}\,\mathsf{b}\, \dd t^2+h^{\frac12}\,{\mathsf{b}^{-1}}\,{\dd r^2}
\right).
\end{equation}
becomes conformally flat, we define the tortoise coordinate $r_\ast$ as
\begin{equation}
\dd r_\ast={h^{\frac12} }\,{\mathsf{b}^{-1}}\,\dd r.
\label{tortoise}
\end{equation}
in terms of which Eq.~\eqref{eq:t_t_part_1} becomes 
\begin{equation}
\dd s_{(t,r)}^2
=
e^{-\Phi/2}h^{-\frac12} \mathsf{b} \,(- \dd t^2+ \dd r_\ast^2).
\end{equation}
Now, the ingoing Eddington-Finkelstein coordinate $v=t+r_\ast$ allows us to write
\begin{equation}
\dd t=\dd v- \dd r_\ast= \dd v-{h^{\frac12}}{\mathsf{b}^{-1} } \dd r\,,
\end{equation}
in such a way that the full metric \eqref{eq:10Dansatz_Einstein} in ingoing EF coordinates becomes
\begin{equation}
    \dd s_{\rm E}^2 =e^{-\Phi/2}\left[h^{-\frac12}\left(- \mathsf{b}\,\dd v^2+2h^{\frac12}\,\dd v\,\dd r + \dd x_1^2 + \dd x_ 2^2\right)+h^{\frac12} \left(e^{2f}\dd\Omega_4^2+e^{2g}\left[\left(E^1\right)^2+\left(E^2\right)^2\right] \right)\right]\,,\\
\end{equation}

The wormholes we consider (see Eq.~\eqref{eq:definition_complexity}) are codimension-one surfaces $\WH$ anchored at the same boundary time, $(t_L=t_R)$. They wind entirely around the internal geometry, extend along the two gauge theory directions $x_1$ and $x_2$, and have a non-trivial profile in the $(v,r)$ plane, specified by 
\begin{equation} v=v(\parwh),\qquad r=r(\parwh). \end{equation}
Here $\parwh$ is an arbitrary parameter along $\WH$. We denote by
$\vol$ the volume of the gauge theory directions, and by $\volcp = 32\pi^3/3$ the volume of $\CP^3$. Consequently, the volume of $\WH$ which we are instructed to extremize is
\begin{equation}
\label{eq:volume_to_minimise}
\begin{aligned}
   V= \int_{\WH}*1
&=
\vol\volcp \int \dd \parwh \, e^{-9\Phi/4}h^{3/4}e^{4f+2g}
\sqrt{-\mathsf{b}\,\dot v^{\,2}+2 h^{1/2}\,\dot v\,\dot r} \\
&= \vol\volcp \int \dd \parwh \, \W\,\sqrt{-\mathsf{b}\,\dot v^{\,2}+2 h^{1/2}\,\dot v\,\dot r},
\end{aligned}
\end{equation}
where dots denote derivatives with respect to $\parwh$ and we have defined $\W \equiv e^{-9\Phi/4}h^{3/4}e^{4f+2g} $ for  ease of notation.

Note that the integrand
\begin{equation}
    \mathcal{L} = \W\,\sqrt{- \mathsf{b} \,\dot v^{\,2}+2 h^{1/2}\,\dot v\,\dot r}
\end{equation}
does not depend explicitly on $v$, which implies that there is a conserved quantity,
\begin{equation}
\mathcal E
\equiv
-\,\frac{\partial \mathcal L}{\partial \dot v} = W\,
\frac{\mathsf{b}\,\dot v-h^{1/2}\,\dot r}
{\sqrt{-\mathsf{b}\,\dot v^{\,2}+2h^{1/2}\,\dot v\,\dot r}}.
\label{conserved-energy}
\end{equation}
Moreover, the functional is reparametrization invariant, and therefore we are free to choose the gauge in which the integrand is simply
\begin{equation}
W\sqrt{- \mathsf{b}\,\dot v^{\,2}+2h^{1/2}\,\dot v\,\dot r}=1.
\label{gauge-choice}
\end{equation}
With this choice Eqs.~\eqref{conserved-energy} and \eqref{gauge-choice} can be combined to obtain an expression for $\dot{r}$ that does not involve $\dot{v}$,
\begin{equation}
\dot r^{\,2}
=
\frac{\mathcal E^2+ \mathsf{b}\,W^2}{h\,W^4}.
\label{rdot-eq}
\end{equation}
This has a nice geometrical consequence: the extremal surface reaches a turning point where
\begin{equation}
\dot r\big|_{r=r_m}=0
\qquad\Longrightarrow\qquad
\mathcal E^2+\mathsf{b}(r_m)\,W(r_m)^2=0\,,
\label{turning-point}
\end{equation}
which is possible only for $\mathsf{b}(r_m)<0$, implying that such a turning point lies inside the horizon. 
This also means that $\mathcal{E}\leq0$, see Eq.~\eqref{conserved-energy}. Moreover, the absolute value of $\mathsf{b}(r)\,W(r)^2$ develops a maximum at $r=\rsat$ inside the horizon, which is a limiting spacelike slice that the wormhole does not penetrate. As a consequence,
\begin{equation}\label{eq:rangeE}
    -|b(\rsat)|^{\frac{1}{2}}W(\rsat)\leq\mathcal{E}\leq0\,.
\end{equation}

Next, we use Eq.~\eqref{rdot-eq} to write Eq.~\eqref{eq:volume_to_minimise} as
\begin{equation}\label{eq:volume_E}
    {V(\mathcal{E})} =  2 \volcp \vol \int_{r_m}^{r_{\max}}
\dd r\,
\frac{h^{1/2}\,W^2}
{\sqrt{\mathcal E^2+\mathsf{b} \,W^2}}\,.
\end{equation}
Note that we restrict ourselves to the $\dot r >0$ branch, taking advantage of the symmetry of the problem. Moreover, the upper limit $r_{\max}$ has been introduced to regulate the divergence of the integral close to the boundary. To get a finite result when $r_{\max}\to\infty$, we consider the difference in volumes\footnote{Alternatively, we could have analyzed the divergent structure of $V(\mathcal{E})$ and introduced finite counterterms.} between the wormhole whose conserved quantity is $\mathcal{E}$ and the one with $\mathcal{E} = 0$ (for which $r_m=\rH$, see Eq.~\eqref{turning-point}). From Eq.~\eqref{eq:definition_complexity}, the complexity is then given by
\begin{eqnarray}\label{eq:complexity_formula}
\CCV(\mathcal{E}) &\equiv& {\frac{V(\mathcal{E})-V({0})}{G_{10} \lm}} \nonumber\\[2mm]
    &=&\frac{2 \volcp \vol}{G_{10}\lm} \left[\int_{\rH}^{\infty}
\dd r\left(
\frac{h^{1/2}\,W^2}
{\sqrt{\mathcal E^2+\mathsf{b} \,W^2}}-
\frac{h^{1/2}\,W}
{\sqrt{\mathsf{b}}}
\right)
+\int_{r_m}^{\rH}
\dd r\,
\frac{h^{1/2}\,W^2}
{\sqrt{\mathcal E^2+\mathsf{b} \,W^2}}
\right],
\end{eqnarray}
with $\lm$ a convenient length scale to make $\CCV$ dimensionless. In our conventions, the ten dimensional Newton's constant is given by
\begin{equation}\label{eq:Newton_10}
    16\pi G_{10} = (2\pi)^7g_s^2\ell_s^8.
\end{equation}
In order to be able to express the result purely in terms of gauge theory quantities we choose 
\begin{equation}
    \lm=g_s^{{1}/{4}}\ell_s\,.
\end{equation}

Now, to find the time $t_R(\mathcal{E})$ at which the embedding touches the boundary, let us use Eqs.~\eqref{conserved-energy} and~\eqref{rdot-eq} to write
\begin{equation}
\frac{\dd v}{\dd r}
=
\frac{h^{1/2}}{\textsf{b}}
\left(
1+\frac{\mathcal E}{\sqrt{\mathcal E^2+\mathsf{b}\,W^2}}
\right).
\label{dvdr}
\end{equation}
Since $t=v-r_\ast$ (i.e. $t+r_\ast = v$), the time $t_R(\mathcal E)$ can be determined from
\begin{equation}\label{tR-with-cutoff}
t_R(\mathcal E)+r_\ast(r_{\max})-r_\ast(r_{m})=
\int_{r_m}^{r_{\max}}\dd r\,
\frac{h^{1/2}}{\mathsf{b}}
\left(
1+\frac{\mathcal E}{\sqrt{\mathcal E^2+\mathsf{b}\,W^2}}
\right)\equiv I(\mathcal{E}).
\end{equation}
In this case, the integral $I(\mathcal E)$ on the right hand side is perfectly finite, even if $r_{\max}\to\infty$. At the horizon, the simple pole of the denominator of the integrand originated by $\mathsf{b}$ cancels with a simple zero that the piece inside the parenthesis develops at $r=\rH$ (recall that $\mathcal{E}<0$). At the boundary, the integrand vanishes sufficiently fast. However, computing the difference in the tortoise coordinate on the left hand side requires integrating Eq.~\eqref{tortoise} across the horizon, which is problematic since the integrand contains a simple pole at the horizon. For this reason, we consider time differences instead. Choosing two different values of the conserved quantity, ${\mathcal E}^{(1)}$ and ${\mathcal E}^{(2)}$, we define
\begin{equation}
    \Delta t_R({\mathcal E}^{(1)},{\mathcal E}^{(2)})\equiv t_R({\mathcal E}^{(1)})- t_R({\mathcal E}^{(2)})=r_*(r_m^{(1)}) - r_*(r_m^{(2)}) + I(\mathcal{E}^{(1)}) - I(\mathcal{E}^{(2)})\,.
\end{equation}
Note that in the time difference the contribution from $r_\ast(r_{\max})$ disappears. In practice, it is useful to fix $\mathcal{E}^{(2)}$ to some prescribed value and compute the time difference with respect to it. At the end of the computation, we make a shift to all our solutions with effectively fixes $\mathcal{E}^{(2)} = 0$, and we are left with
\begin{equation}\label{eq:time_difference}
    \tau(\mathcal{E}) \equiv \Delta t_R({\mathcal E},0) = \Delta t_R({\mathcal E},{\mathcal E}^{(2)}) +\Delta t_R({\mathcal E}^{(2)},0) >0\,.
\end{equation}
which is the quantity that we will show in the plots.
\begin{figure}[t]
	\begin{center}
    \includegraphics[width=0.55\textwidth]{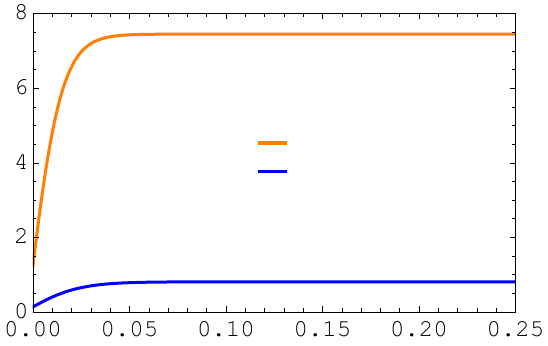} 
   \put(-300,170){\small $\displaystyle  \frac{1}{\alpha^{9/8}  \Lambda_c }\times\frac{1}{\vol}\, \frac{\dd \CCV}{\dd \tau}\times 10^{-5}$}
    \put(-130,-15){\small $\tau \, \Lt $}
    \put(-112,90){\small $1.68\times 10^4\,  \alpha\Lt^3\vol $}
    \put(-112,75){\small $2.18\times 10^3\,  \alpha\Lt^3\vol $}
	\caption{\small 
    Complexity growth as a function of time for two different black brane solutions whose energy is indicated in the plot, for the theory with $b_0 = \frac25$.}\label{fig:complexity_time_generic}
	\end{center}
\end{figure}
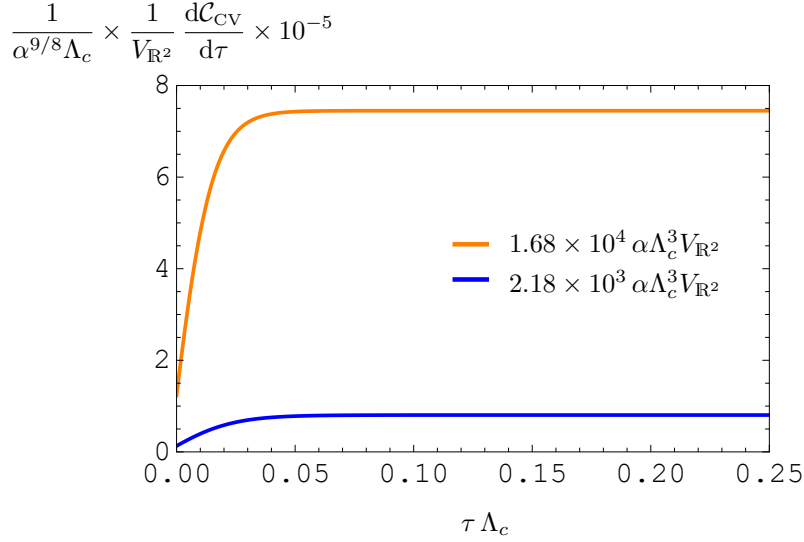

At this point, the time evolution of the complexity is determined by Eqs.~\eqref{eq:complexity_formula} and~\eqref{eq:time_difference}, which can be used to construct parametric plots. Instead, let us focus on the complexity growth, for which a simple expression in terms of $\mathcal{E}$ can be obtained. Indeed, returning to Eq.~\eqref{eq:volume_E} and using Eq.~\eqref{tR-with-cutoff}, we can write 
\begin{equation}
    \frac{V(\mathcal{E})}{2 \volcp \vol} =  \int_{r_m}^{r_{\max}}
\dd r\,
\frac{h^{1/2}}{\mathsf{b}}
\left(
{\sqrt{\mathcal E^2+\mathsf{b} \,W^2} + \mathcal{E}}\right) \, - \,  \mathcal{E} \cdot\Big(t_R(\mathcal E)+r_\ast(r_{\max})-r_\ast(r_{m})\Big)\,.
\end{equation}
Given that $\mathcal{E}$ is constant on $\WH$, the chain rule leads to
\begin{eqnarray}
    \frac{1}{2 \volcp \vol} \frac{\dd V(\mathcal{E})}{\dd t_R}&=&  \frac{\dd\mathcal{E}}{\dd t_R}\int_{r_m}^{r_{\max}}
\dd r\,\frac{h^ {1/2}}{\mathsf{b}}
\left(\frac{\mathcal{E}}
{\sqrt{\mathcal E^2+\mathsf{b} \,W^2}}+1\right)\nonumber\\[2mm]
&&- \frac{\dd\mathcal{E}}{\dd t_R} \cdot\Big(t_R(\mathcal E)+r_\ast(r_{\max})-r_\ast(r_{m})\Big)-\mathcal{E}\,.
\end{eqnarray}
The two first terms cancel due to Eq.~\eqref{tR-with-cutoff}. Since $t_R$ and $\tau$ are related by a constant shift,
\begin{equation}
\frac{\dd \CCV}{\dd \tau} = - \frac{2 \volcp \vol}{{G_{10}\lm}}\mathcal{E}\,.
\end{equation}

We plot this quantity in Fig.~\ref{fig:complexity_time_generic} for a particular choice of $b_0$, in terms of the gauge theory quantities defined in Eq.~\eqref{eq:def_Lt_alpha}. We see that the time derivative of the complexity is monotonically increasing and saturates to a constant value at late times. This value to which it saturates is the quantity we will examine. Note that this late-time growth of the complexity ---which we denote by $\growth$--- is given by the maximum possible value of $|\mathcal{E}|$, obtained at $r_m  =\rsat$ (see Eq.~\eqref{eq:rangeE}),
\begin{equation}\label{eq:formula_growth}
    \growth = \lim_{\tau\to\infty} \frac{\dd \CCV}{\dd\tau} = 
    \frac{2 \volcp \vol}{{G_{10}\lm}}\, |\mathsf{b}(\rsat)|^ {{1}/{2}}\,W(\rsat)
    \,.
\end{equation}

\subsection{Late time growth of complexity}

\begin{figure}[t]
	\begin{center}
    \includegraphics[width=0.55\textwidth]{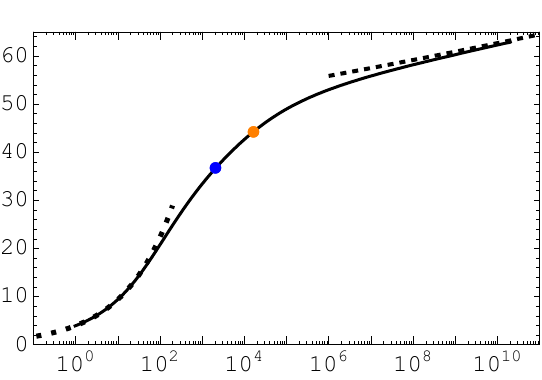} 
    \put(-305,95){$\displaystyle\frac{\growth}{\ene}\times\alpha^{-1/8}$}
    \put(-140,-20){$ \displaystyle \frac{\ene}{\alpha\Lt^3\vol}$}
	\caption{\small 
    Complexity growth normalized by the energy as a function of the energy for the theory $b_0 = \frac{2}{5}$. For illustration, the  blue and orange dots correspond to the values extracted from the late-time growth for the two states shown in Fig.~\ref{fig:complexity_time_generic}. The solid curve shows the data computed numerically. The black dashed curve is obtained by considering only the leading order in the UV region, where the geometry becomes that of a D2-brane, Eq.~\eqref{eq:GoverE_D2}. The black dotted curve is the approximation at low energies, obtained by fitting the last five computed points, Eq.~\eqref{eq:fit_low_energies}.}
    \label{fig:complexity_growth}
	\end{center}
\end{figure}

As it was found in Refs.~\cite{Stanford:2014jda,Carmi:2017jqz}, the late-time growth of circuit complexity is proportional to the energy of the system in a CFT.
We can check that this is indeed the case in the OP CFT, whose dual geometry was given in Eq.~\eqref{eq:OPmetric}. In this case,
\begin{equation}
    \mathsf{b}(r)\,W(r)^2 =q_c^{3/4}\lvert Q_k\rvert^{15/4}\cdot \frac{
81\cdot 2^{3/4}\cdot 3^{7/8}\,
}{
25\cdot 5^{3/8}
}\cdot r^{3}\left(r^{3}-\rH^{3}\right)
\end{equation}
which has a maximum at $\rsat = \rH/2^{{1}/{3}}$. Then, the late-time growth of the complexity \eqref{eq:formula_growth} on the fixed point is
\begin{equation}
    \growth_{\text{OP}} = \frac{\vol}{G_{10}\lm}q_c^{3/8}|Q_k|^{15/8}\times
    \frac{96\cdot 2^{3/8}\cdot 3^{7/16}\,\pi^{3}}
{5\cdot 5^{3/16}}\,\rH^{3}\,.
\end{equation}
In addition, the energy in this case is essentially that of a black brane in AdS$_4$,
\begin{eqnarray}
    \frac{\ene_{\text{OP}}}{\vol} = \frac{108 \pi^{2}}{25}\times \frac{ |Q_k|^{2}\rH^{3}}
{G_{10}}\,
\end{eqnarray}
in our conventions. As a consequence, the ratio between the growth and the energy,
\begin{equation}\label{eq:GoverE_OP}
    \frac{\growth_{\text{OP}}}{\ene_{\text{OP}}} = \frac{8\cdot 2^{3/8}\cdot 5^{13/16}\,\pi}
{3\cdot 3^{9/16}} \, \frac{q_c^{3/8}}{\lm|Q_k|^{1/8}}\simeq 21.65  \, \frac{q_c^{3/8}}{\lm|Q_k|^{1/8}} = 57.10 \,\alpha^{1/8}\,,
\end{equation}
is constant. Note that in the last equality we have translated to gauge theory quantities using Eqs.~\eqref{eq:constants_Qk_qc},~\eqref{eq:tHooft} and~\eqref{eq:def_Lt_alpha}, together with Eq.~\eqref{eq:Newton_10}. We utilize these same equations to present our results in what follows.

A similar analysis can be done considering the asymptotic D2-geometry~\eqref{eq:D2metric}. With this background (recall that $r = |Q_k|/u$), we obtain
\begin{equation}
    \mathsf{b}(r)\,W(r)^2 = \frac{q_c^{3/4}}{\lvert Q_k\rvert^{3/8}}\cdot\frac{
\left(1-b_0^{2}\right)^{3/8}
}{
32\cdot 2^{3/4}\cdot 15^{3/8}
}\cdot r^{41/8}
\left(r^5 + \mathsf{b}_5\lvert Q_k\rvert^{5}\right)\,,
\end{equation}
for which the maximum is at
\begin{equation}
    \rsat = \frac{41^{1/5}\,|\mathsf{b}_5|^{1/5}}{3^{4/5}}\,\lvert Q_k\rvert\,,
\end{equation}
resulting in
\begin{equation}
    \label{eq:growth_D2}
    \growth_{\text D2} =
    \frac{\vol}{G_{10}\lm}
   q_c^{3/8}\lvert Q_k\rvert^{39/8}\cdot 
    \frac{
16\cdot 2^{5/8}\cdot 5^{5/16}\cdot 41^{41/80}\,\pi^{3}
\left(1-b_0^{2}\right)^{3/16}
}{
243\cdot 3^{19/80}
}|\mathsf{b}_5|^{81/80}\,.
\end{equation}
To obtain the ratio between this growth and the energy, we can use Eq.~\eqref{eq:energy_formula} ignoring the contribution from $f_4$ and $f_5$, in which case we approximate
\begin{equation}
    \mathsf{b}_5
=
-\frac{411}{7} -\frac{2}{7} \cdot \frac{\ene}{\alpha\Lt^3\vol}\,.
\end{equation}
Then, expanding for large energies, we conclude
\begin{equation}\label{eq:GoverE_D2}
    \frac{\growth_{\text{D2}}}{\ene_{\text{D2}}} =
\frac{512\cdot \,2^{2/7}\,5^{5/16}\,41^{41/80}\,
(1-b_0^2)^{3/16}\,
\pi^{3/2}}
{567\,3^{9/80}}\, \alpha^{1/8}
\left(\frac{\ene}{\alpha\Lt^3\vol}\right)^{1/80}
+O\left(\frac{1}{\ene^{79/80}}\right)\,.
\end{equation}

Away from these approximations, the total energy is given in Eq.~\eqref{eq:energy_formula} in terms of boundary data, while the growth is given by Eq.~\eqref{eq:formula_growth}, with $\rsat$ found numerically. The ratio of these quantities is depicted in Fig.~\ref{fig:complexity_growth}, where a non-trivial dependence is observed. At low energies the ratio vanishes, which is a manifestation that theory is developing a gap. For instance, fitting the last numerical data points computed for $b_0=\frac{2}{5}$ we find
\begin{equation}
    \label{eq:fit_low_energies}
    \frac{\growth}{\ene}\simeq 1.367\,  \alpha^{1/8}\left(\frac{\ene}{\alpha\Lt^3\vol}\right)^{0.377}\,.
\end{equation}
At high energies, in contrast, the ratio $\growth/\ene$ grows without bound as dictated by Eq.~\eqref{eq:GoverE_D2}. 

This behavior is rather generic, and in most of the cases (except for small values of $b_0$), the ratio $\growth/\ene$ increases monotonically with the energy as we will see next. Having understood the general behavior, let us now discuss how $\growth$ probes the different IR regimes of theories with multi-scale dynamics.

\subsection{Walking dynamics of complexity growth}

\begin{figure}[t]
	\begin{center}
    \includegraphics[width=0.49\textwidth]{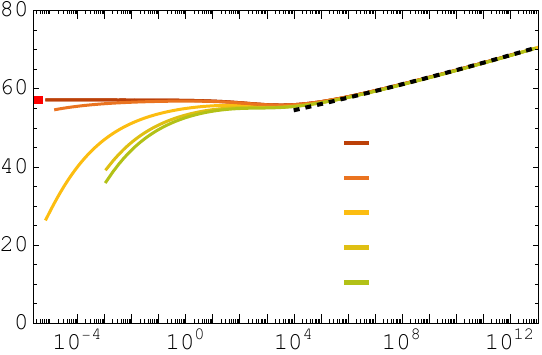} 
   \put(-220,155){\small $\displaystyle\frac{\growth}{\ene}\times\alpha^{-1/8}$}
    \put(-130,-18){\small  $ \displaystyle \frac{\ene}{\alpha\Lt^3\vol}$}
    \put(-60,85){\small $b_0 = 0 \, (\Be^\infty)$}
    \put(-60,70){\small $b_0 = 0.0191$}
    \put(-60,55){\small $b_0 = 0.0605$}
    \put(-60,40){\small $b_0 = 0.0808$}
    \put(-60,25){\small $b_0 = 0.0876$}
    \hfill
    \includegraphics[width=0.49\textwidth]{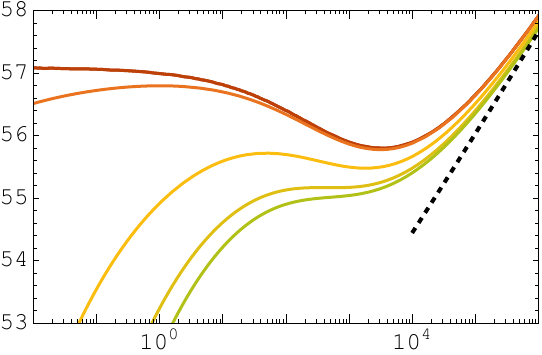} 
      \put(-220,155){\small $\displaystyle\frac{\growth}{\ene}\times\alpha^{-1/8}$}
    \put(-130,-18){\small $ \displaystyle    \frac{\ene}{\alpha\Lt^3\vol}$}
	\caption{\small 
    Complexity growth for different values of $b_0$ in the walking region. The red rectangle on the left indicates the value at OP, computed analytically in Eq.~\eqref{eq:GoverE_OP}. 
    The black dashed curve is obtained by considering only the leading order in the UV region, where the geometry becomes that of a D2-brane, Eq.~\eqref{eq:GoverE_D2} (in this plot we set $b_0=0$ for this approximation).}\label{fig:complexity_growth_OP}
	\end{center}
\end{figure}
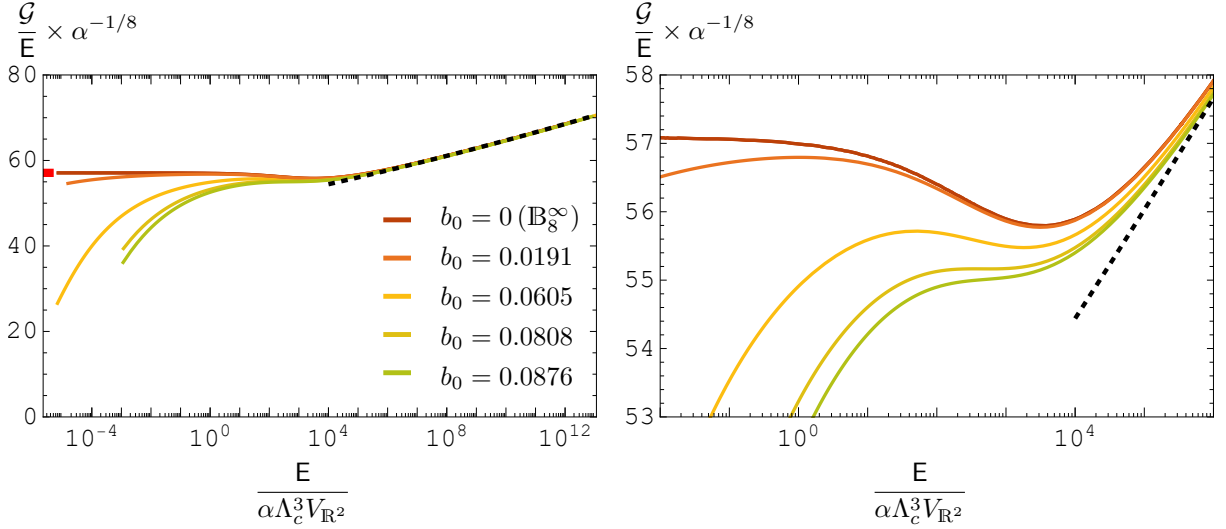

In the limit $b_0\to0$, corresponding to the leftmost part of Fig.~\ref{fig:triangle}, the theories flow close to the OP CFT. The complexity growth to energy ratio for theories in this regime is shown in Fig.~\ref{fig:complexity_growth_OP}. The effect of the nearby CFT is manifested in the flattening of the ratio over an increasingly wide range of scales as $b_0$ approaches zero. At high energies, the metrics follow the D2-brane behavior dictated by Eq.~\eqref{eq:GoverE_D2}, until the effect of the new IR scales takes over as the energy is lowered. For the $\Be^\infty$ metric ---which has $b_0=0$ and whose IR is precisely the OP fixed point--- a minimum appears, after which $\growth/\ene$ rises again towards the value that it has to reach at the CFT, given by Eq.~\eqref{eq:GoverE_OP}. 

Theories flowing close to the CFT are very similar to $\Be^\infty$ above the energy corresponding to the approximate scale invariance. Thus, the corresponding curve follows the trend dictated by $\Be^\infty$ until the effect of the mass gap develops. This happens at lower and lower energies, as $b_0$ is taken to zero. We believe that in all cases $\growth/\ene$ vanishes with the energy whenever the mass gap is present, even though it is numerically challenging to access this regime for flows very close to the CFT. In a way, the ``walking dynamics'' introduce a physical hierarchy that the numerical code needs to resolve. 

\subsection{Complexity growth near confinement}

\begin{figure}[t]
	\begin{center}
    \includegraphics[width=0.50\textwidth]{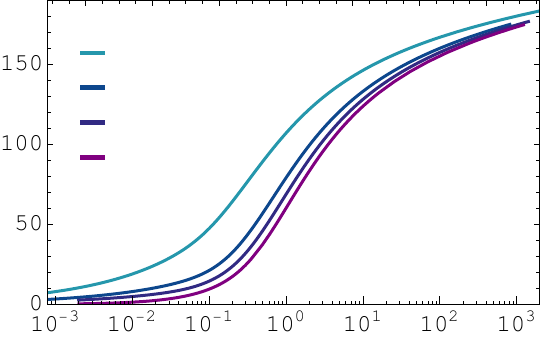} 
   \put(-220,155){\small $\displaystyle\frac{\growth}{\ene}\times\frac{\lm |Q_k|^{{1}/{8}}}{q_c^{{3}/{8}}}$}
    \put(-138,-18){\small $ \displaystyle \ene \, \times\,  \frac{G_{10}}{|Q_k|^5\vol}$}
    \put(-180,117){\small $b_0 = 0.7902$}
    \put(-180,102){\small $b_0 = 0.9201$}
    \put(-180,87){\small $b_0 = 0.9640$}
    \put(-180,72){\small $\Bconf$}
	\caption{\small Complexity growth for different values of  $b_0$ approaching the confining geometry.
    In this plot, $|Q_k|$ has been rescaled according to Eq.~\eqref{eq:rescaling_Qk}, so that $\Bconf$ is recovered in the limit $b_0\to1$.}\label{fig:complexity_growth_confinement}
	\end{center}
\end{figure}
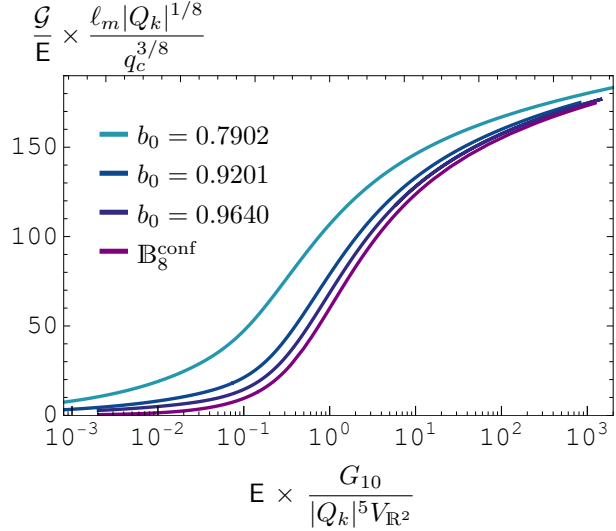

In the opposite limit, $b_0\to 1$, the geometries approach a confining geometry. As shown in Fig.~\ref{fig:complexity_growth_confinement}, the curves for $\growth/\ene$ also approach that of the confining geometry. Nonetheless, the presence of a confining IR does not introduce any qualitatively different feature in the complexity: the ratio still approaches the D2 brane regime at high energies and vanishes at low energies. In a way the presence of the horizon washes out the properties of the ground-state.

Actually, recall that above a certain value of $b_0$, the thermal phase diagram becomes very rich, with metastable and dynamically unstable branches ---see Fig.~\ref{fig.DifferentCasesThermalPT}. 
For instance, the curves depicted in
Fig.~\ref{fig:complexity_growth_confinement}
correspond to values of $b_0$ for which a first order phase transition is present, implying that these curves contain states that are thermodynamically metastable (at intermediate energies); and unstable (at low energies),
see Fig.~{\ref{fig.DifferentCasesThermalPT}}.
Yet the CV growth-to-energy ratio lacks any distinctive features regarding the thermal structure of the system, as the monotonic decrease is qualitative the same as the one found in Figs.~\ref{fig:complexity_growth} and~\ref{fig:complexity_growth_OP}, in which all the states considered are stable. Put differently, our result is that ${\growth}/{\ene}$ shows no universal, sharp signature of thermal phase transitions \emph{when regarded as an observable of individual black-brane solutions parametrized by their energy} (i.e., microcanonical ensemble). It should not be read as a claim that the equilibrium CV growth ---which instead, would be computed considering only the globally preferred saddle at a given fixed temperature (canonical ensemble)--- is insensitive to phase transitions. In this same sense, the CV growth-to-energy ratio does not distinguish whether the transition to the horizonless solution ---happening in the limit $\ene\to 0$--- is reached along a stable branch.

\section{Complexity of locally excited states}
\label{sec:Krylov}

We now turn from the computational complexity of thermal thermofield-double states to the Krylov spread complexity of localized excitations above the vacuum.  These are distinct state-dependent constructions: the CV observable of Section~\ref{sec:CV} is encoded in a maximal hypersurface of a black-brane geometry, whereas the observable studied in this section---first defined in Ref.~\cite{Balasubramanian:2022tpr}---is inferred from the proper radial momentum of a probe in the corresponding ground-state geometry.  We therefore do not expect a numerical equality between them.  The purpose of the comparison is instead to determine which features of the common RG flow survive these two different bulk filters.  In particular, we find below that both observables detect the near-conformal walking region, while the local probe is substantially more sensitive to the infrared cap and to the nonuniform confining limit.

We follow Refs.~\cite{Caputa:2024sux, Fan:2024iop, He:2024pox, Li:2025fqz}, where the holographic quantity dual to Krylov spread complexity was proposed. Essentially, the idea is to consider the geodesic motion of a massive particle, following a radial trajectory in $r = r(t)$ in the $\Be$ geometries \eqref{10Dansatz}. Moreover, because we are now inspecting the zero temperature ground states, we set $\mathsf{b}=1$.

We consider two kinds of probes. To begin with, we  apply the prescription in Ref.~\cite{Caputa:2024sux} directly in the type IIA backgrounds~\eqref{eq:10Dansatz_Einstein}, considering the radial geodesic motion of a point-like massive particle. In doing so, we  discover divergences in the growth of complexity, which we  attribute to the singular nature of the ten-dimensional geometry. Such `particles' are not part of the spectrum of type IIA supergravity. For this reason we also study the geodesic motion of D0-branes, which do belong to the type IIA excitations and are thus the natural objects to investigate. To complement our findings, in  Appendix \ref{app11d} we investigate the M-theory lift of these computations.

All cases can be treated in a very similar way. For this reason we give first general formulae and specialize them to each kind of object separately. Other excitations, such as different types of branes or strings could be studied too~\cite{Nastase:2026lhz,Chatzis:2026ekd}.

\subsection{General formulae}
\label{sec:generalformulae}
The motion of an in-falling object can be parametrized by $r(t)$, assuming the motion is purely radial. In particular, we fix all other coordinates to constant values. The corresponding induced metric on the world-line of the particle is 
\begin{equation}\label{eq:induced_metric}
 \dd s_{\text{{\tiny ind}}}^2= \left(-A(r) + A(r) B(r)\dot{r}^2\right)\dd t^2,
\end{equation}
and the dynamics is thus governed by the world-line action
\begin{equation}
 S =-m \int \dd t \sqrt{A(r)\left( 1-B(r)\dot{r}^2\right)}
\, \equiv \, \int \dd t \, L\,.
  \label{action-KS}   
\end{equation}
Here the last equality defines the Lagrangian $L$, the dot denotes derivative with respect to time, and $m$ is the mass of the particle (for a D0-brane it will be its tension). The concrete expressions of $A(r)$ and $B(r)$ for the particle and the D0-brane will be given later. 

From the action, we obtain an equation of motion for $r$,
\begin{equation}
 \frac{\dd}{\dd t}\Bigg( \frac{A B \dot{r}}{L}\Bigg)= \frac{-A'(1- B\dot{r}^2) +A B'\dot{r}^2}{2L},
 \label{eq-mot-r}  
\end{equation}
with primes indicating derivatives with respect to $r$. Since the integrand in Eq.~\eqref{action-KS} does not depend on time, the Hamiltonian
\begin{equation}
H= \dot r \frac{\partial L}{\partial \dot r} - L= m \frac{A}{\sqrt{A(1- B \dot{r}^2)}} \label{lionel} 
\end{equation}
is conserved. We choose as initial conditions that the particle is  released at $t=0$ from $r(0)=r_i$, where it is initially at rest (namely, $\dot r(0) = 0$). With this choice, the Hamiltonian~\eqref{lionel} is simply
\begin{equation}\label{eq:Hmailtonian_particle}
H= m\sqrt{A(\rUV)}\,,
\end{equation}
and  a first integral of the equation of motion is found from Eq.~\eqref{lionel},
\begin{equation}\label{eq:rdot_Krylov}
 \dot{r}=\pm \sqrt{\frac{1}{B(r)}\left[1-\frac{A(r)}{A(\rUV)} \right]}\,.
\end{equation}

To develop some intuition, it is useful to recast the problem in terms of the motion of the particle in an effective potential, rather than as geodesic motion in a curved spacetime. To this end, we write the Hamiltonian in Eq.~\eqref{lionel} in terms of the momentum conjugate to $r$, namely
\begin{equation}
    |P_r|=\left|\frac{\partial L}{\partial\dot{r}}\right|= m \sqrt{\frac{A(r) B(r)^2~\dot{r}^2}{1- B(r) ~\dot{r}^2}}\,.
\end{equation}
Then
\begin{equation}
H^2= m^2 A(r) + \frac{P_r^2}{B(r)}. \label{PrH2}
\end{equation}
Consequently, in {\it analogy} with the Newtonian energy balance, the term $m^2A(r)$ plays the role of an effective potential, while $B(r)$ controls the radial effective mass of the particle.

In Ref.~\cite{Caputa:2024sux} ---see also Refs.~\cite{He:2024pox, Fan:2024iop}--- the time derivative of the Krylov complexity $\CK(t)$ was identified with the proper momentum of the particle, which we denote below by $P_{\overline{y}}$. Although our geometries are not pure AdS, we nevertheless adopt this identification and discuss the differences between treating the probe as an ordinary point particle and as a D0-brane.

The proper-coordinate $\bar{y}$ is defined through the radial part of the metric in which the probe follows a geodesic. This is the Einstein frame for a massive point particle (Sec.~\ref{krylovparamassive10d}), the string frame for a D0-brane (Sec.~\ref{sectionD0-IIA}), and the eleven-dimensional metric for a massive particle in M-theory, as discussed in Appendix~\ref{appendix-mtheory-particle}. The differential of the proper coordinate $\dd\bar{y}$ is given by
\begin{equation}
\dd s^2= A(r) B(r) \dd r^2= \dd {\bar{y}}^2 \quad \Rightarrow\quad  
\frac{\dd\dot{r}}{\dd\dot{\bar{y}}}=\frac{\dd r}{\dd \bar{y}}\pm \frac{1}{\sqrt{AB}}.\label{cambiovariable}
\end{equation}
In terms of this new coordinate, the Lagrangian in Eq.~\eqref{action-KS} reads
\begin{equation}
    L=-m \sqrt{A(\bar{y})-\dot{\bar{y}}^2}\,.
\end{equation}
Then
\begin{eqnarray}
& & \dot{\CK}(t) =- P_{\bar{y}}=-\frac{\partial L}{\partial \dot{r}} \frac{d\dot{r}}{d\dot{\bar{y}}}= -
\frac{m A B  ~\dot{r}}{\sqrt{A (1- B \dot{r}^2)} }\times\frac{1}{\sqrt{A B}}= -m \dot{r} \sqrt{\frac{B}{1- B\dot{r}^2}}~.\label{complexityBBgeneric}
\end{eqnarray}
There is a subtle point with Eq.~\eqref{complexityBBgeneric} and the double sign in Eqs.~\eqref{eq:rdot_Krylov} and~\eqref{cambiovariable}. When the particle falls from $r_i$ towards smaller values of $r$ we have $\dot{r}<0$ and the complexity should grow. This is the idea in Refs.~\cite{Susskind:2018tei, Susskind:2019ddc, Ageev:2018msv}. Conversely, as the probe particle bounces back and starts climbing up the radial coordinate back to $r_i$ the complexity should decrease. Hence we need to be careful with the sign in Eq.~\eqref{complexityBBgeneric}.

Note also that the complexity $\CK$ has no units. Usually, there is a conversion factor between $\dot \CK(t)$ and the momentum. In this work, we ignore this factor.
Putting the expressions above together (and using that the particle starts falling from $r_i$ with zero initial velocity), we find
\begin{equation}
 \dot{\CK}(t)=-\text{sign}(\dot{r})\times  m\sqrt{\frac{A(\rUV)}{A(r)} -1}.   \label{eq:CK_growth}
\end{equation}
The sign of $\dot{r}$ appears as a consequence of the discussion below Eq.~\eqref{complexityBBgeneric}. For this simple expression to be useful, $r(t)$ must be found first. In Ref.~\cite{Nunez:2026kwr}, other expressions for $\CK(t)$ and ${\ddot\CK(t)}$ are derived.

Let us now evaluate these formulae for an ordinary particle and a D0-brane, and highlight the differences.

\subsection{Krylov spread complexity from a free-falling particle}\label{krylovparamassive10d}
Consider the geometries in Eq.~\eqref{eq:10Dansatz_Einstein}. For an embedding of the form $r(t)$, the induced metric on the world-line of a particle reads
\begin{equation}
\dd s_{\text{\tiny p}}^2= e^{-{\Phi}/{2}}h^{-{1}/{2}}~ \left[-1+ {h}\dot{r}^2 \right] \dd t^2.
\end{equation}
Note that the particle couples to the Einstein-frame metric, which explains the overall factor of the dilaton.
In this case, the action is
\begin{equation}
 S
=-m \int \dd t \sqrt{ e^{-{\Phi}/{2}}h^{-{1}/{2}}\,\left( 1-{h\, \dot{r}^2}\right)}
  \label{action-KS-particle}   
\end{equation}
which is of the form of Eq.~\eqref{action-KS}, after we identify
\begin{equation}
A =  e^{-{\Phi}/{2}}h^{-{1}/{2}}~~ \text{and} ~~B={h}.\label{definitions}
\end{equation}
The function $A(r)$ is a monotonically increasing positive function\footnote{In Fig.~\ref{fig:ruleA}, we plot $A(r)$ for $b_0=0.4390$.} of the radial coordinate that vanishes at the end of the space, $r=\rs$. As a consequence, the Hamiltonian~\eqref{eq:Hmailtonian_particle} can take any positive value. Recall that this value corresponds to the initial ``potential'' energy of the particle, $H = mA({r_i})^{1/2}$, as it is released from $r = r_i$, where it is initially at rest.

\begin{figure}[t]
	\begin{center}
    \includegraphics[width=0.46\textwidth]{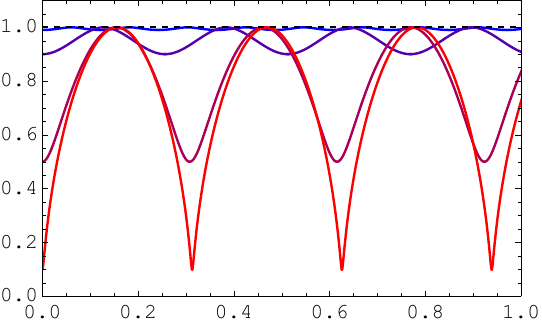} 
   \put(-225,70){$\displaystyle \frac{\rs}{r}$}
    \put(-120,-15){$t \, \Lt$}
    \hfill
    \includegraphics[width=0.46\textwidth]{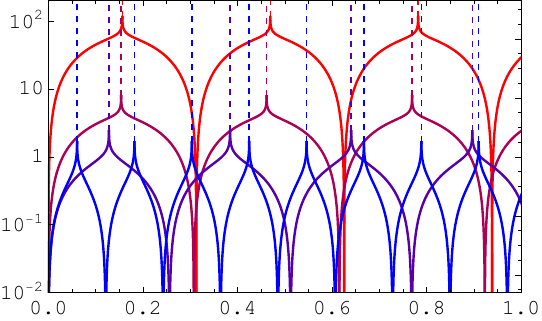} 
   \put(-225,70){$\displaystyle \frac{|\dot\CK|}{m}$}
    \put(-120,-15){$t \, \Lt$}
	\caption{\small 
    Trajectories (left) and Krylov spread complexity growth (right) as a function of time for the theory with $b_0\simeq0.4390$. The different colors stand for different initial positions $r_i$: from blue to red, $\rs/r_i = 0.99$, $0.9$, $0.5$, $0.10$. On the left panel, $\rs/r = 0$ in the vertical axis corresponds to the UV, while $\rs/r = 1$ is the IR, indicated with a horizontal dashed black line. On the right panel, the divergence in the growth of complexity is indicated with dashed lines.}\label{fig:trajectories}
	\end{center}
\end{figure}
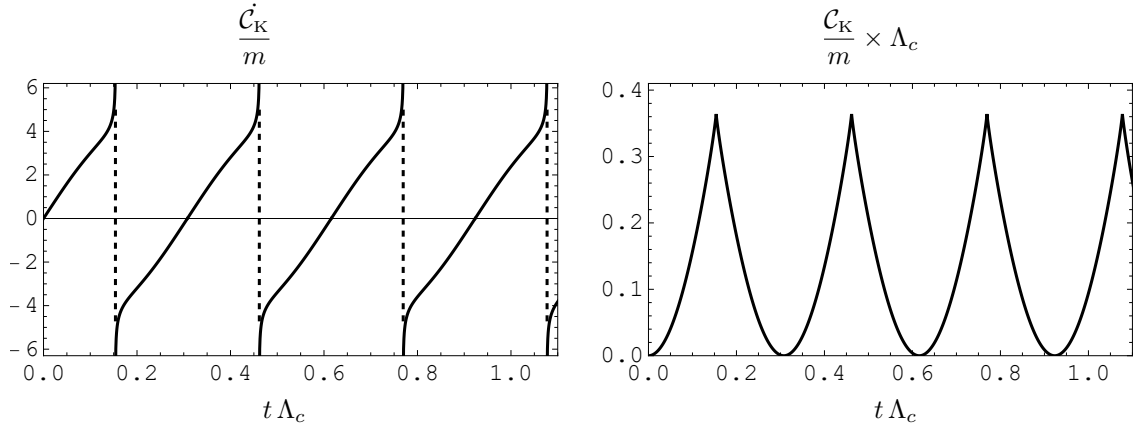
\begin{figure}[t]
    \noindent
    \includegraphics[width=1\textwidth]{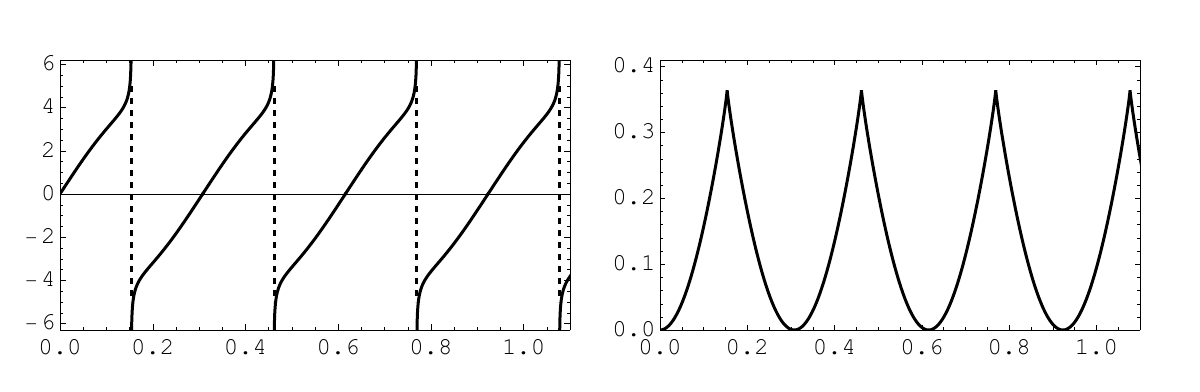} 
    \put(-120,0){\small $t \, \Lt$}
    \put(-350,0){\small $t \, \Lt$}
    \put(-140,140){\small $\displaystyle \frac{\CK}{m}\times \Lt$}
    \put(-360,140){\small $\displaystyle \frac{\dot\CK}{m}$}
	\caption{\small 
    Complexity growth (left) and complexity (right) for a massive particle released from $r = 2\rs$ in the ten-dimensional geometries, and $b_0 = 0.4390$. Other choices of $r_i$ lead to the same qualitative evolution.}\label{fig:complexity_and_derivative_10D}
\end{figure}

Crucially, from the IR behavior of the functions given in Eq.~\eqref{eq:parameters_IR_B8} it follows that close to the end of the space 
\begin{equation}\label{eq:A_B_particle}
    A\sim (r-\rs)^{1/8}\,,\quad B\sim (r-\rs)^{-1}\,.
\end{equation}
Therefore, the probe particle travels through the bulk and reaches the end of the space located at $r=r_s$, where $A(r)$ vanishes and thus all the energy is in kinetic form. In fact, using Eq.~\eqref{eq:rdot_Krylov} one finds that
\begin{equation}
B(r)~\dot{r}^2= 1-\frac{A(r)}{A(r_i)}\,,
\end{equation}
and the ``kinetic term'' in Eq.~\eqref{PrH2},
\begin{equation}
\frac{P_r^2}{B(r)}= m^2\Big( A(r_i) - A(r)\Big),
\end{equation}
is maximized close to $r=r_s$, where $A(r_s)=0$. 
When the end of the space is reached, the particle  bounces back, climbing back to $r=r_i$, from where it falls again. This oscillatory motion is illustrated in Fig.~\ref{fig:trajectories}~(left), where we plot several trajectories for the theory with $b_0 = 0.4390$. Analogous oscillatory motion has been observed previously in confining backgrounds \cite{Fatemiabhari:2025usn, Fatemiabhari:2026goj, Zoakos:2026obl}.

The oscillatory behavior is also present in the Krylov spread complexity growth, as shown in Fig.~\ref{fig:trajectories}~(right). This quantity starts at zero at $r=r_i$ and diverges in the IR, where $A(r)$ vanishes--see Eq.~\eqref{eq:CK_growth}. We attribute the divergence of $\dot{\CK}$ at the instants when the particle reaches the end of the space to the singular nature of the ten-dimensional background. Indeed, as mentioned in the Introduction, the geometries with $b_0\in(0,1)$ are {\it singular} in type IIA. This singularity originates from the divergence of the warp factor and is reflected, for example, in the divergence of curvature invariants at the end of the space. We therefore prescribe that, upon reaching the singularity, the particle ricochets off it, which amounts to flipping the sign of $\dot r$. 

This reflective boundary condition should be understood as an effective prescription that incorporates the influence of the singularity on the evolution. Note that, in order to display several orders of magnitude, in Fig.~\ref{fig:trajectories}~(right) we show the absolute value of the growth rate ---which changes sign at each bounce, see Fig.~\ref{fig:complexity_and_derivative_10D}~(left)--- using a logarithmic scale. The actual complexity can then be obtained from the growth rate through a straightforward integration, as shown in Fig.~\ref{fig:complexity_and_derivative_10D}~(right).

The presence of the singularity naturally raises questions about the interpretation of our results. We return to this issue at the end of this Section, but we can already anticipate that the main conclusion ---the emergence of oscillations in the complexity--- is robust and does not rely on the singular nature of the ten-dimensional background.

\begin{figure}[t]
	\begin{center}
    \includegraphics[width=1.05\textwidth]{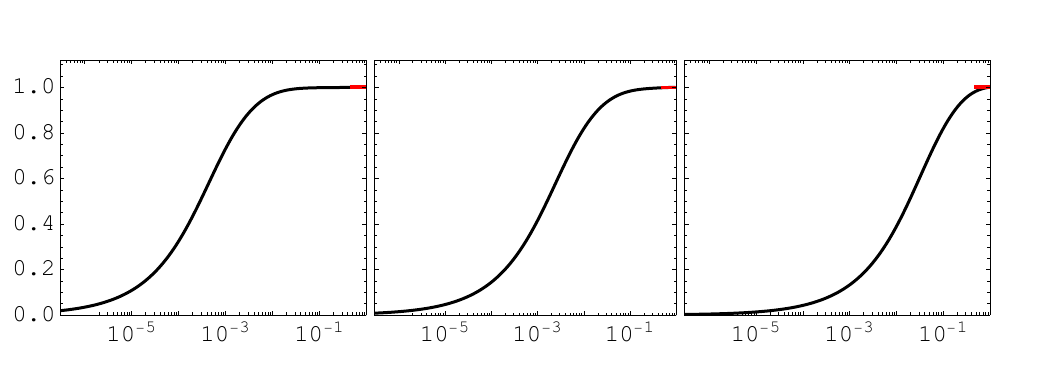} 
   \put(-505,120){\small$\displaystyle  \frac{T}{\Tuv}$}
    \put(-125,0){\small $ 1-\rs/\rUV$}
    \put(-260,0){\small $ 1-\rs/\rUV$}
    \put(-400,0){\small $ 1-\rs/\rUV$}
    \put(-160,130){\small $b_0 = 0.4390$}
    \put(-300,130){\small $ b_0 = 0.1915$}
    \put(-445,130){\small $ b_0 = 0.1076$}
	\caption{\small 
    Period of the oscillations of a massive particle as a function of the position from which it is released. The three different curves correspond to the three different values of $b_0$ indicated on each plot. Note that we are normalizing the period by $\Tuv$, given by Eq.~\eqref{eq:Tmax_particle}, which corresponds to the value at the boundary as indicated by the red tick. Distinct choices of $b_0$ produce the same qualitative behavior ---the period decreases monotonically with $r_i$--- as opposed to what we will find in the next Section---see for instance Fig.~\ref{fig:periodD0}.
    }\label{fig:period}
	\end{center}
\end{figure}
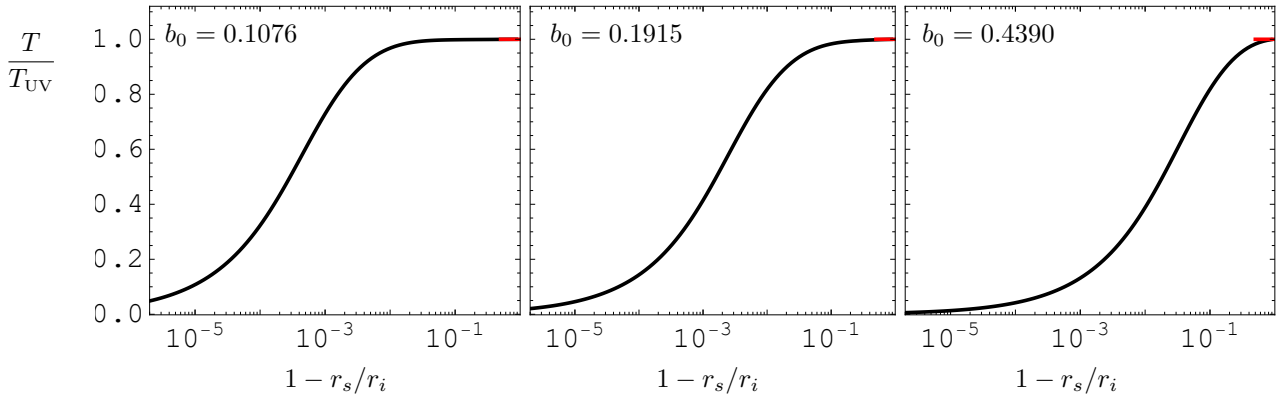

The period of the motion is twice the time to fall from the initial position $r_i$ to the end of the space at $\rs$,
\begin{equation}
T=2 \int_{\rs}^{r_i}\frac{\dd r \sqrt{B(r)}}{\sqrt{1- \frac{A(r)}{A(r_i)}}}. \label{periodoparticulaIIA}   
\end{equation}
This quantity is represented in Fig.~\ref{fig:period} for three different values of $b_0$, as a function of the initial position of the particle. For particles released close to the IR (i.e. $r_i\simeq \rs$), the period vanishes, since the integrand diverges as $(r-\rs)^{-1/2}$, see Eq.~\eqref{eq:A_B_particle}. In the opposite limit, when $r_i$ is taken to the boundary, $A(r_i)\to \infty$ and we obtain
\begin{equation}\label{eq:Tmax_particle}
\Tuv=2 \int_{\rs}^{\infty}{\dd r \sqrt{B(r)}}=2 \int_{\rs}^{\infty}{\dd r \sqrt{h(r)}}.    
\end{equation}
For massive radially in-falling particles in ten dimensions, $\Tuv$ constitutes an upper bound for the period of the oscillation, since for all the values of $b_0$ the period behaves qualitatively the same: it is maximized when the particle is released from the boundary, and decreases monotonically to zero as the initial position $r_i$ is pushed towards the end of the space.

\begin{figure}[t]
    \noindent
    \includegraphics[width=1.1\textwidth]{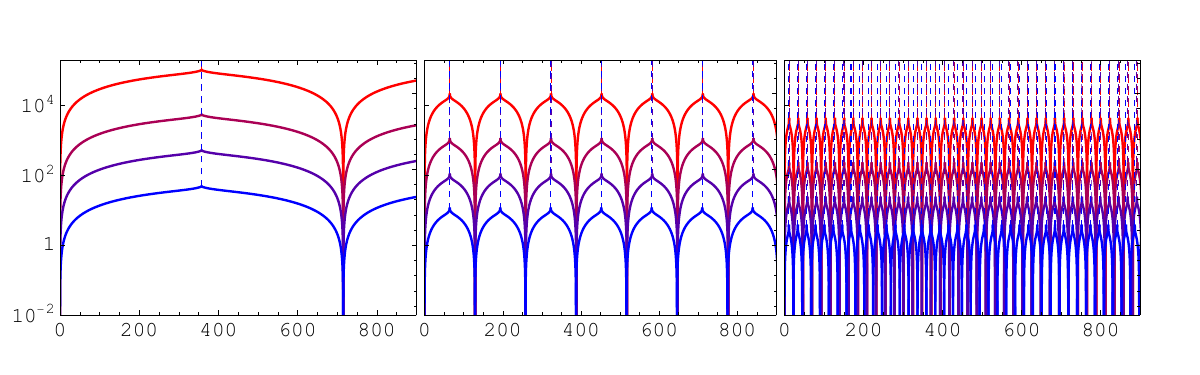} 
    \put(-510,75){\small $\displaystyle \frac{|\dot\CK|}{m}$}
    \put(-120,0){\small $t \, \Lt$}
    \put(-280,0){\small $t \, \Lt$}
    \put(-430,0){\small $t \, \Lt$}
    \put(-120,135){\small $b_0=0.1076$}
    \put(-280,135){\small $b_0=0.0605$}
    \put(-430,135){\small $b_0=0.0340$}
	\caption{\small 
    Complexity growth for the different values of $b_0$ indicated on the plots, as  a function of time. From blue to red, the value of the initial position is given by $\rs/r_i = 0.99$, $0.9$, $0.5$, $0.10$. The maximum period of the oscillation grows as the conformal theory is approached. Dashed lines stand for the divergence in the complexity growth.}\label{fig:oscillations_close_to_CFT}
\end{figure}
\begin{figure}[t]
	\begin{center}
    \includegraphics[width=0.48\textwidth]{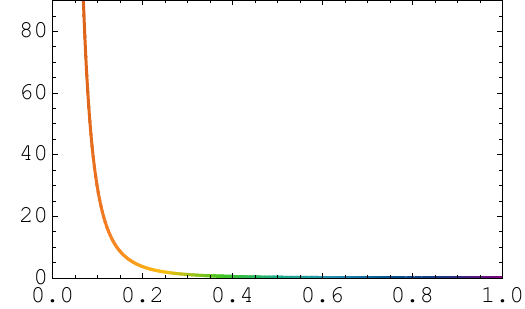} 
   \put(-220,135){\small $\displaystyle \,  \Tuv \times \Lt$}
    \put(-105,-10){\small $b_0$}
    \hfill
    \includegraphics[width=0.48\textwidth]{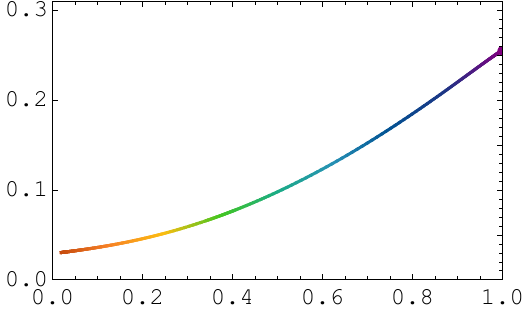} 
   \put(-220,140){\small $\displaystyle \,  \Tuv \,\times \frac{ \Lt b_0^3}{(1-b_0)^2}$}
    \put(-105,-10){\small $b_0$}
	\caption{\small (Left) Maximum period of the oscillation ---given by the period of a particle released from the boundary, see Eq.~\eqref{eq:Tmax_particle}--- as a function of $b_0$. The color of the curve illustrates the position of the corresponding theory in Fig.~\ref{fig:triangle}. 
    (Right) Same quantity, normalized by $b_0^3/(1-b_0)^2$. This normalization keeps both limits (near the CFT and near confinement) finite.} 
    \label{fig:period_walking}
	\end{center}
\end{figure}
\vspace{2mm}
Finally, let us inspect briefly the two different limits. Theories with $b_0 \gtrsim 0$ flow close to the OP CFT: the main effect of the proximity of the CFT is that the period of the oscillations becomes larger and larger as shown in Fig.~\ref{fig:oscillations_close_to_CFT}. Actually, for the $b_0 = 0$ geometry ---dual to the theory whose IR is the OP fixed point and dubbed $\Binf$ in Fig.~\ref{fig:triangle}--- the oscillations cease to exist, as the particle takes an infinite boundary time $t$ to reach the Poincar\'e horizon. In the opposite limit, for $b_0\lesssim 1$, the theories approach the confining $\Bconf$ theory provided the CS level $k$ vanishes as dictated by the rescaling of $Q_k$ in Eq.~\eqref{eq:rescaling_Qk}. As a consequence, if the rescaling is not performed, the periods collapse to zero as $b_0\to 1$. This rich phenomenology is portrayed in Figure~\ref{fig:period_walking}.

As we pointed out, the results in this Section should be taken with care due to the presence of an IR singularity in the type IIA description. To assess how general our results are, we present the perfectly regular eleven-dimensional uplift of the geometries in Appendix~\ref{app:uplift} and study the motion of a massive particle in them in Appendix~\ref{appendix-mtheory-particle}. The features of the oscillating motion and complexity found above are robust; with the only important difference that the divergence in $\dot{\CK}$ is absent there. This supports our claim that the divergence is caused by the singularity probed by the massive particle in ten dimensions.

In the next Section, we  extend our analysis in type IIA and consider the natural ``particles'' present in the spectrum of the theory: D0-branes\footnote{Although the dual field theories are not known explicitly, the ABJM-like $F_2$ flux suggests that D0-brane charge is represented in the UV SYM-CSM theories by dressed monopole operators, as in ABJM.} (see for instance Ref.~\cite{Chatzis:2026ekd}). Oscillations  also appear for D0-branes, while the complexity growth does not suffer the divergence that we attributed to the presence of the singularity. Crucially, as we review in Appendix~\ref{appendix-D0-from-11D}, D0-branes are uplifted to massless excitations with momentum along the M-theory circle in eleven dimensions. This produces an effective coupling to the dilaton in ten dimensions that prevents them from approaching the singularity. Hence, a consistent picture emerges: the presence of the oscillations is robust and persist when the singularity is not probed, or is absent, like in the lift to M-theory of this family of backgrounds.

\subsection{Krylov spread complexity from a free-falling D0-brane}
\label{sectionD0-IIA}
\begin{figure}[t]
	\begin{center}
    \includegraphics[width=0.48\textwidth]{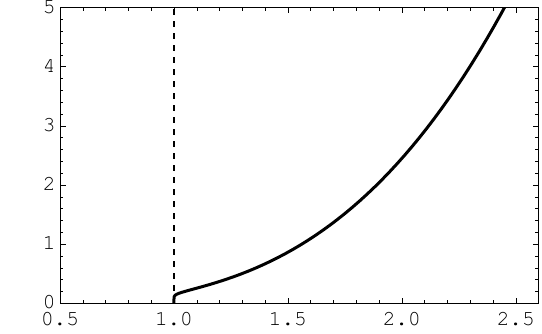} 
   \put(-220,150){\small $\displaystyle \frac{10^{-2}q_c^{5/4}}{|Q_k|^{15/4}}\,A$}
    \put(-120,-15){$r/\rs$}
    \hfill
    \includegraphics[width=0.48\textwidth]{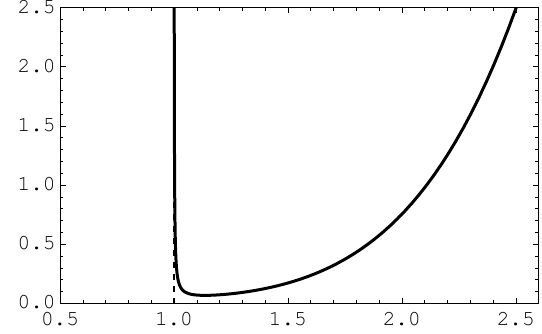} 
   \put(-220,150){\small $\displaystyle \frac{10^{-4}q_c^2}{|Q_k|^6}\,A$}
    \put(-120,-15){\small $r/\rs$}
	\caption{\small 
    Function $A(r)$ for the massive particle (left) and for the D0-brane (right), in the theory with $b_0 = 0.4390$. For other values of $b_0$ the behavior is qualitatively the same, as discussed in the main text. In both cases, the position of the end of space is indicated by the black dashed vertical line.}\label{fig:ruleA}
	\end{center}
\end{figure}
Let us analyze the geodesic motion of a radially in-falling D0-brane. This couples to the ten-dimensional string metric, and its world-line reads
\begin{equation}
\dd s_{\text{\tiny D0}}^2= h^{-{1}/{2}}~ \left[-1+ {h}\dot{r}^2 \right] \dd t^2.
\end{equation}
The world-line action is now
\begin{equation}
 S =- \TD \int \dd t \, e^{-{\Phi}}\sqrt{ h^{-{1}/{2}}\,\left( 1-{h\, \dot{r}^2}\right)}\,.
  \label{action-KS-D0}   
\end{equation}
Note that the Wess-Zumino term does not contribute, for all the internal angles are kept constant. We can consequently use the formulae derived in Section~\ref{sec:generalformulae} with the identifications
\begin{equation}\label{definitionsD0}
A =  e^{-{2\Phi}}h^{-{1}/{2}}\,,\qquad B = {h}\qquad \text{and}\qquad m = \TD\,.
\end{equation}
Note that $B$ is the same as for the case of the falling massive particle. In contrast, the  different factor of the dilaton appearing in $A(r)$ ---compare for instance Eq.~\eqref{definitionsD0} with Eq.~\eqref{definitions}--- does have important consequences. Indeed, the function $A(r)$ now diverges in the IR, rather than vanishing as in the previous case. This is illustrated in Fig.~\ref{fig:ruleA}. In fact, from Eq.~\eqref{eq:parameters_IR_B8}  we conclude that for the D0-probe we have
\begin{equation}
    A\sim (r-\rs)^{-1}\,,\quad B\sim (r-\rs)^{-1}\,.
\end{equation}

Given that $A(r)$ is now a non-monotonic function that diverges at the boundary and in the IR, for every chosen value of $r_i$ there will be another value of the radial coordinate $r_f$ such that
\begin{equation}
    H = \TD\sqrt{A(r_f)} = \TD\sqrt{A(r_i)}\,.
\end{equation}
As a consequence of Eq.~\eqref{eq:rdot_Krylov}, at $r=r_f$ the D0-brane must stop.  As a result, it does not reach the end of the space, since $r=r_s<r_f<r_i$. Put differently, the positions $r=r_i$ and $r=r_f$ correspond to the points where all the energy in Eq.~\eqref{PrH2} is in the form of potential energy. 

\begin{figure}[t]
	\begin{center}
    \includegraphics[width=0.46\textwidth]{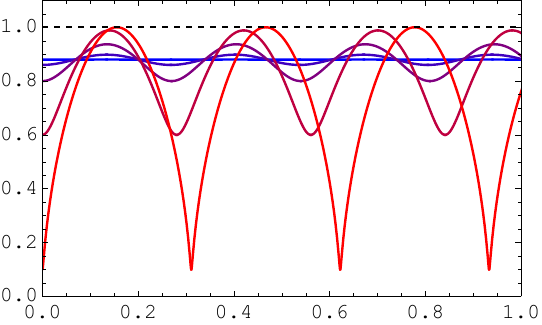} 
   \put(-225,70){$\displaystyle \frac{\rs}{r}$}
    \put(-120,-15){$t \, \Lt$}
    \hfill
    \includegraphics[width=0.46\textwidth]{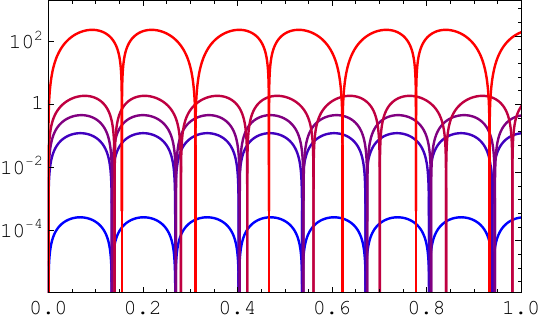} 
   \put(-230,70){$\displaystyle \frac{|\dot\CK|}{\TD}$}
    \put(-120,-15){$t \,\Lt$}
	\caption{\small 
    (Left) Trajectories for the D0-brane. Note that, contrary to the previous case, now the D0-brane does not reach the IR, represented by the black dashed line.
    (Right) Krylov complexity growth as a function of time for the theory with $b_0\simeq0.4390$, obtained from the geodesic motion of a D0-brane. The different colors stand for different initial positions $r_i$: from blue to red, $\rs/r_i = 0.8799 (\simeq \rs/\rmin)$, $0.86$, $0.8$, $0.6$, $0.10$. On the left panel, $\rs/r = 0$ in the vertical axis corresponds to the UV, while $\rs/r = 1$ is the IR.
    }\label{fig:trajectoriesD0}
	\end{center}
\end{figure}
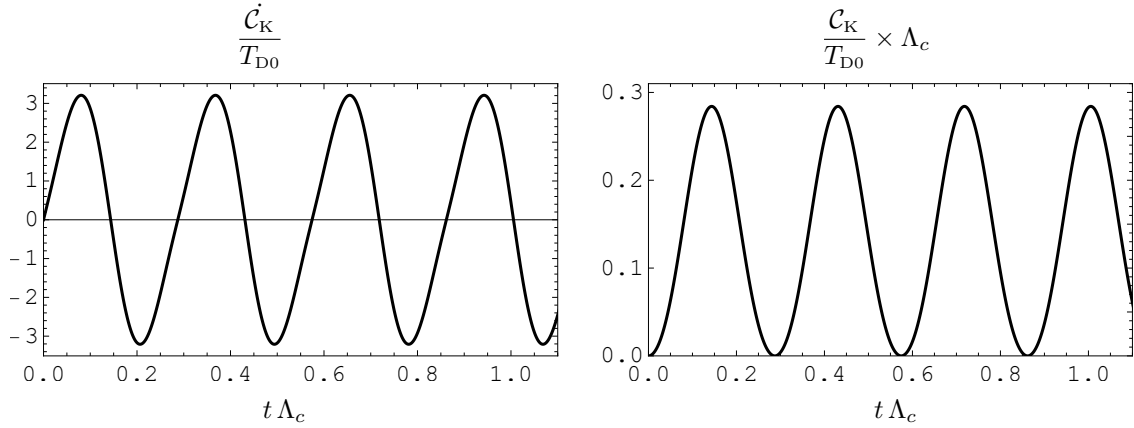
\begin{figure}[t]
    \noindent
    \includegraphics[width=1\textwidth]{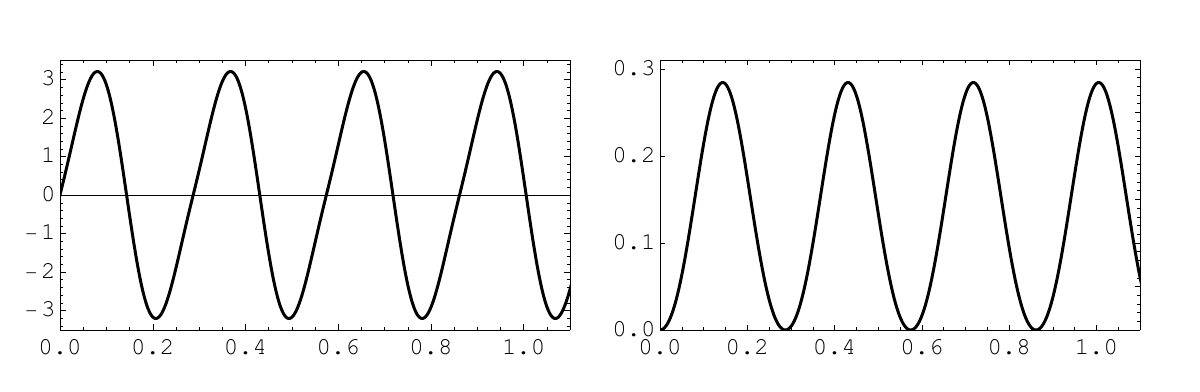} 
    \put(-120,0){\small $t \,  \Lt$}
    \put(-350,0){\small $t \,  \Lt$}
    \put(-140,140){\small $\displaystyle \frac{\CK}{\TD}\times \Lt$}
    \put(-360,140){\small $\displaystyle \frac{\dot\CK}{\TD}$}
	\caption{\small 
    Complexity growth (left) and complexity (right) for a D0-brane released from $r = 2\rs$ and $b_0 = 0.4390$. The qualitative behavior for other choices of $r_i$ is qualitatively the same. }\label{fig:complexity_and_derivative_D0}
\end{figure}
In Fig.~\ref{fig:trajectoriesD0} we show different trajectories. As was the case for massive particles, D0-branes also exhibit oscillations, which propagate to the complexity growth, whose absolute value is shown in the same figure. For D0-branes, however, the mechanism underlying these oscillations is different. Rather than being simply reflected at the end of the space, at $r=r_f$ the D0-brane reaches a point of vanishing momentum, where its potential energy is converted back into kinetic energy, causing it to move towards the boundary. Consequently, the momentum changes sign and the brane starts moving towards larger values of $r$, as shown in Fig.~\ref{fig:complexity_and_derivative_D0}. Eventually, it reaches $r=r_i$ again and the process repeats.

As a consequence, the evolution is completely smooth in this case: not only is the divergence in the growth absent, but $\dot{\CK}(t)$ also remains continuous. In a sense, D0-branes provide a regular description of the evolution close to the singular region due to their coupling to the dilaton ---see also Appendix~\ref{appendix-D0-from-11D}.

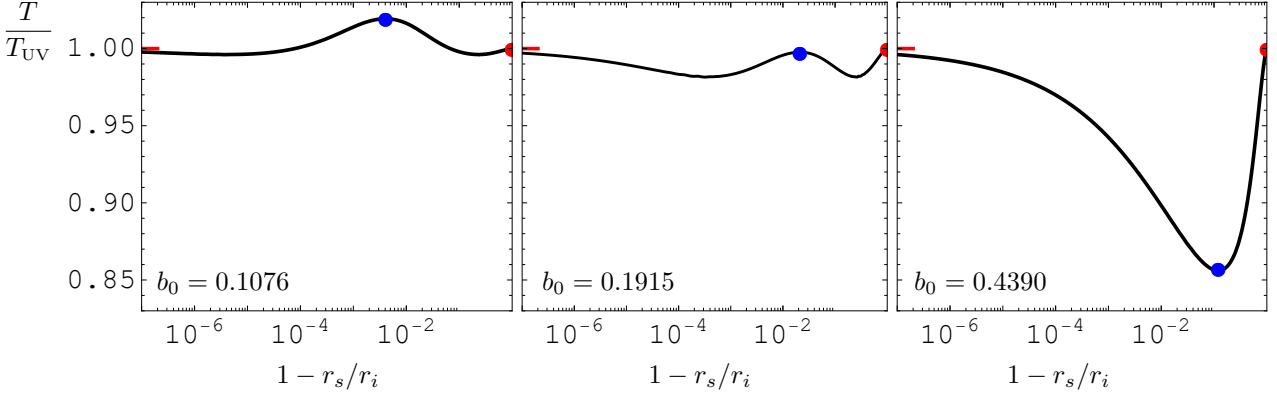
\begin{figure}[t]
	\begin{center}
    \includegraphics[width=1.05\textwidth]{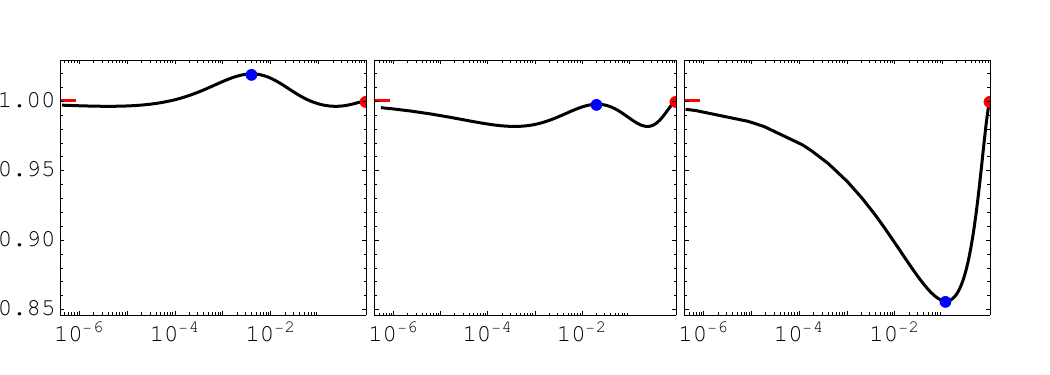} 
   \put(-503,130){\small $\displaystyle \frac{T}{\Tuv}$}
    \put(-125,0){\small $ 1-\rs/\rUV$}
    \put(-260,0){\small $ 1-\rs/\rUV$}
    \put(-400,0){\small $ 1-\rs/\rUV$}
    \put(-160,35){\small $ b_0 = 0.4390$}
    \put(-300,35){\small $ b_0 = 0.1915$}
    \put(-445,35){\small $ b_0 = 0.1076$}
	\caption{\small 
    Period of the oscillations of a D0-brane as a function of the position from which it is released. The three different curves correspond to the three different values of $b_0$ indicated on each panel. Note that we are normalizing the period by $\Tuv$, given by Eq.~\eqref{eq:Tmax_particle}. From the plots we see that $T=\Tuv$ when the D0-brane is released from the boundary or from the IR (as indicated in red). The value of the period obtained at $\rmin$---given by Eq.~\eqref{eq:Tmin_D0}--- is indicated with a blue dot.}\label{fig:periodD0}
	\end{center}
\end{figure}

The period of the trajectory of a particle released from $r=r_i$ is in this case
\begin{equation}\label{eq:time_D0brane}
T=2 \int_{r_f}^{r_i}\frac{\dd r \sqrt{B(r)}}{\sqrt{1- \frac{A(r)}{A(r_i)}}}. 
\end{equation}
This quantity develops a non-trivial structure, as shown in Fig.~\ref{fig:periodD0}. Let us inspect two interesting regimes. First, observe that the closer to the boundary the D0-brane is released, the higher it  climbs the potential well generated by the IR divergence of $A(r)$. Consequently, when released exactly from the boundary ($r_i\to \infty$), the period is again $\Tuv$, given by Eq.~\eqref{eq:Tmax_particle}.

Because $A(r)$ is a continuous positive function that diverges both in the UV and in the IR, it must develop (at least) one minimum at a certain value $r=\rmin$. Plotting the function as in Fig.~\ref{fig:ruleA} (right), we find only one minimum. If a particle is released  close to the minimum, the positions $r_i$ and $r_f$ are close to $\rmin$ and, in particular, close to each other. Demanding that the factor
\begin{equation}\label{eq:expansion_ArAi}
    1-\frac{A(r)}{A(r_i)} = \left(1-\frac{A(\rmin)}{A(r_i)}\right)-\frac{A''(\rmin)}{2A(r_i)}(r-\rmin)^2 + O\left((r-\rmin)^3\right)
\end{equation}
vanish under the square root in Eq.~\eqref{eq:rdot_Krylov}, the turning points $r_f$ and $r_i$ are, to this order in the expansion about $\rmin$, given by
\begin{equation}
    r_{i,f} = \rmin \pm \sqrt{2\,\frac{A(r_i)-A(\rmin)}{A''(\rmin)}} \equiv \rmin \pm \varepsilon\,.
\end{equation}
The last equality defines $\varepsilon$. Now, to take appropriate care of the integral in Eq.~\eqref{eq:time_D0brane} in this limit, it is not enough to expand the integrand about $r=\rmin$, since the denominator diverges at $r = r_i$ and $r= r_f$. Instead, when $r_i$ and $r_f$ are close to each other, for $r\in(r_f,r_i)$ we approximate 
\begin{equation}\label{eq:approx_Tmin}
    1-\frac{A(r)}{A(r_i)}\simeq 4\left(1-\frac{A(\rmin)} {A(r_i)}\right)\frac{(r_i-r)(r-r_f)}{(r_f-r_i)^2}\,.
\end{equation}
The reason why this is a good approximation in this limit is that the right hand side of Eq.~\eqref{eq:approx_Tmin} is a polynomial that vanishes at $r=r_{i,f}$ and whose value is precisely $1-A(\rmin)/A(r_i)$ at $r=\rmin$. Within this approximation, Eq.~\eqref{eq:time_D0brane} becomes 
\begin{equation}
T\simeq\frac{(r_i-r_f)\sqrt{B(\rmin)}}{\left(\displaystyle1-\frac{A(\rmin)} {A(r_i)}\right)^{1/2}}\int_{r_f}^{r_i}\frac{\dd r }{\sqrt{(r_i-r)(r-r_f)}} = \pi\, \frac{(r_i-r_f)\sqrt{B(\rmin)}}{\left(\displaystyle1-\frac{A(\rmin)} {A(r_i)}\right)^{1/2}}. \label{periodD0dibu}
\end{equation}
Now, using Eq.~\eqref{eq:expansion_ArAi} we obtain
\begin{equation}\label{eq:Tmin_D0}
    \Tmin= 2\pi\sqrt{2}\sqrt{B(\rmin)\frac{A(\rmin)}{A''(\rmin)}}\,
\end{equation}
in the limit $\varepsilon\to 0$.

For sufficiently large values of $b_0$ ---see for instance the case $b_0 = 0.4390$ in Fig.~\ref{fig:periodD0} (right)--- the value of $\Tmin$ provides a minimum value of the period, while $\Tuv$ constitutes an upper bound. At some point $\Tmin$ ceases to be a minimum and becomes a local maximum of the potential ---see Fig.~\ref{fig:periodD0} (middle)--- which keeps increasing as $b_0$ decreases. Remarkably, for theories close to the CFT, the period $\Tmin$ becomes an absolute maximum, as in Fig.~\ref{fig:periodD0} (left). Note that, for these small values of $b_0$, the position $r=\rmin$ can be understood as the position where the geometry resembles AdS the most, since in AdS we have $A''(r) = 0$.

\begin{figure}[t]
	\begin{center}
    \includegraphics[width=0.5\textwidth]{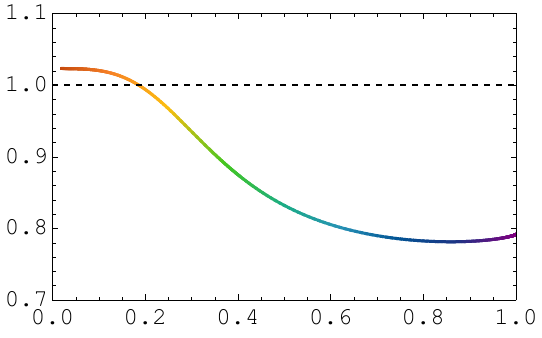} 
    \put(-260,100){$\displaystyle \frac{\Tmin}{\Tuv}$}
    \put(-130,-15){$b_0$}
	\caption{\small Ratio between the periods $\Tmin$ and $\Tuv$ of a D0-brane. The hue of the curve illustrates the position of the corresponding theory in Fig.~\ref{fig:triangle}. Note that in this case the periods are roughly the same.
    } 
    \label{fig:period_walking_D0}
	\end{center}
\end{figure}

To make this point even more explicit, in Fig.~\ref{fig:period_walking_D0} we plot the ratio between $\Tmin/\Tuv$ for the whole range of $b_0$. Note that both quantities are roughly the same. In addition, $\Tmin$ becomes the absolute maximum below $b_0= 0.1858$, where the curve on this plot crosses unity.

\begin{figure}[t]
    \noindent
    \includegraphics[width=1.1\textwidth]{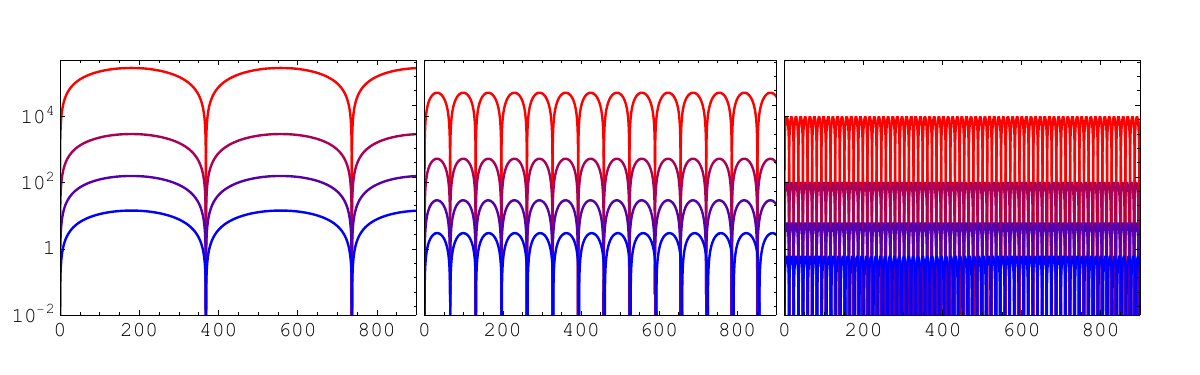} 
    \put(-510,75){$\displaystyle \frac{\dot\CK}{\TD}$}
    \put(-130,0){$t \, \Lt$}
    \put(-280,0){$t \, \Lt$}
    \put(-430,0){$t \, \Lt$}
    \put(-130,135){$b_0=0.1076$}
    \put(-280,135){$b_0=0.0605$}
    \put(-430,135){$b_0=0.0340$}
	\caption{\small 
    Complexity growth obtained from the computation of the falling D0-brane, for the different values of $b_0$ indicated on the plots, as a function of time. From blue to red, the value of the initial position is given by $\rs/r_i = 0.99$, $0.9$, $0.5$, $0.10$. The maximum period of the oscillations grows as the conformal theory is approached. The cusps indicate the times at which turning points of the trajectory occur. }\label{fig:oscillations_close_to_CFT_D0}
\end{figure}
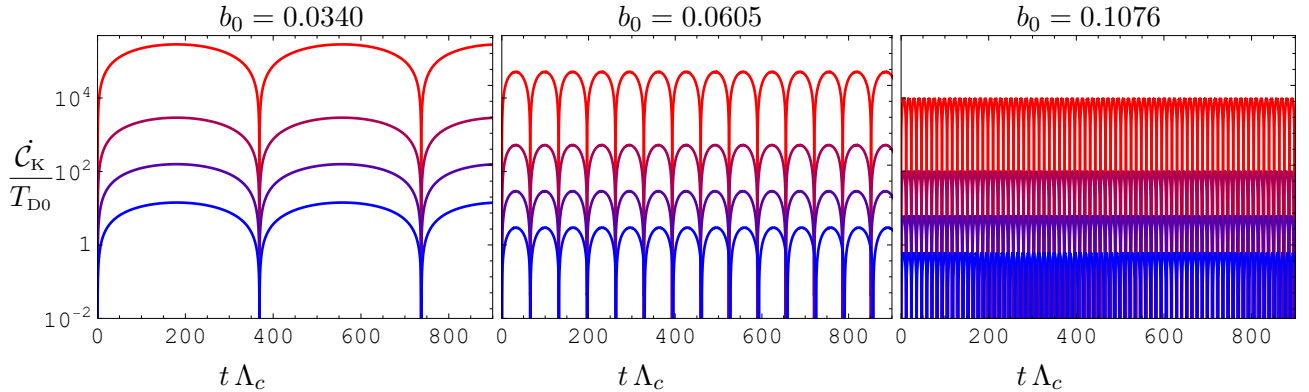

The growth of Krylov spread complexity is also peculiar. Indeed, Eq.~\eqref{eq:CK_growth} shows that $\dot\CK$ vanishes twice during each period, namely at $r=r_i$ and $r=r_f$. Since for the D0-brane $A(r)$ is a positive function bounded from below by $A(\rmin)$, for every choice of initial $r=r_i$ the absolute value of the growth of Krylov complexity is bounded from above by
\begin{eqnarray}
    |\dot\CK| \leq \TD\sqrt{\frac{A(\rUV)}{A(\rmin)} -1}\,,
\end{eqnarray}
as can be seen in Fig.~\ref{fig:trajectoriesD0}. This is in contrast to the previous case of the falling particle, in which the growth of complexity $\dot\CK$ diverges when the particle reaches the IR. In Fig.~\ref{fig:oscillations_close_to_CFT_D0} we show the complexity growth as a function of time for different choices of $b_0$, revealing the widening of the oscillations in the walking regime.

We would like to note a subtle point. When studying a massive particle probing the type~IIA geometry in Sec.~\ref{krylovparamassive10d}, we use the Einstein frame metric to compute geodesics. In Sec~\ref{sectionD0-IIA}, in contrast, the D0-brane probe couples to the string frame metric (through the Born-Infeld-Wess-Zumino action). In Appendix~\ref{appendix-mtheory-particle}, a probe massive particle coupled to the eleven~dimensional background is used to compute the geodesic and the complexity.

Let us close this discussion summarizing the different probes used and the corresponding expressions of the functions $A(r),B(r)$. All the probes studied in this paper are described by Eq.~\eqref{action-KS}, governing geodesic motion in a specific metric. The induced metric along the worldline of the particle is given by Eq.~\eqref{eq:induced_metric}. We find, for the different cases,
\begin{center}
\setlength{\tabcolsep}{3pt}
\renewcommand{\arraystretch}{1.25}
\begin{tabular}{@{}p{0.30\linewidth}p{0.25\linewidth}p{0.15\linewidth}p{0.15\linewidth}l@{}}
\toprule
Probe & Constant-mass frame & $A$ & $AB=g_{rr}$ & Mass\\
\midrule
10d particle (Section \ref{krylovparamassive10d}) & $g_E=e^{-\Phi/2}g_{\rm st}$ & $e^{-\Phi/2}h^{-1/2}$ & $e^{-\Phi/2}h^{1/2}$ & $m$\\
D0-brane (Section \ref{sectionD0-IIA}) & $g_{\rm D0}=e^{-2\Phi}g_{\rm st}$ & $e^{-2\Phi}h^{-1/2}$ & $e^{-2\Phi}h^{1/2}$ & $\TD$\\
11d particle (Appendix \ref{appendix-mtheory-particle}) & $G_{MN}$ & $e^{-2\Phi/3}h^{-1/2}$ & $e^{-2\Phi/3}h^{1/2}$ & $m$\\
\bottomrule
\end{tabular}
\end{center}
In all cases $B={g_{rr}}{|g_{tt}|^{-1}}=h$. Hence,  $B$ is invariant under  a Weyl rescaling, while $A$ is not. The general rule ---implicit in Eq.~(\ref{cambiovariable})--- is that \emph{the proper radial distance is measured with the metric in which the probe is a geodesic of constant mass}. For the 10D particle this is the Einstein frame, for the 11D particle the 11D metric, and for the D0-brane it is the string frame metric. Appendix \ref{appendix-D0-from-11D} shows that the type~IIA results for a D0-probe can be obtained by studying a (massless) probe with momentum along the eleventh-direction in M-theory. This explains why the string frame metric choice for the D0-brane is not an ad-hoc choice but follows from its 11D uplift.

In these sections we have learned that the growth of complexity is very sensitive to the features of the ground-state. We come back to this in the Discussion Section. We close this section with a comparison of our results with a Lattice calculation of Krylov spread complexity in a confining model.

\subsection{Comparing with the transverse field Ising model}
\label{sec:comparison}
The model studied in Ref.~\cite{Jiang:2025wpj} is the one-dimensional Ising chain with transverse and longitudinal fields. The Hamiltonian is
\begin{equation}
 \widehat H=-J\sum_{j=1}^{L}\left(
\widehat\sigma_j^z\widehat\sigma_{j+1}^z+h_x\widehat\sigma_j^x+h_z\widehat\sigma_j^z\right),
 \qquad J>0.
\end{equation}
For $h_z=0$ the model is the \textit{integrable transverse-field Ising model}, which possesses a global $\mathds Z_2$ symmetry generated by 
\begin{equation}
\widehat U=\prod_{j=1}^{L}\widehat\sigma_j^x\,,    
\end{equation}
which reverses all $z$-spins. In this case, for values of $h_x$ in the range $0\leq h_x<1$, an ordered ferromagnetic phase is realized, with two degenerate vacua and elementary domain-wall (kink) excitations separating them. Moreover, at $h_x=1$ it possesses a zero-temperature Ising quantum critical point, in which the system is described in the continuum limit by the $c=\tfrac12$ Ising CFT. For $h_x>1$, a disordered paramagnetic phase is present, with a unique ground state.

Since the two vacua remain exactly degenerate at $h_z=0$, there is no energy cost ---no string tension--- associated with separating a kink from an antikink. Domain walls are then the non-interacting free-fermion quasiparticles obtained via a Jordan--Wigner and Bogoliubov transformation, and can separate and propagate freely throughout the ordered phase.

A longitudinal field $h_z\neq0$ explicitly breaks the $\mathds Z_2$ symmetry and lifts the degeneracy of the two ferromagnetic vacua.  Now a domain of the disfavored vacuum carries an energy cost proportional to its length, $\ell$.  A pair of domain walls, therefore, feel the linear potential
\begin{equation}
 V(x)=\chi |x|,\qquad
 \chi=2Jh_z\,\bar\sigma,
 \qquad \bar\sigma=(1-h_x^2)^{1/8},                              
\end{equation}
where $\bar\sigma$ is the spontaneous magnetization of the $h_z=0$ ordered phase.  
Bound states are present \cite{Kormos:2016osj}, referred to as Ising ``mesons''. In this situation, Bohr-Sommerfeld quantization of their relative motion produces a discrete set of masses $m_n$. Increasing $h_z$ increases the string tension $\chi$ and, in the regime examined in Ref.~\cite{Jiang:2025wpj}, raises the  meson masses.  Confinement therefore becomes dynamically stronger and the  propagation of mesons is slowed down.

Regarding the Krylov spread complexity, for quenches performed wholly inside the ferromagnetic phase, Ref.~\cite{Jiang:2025wpj} finds a sharp confinement signature.  At $h_z=0$, freely moving domain walls produce large, size-dependent oscillations and substantial spreading in Krylov space.  Turning on even a weak $h_z$ suppresses the amplitude, removes most finite-size dependence and reduces the long-time mean approximately as
\begin{equation}
 \text{average of }{\cal C}(t)\propto h_z^{-1}.                                    
\end{equation}
At the same time, the oscillations become faster.  The power spectrum
\begin{equation}
 S_K(\omega)=\left|\int \dd t\,e^{i\omega t}{\cal C}(t)\right|^2,
\end{equation}
shows peaks at the meson masses $m_n$, and also at certain mass differences.  The associated periods of oscillations are $T_n=2\pi/m_n = 2\pi/\omega_n$. The important trend is therefore that when $h_z$ increases,
\begin{equation}
\chi\,,~m_n\,,~\text{and}~\omega_n\,,~\text{increase}\nonumber
\end{equation}
while
\begin{equation}
    T_n\,,~\text{and the amplitude and average of}~{\cal C}(t)~\text{decrease.}\nonumber
\end{equation}
This observation is specific to quenches within the ordered ferromagnetic phase, with $0\leq h_x<1$.  Quenches inside the paramagnetic phase do not produce meson confinement, and  instead produce an increasing complexity with $h_z$. Finally, quenches across the critical point excite a broad spectrum and display a more complicated, weak-confinement crossover \cite{Jiang:2025wpj,Kormos:2016osj}.

These features can be compared with the findings of  our work. The common physical message is that an infrared mass scale supplies an ``oscillation clock''.  The mechanisms and observables, however, are not identical.

In the lattice-Ising calculation, the many-body quantity ${\cal C}(t)$ itself oscillates after a quench, and its Fourier peaks directly resolve energy differences, notably the meson masses.  In our analysis, the holographic prescription identifies the \emph{rate} of spread complexity with the proper momentum of a radial probe,
\begin{equation}
 \dot\CK(t)=- P_{\bar y}(t).
\nonumber
\end{equation} 
The two frequencies are related infrared diagnostics, but they are not  the same spectral observable. For a point particle or a D0-brane released from $r_i$, we obtained expressions for the period of oscillations, both exact and approximate ones, see Eqs.~\eqref{periodoparticulaIIA}, \eqref{periodD0dibu} and~\eqref{eq:Tmin_D0}, together with Figs.~\ref{fig:period_walking} and~\ref{fig:periodD0}.

In addition, there is a key difference.  In the Ising ordered phase, $h_z$ is directly a ``confining coupling'', as  it creates the linear potential, and increasing $h_z$ means increasing string tension.  In the $\Be$ family, however, each solution with $0<b_0<1$ is gapped but \emph{screened and non-confining}: the nonzero Chern--Simons level gives gauge excitations an effective mass and probe charges do not exhibit an asymptotically linear potential.  Nevertheless, the capped infrared geometry yields bounded probe motion and periodic $\dot{\CK}(t)$.  Therefore, in this family of backgrounds, while the presence of oscillations diagnose an emergent infrared scale, by themselves they do not prove or imply Wilson-loop confinement. We elaborate on the relation of the period of the oscillations with IR quantities in the $\Be$ theories in the next Section.

To sum up, our results share similarities with the results found in the lattice study of the spread Krylov complexity in a confining model \cite{Jiang:2025wpj}. Qualitatively, they both feature oscillations whose characteristic frequency rises (equivalently, the period shortens) as the near-conformal regime is abandoned and a larger physical infrared scale develops. Besides, they differ in the nature of the parameter that controls these effects: while $h_z$ is a genuine string-tension parameter, $b_0$ interpolates among screened, walking and confining geometries and cannot be treated as ``confinement strength''.

\section{Discussion}
\label{sec:discussion}

\subsection*{Summary and main results}

Recent works have studied complexity in confining backgrounds. The identification of the growth rate with the proper momentum of a probe particle has been applied to Anabal\'on--Ross-like solitons~\cite{Anabalon:2021tua,Chatzis:2024kdu,Chatzis:2024top,Chatzis:2025hek,Anabalon:2024che,Fatemiabhari:2024aua,Anabalon:2026yxk, Nunez:2023nnl, Nunez:2023xgl} and, more systematically, to Witten's D4 model~\cite{Witten:1998zw}, the Klebanov--Strassler (KS) cascade and its baryonic branch~\cite{Klebanov:2000hb}, and the wrapped-D5-brane background~\cite{Maldacena:2000yy} in~\cite{Fatemiabhari:2025usn,Fatemiabhari:2026goj}. All of these theories are \textit{genuinely} confining, in the sense that their Wilson loops exhibit an area law. They also possess a discrete spectrum of excitations and, in most cases, a mass gap.

These developments motivated us to undertake a systematic study of complexity measures in the $\Be$ family of geometries. As reviewed in Section~\ref{sec:B8_family}, these backgrounds are dual to a family of three-dimensional ${\cal N}=1$ quiver theories with gauge group U$(N)_k\times$U$(N+M)_{-k}$, characterized by a parameter $b_0\in[0,1]$. This parameter interpolates smoothly between a theory with a conformal infrared fixed point, the Ooguri--Park CFT, at $b_0=0$; and a confining theory at $b_0=1$. For $0<b_0<1$, the eleven-dimensional geometry caps off smoothly and supports a discrete tower of massive normal modes~\cite{Elander:2018gte}. Strictly speaking, however, the theory has no mass gap, because a spontaneously broken global symmetry gives rise to a Goldstone mode~\cite{Faedo:2025nqc}. At the same time, nonzero Chern--Simons interactions screen external charges, so that the heavy-quark potential is not asymptotically linear and the theory is therefore not confining. This rich infrared structure makes the $\Be$ family particularly well suited to disentangling three logically distinct properties of theories with an infrared scale: confinement, a mass gap, and a discrete tower of excitations.

We started analyzing the complexity growth of thermal states in Section~\ref{sec:CV} by means of the complexity=volume prescription. For this observable, we found that complexity is sensitive to the presence of the CFT in theories whose RG flow passes close to it, developing a walking behavior manifested in the flattening of the growth-to-energy ratio, $\growth/\ene$. Aside from this remarkable feature in the regime $b_0\gtrsim 0$, the ratio $\growth/\ene$ exhibits a qualitatively similar dependence on energy across all theories. In most cases, it increases monotonically with energy, except in the region where the influence of the CFT becomes significant. At high energies, it follows a weak power law dictated by the asymptotic D2-brane region, while at low energies it vanishes as the IR scale becomes dominant. The lack of structure we have just pointed out suggests that the complexity growth of thermal states, while sensitive to conformal physics, probes neither the (non)confining nature of the ground states nor the rich thermal phase structure of the system.

We then analyzed the complexity of locally excited states in Section~\ref{sec:Krylov}. Using the proper-momentum prescription, we considered two types of probes: massive particles and D0-branes. In both cases, and for $0<b_0<1$, the infrared scale induces oscillations in the complexity growth. For massive particles, however, the growth rate $\dot{\CK}$ diverges when the particle reaches the singularity.\footnote{In Appendix~\ref{appendix-mtheory-particle}, we perform analogous computations from the eleven-dimensional perspective, supporting this interpretation.} This divergence is absent for D0-branes, which experience a dilaton-induced effective potential that prevents them from reaching the singularity. Instead, the smooth bounce of the D0-brane causes the complexity growth to vanish twice during each period, providing a characteristic signature of this probe. At $b_0=0$, by contrast, the oscillations disappear because the infrared theory is conformal.

We therefore conclude that the complexity growth of locally excited states is strongly influenced by the properties of the ground state. On the one hand, the presence of the CFT produces a maximum in the oscillation period at an intermediate release position of the D0-brane,\footnote{The same behavior occurs for a radially in-falling massive particle in eleven dimensions; see Fig.~\ref{fig:period11D}.} as shown in Fig.~\ref{fig:periodD0}. Following Refs.~\cite{Leichenauer:2013kaa,Aguilar-Gutierrez:2025kmw}, the radial position from which the particle is released is dual to the degree of smearing of the boundary operator, with the boundary limit corresponding to local operator insertions. In this language, our results suggest that quasi-conformal (walking) dynamics qualitatively modify the dependence of the oscillation period on the smearing scale. Whereas generic flows exhibit the longest oscillation period for local operators, flows passing close to a conformal fixed point instead maximize the period for operators smeared over intermediate scales, as in Fig.~\ref{fig:periodD0}~(left). On the other hand, these observables are highly sensitive to the $b_0\to1$ limit, where genuine confinement emerges. In this limit, the type IIA geometry becomes regular, with striking consequences for the probe dynamics. For example, when $Q_k\neq0$, the complexity growth diverges for a massive particle and vanishes twice per period for a D0-brane, whereas neither behavior occurs when $Q_k=0$, see Appendix~\ref{app:Bconf} for details. Thus, although several quantities---including the spectrum~\cite{Elander:2018gte}, phase diagram~\cite{Elander:2020rgv}, and entanglement measures~\cite{Jokela:2020wgs}---approach this limit smoothly, the complexity growth of locally excited states does not. This suggests that it may provide a sensitive diagnostic of confinement, although further studies in other settings are required to establish the generality of this conclusion.

\subsection*{Comparison to previous works}

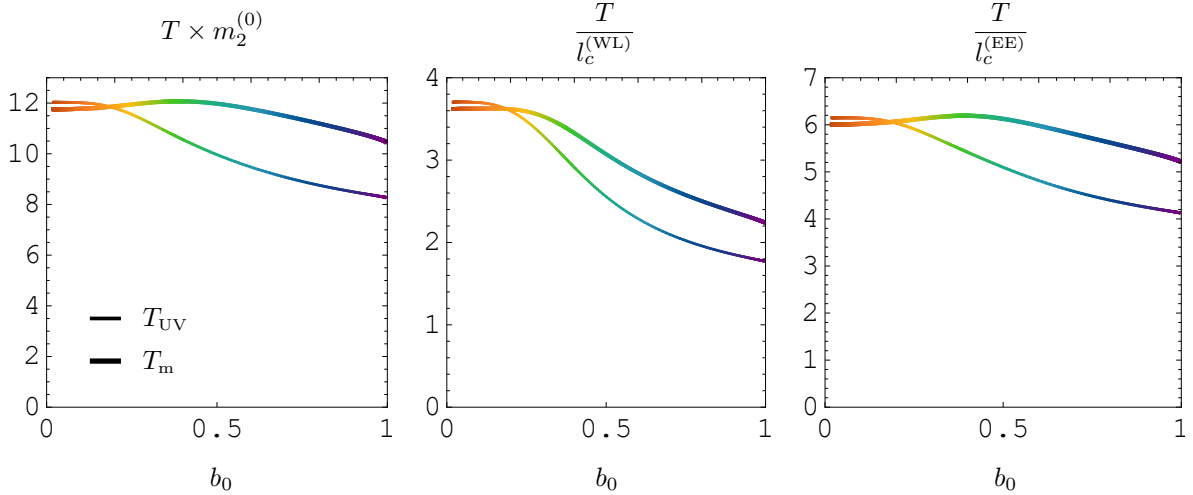
\begin{figure}[t]
	\begin{center}
    \includegraphics[width=1\textwidth]{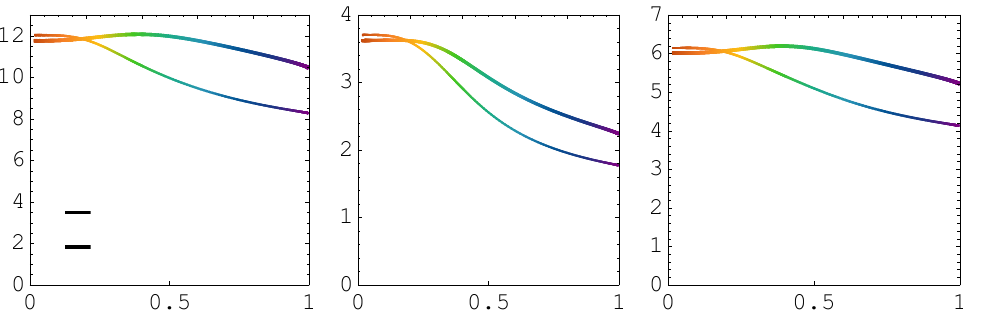} 
    \put(-400,153){\small $\displaystyle \, T\times m^{(0)}_{2}$}
    \put(-407,45){\small $\displaystyle \, \Tmin$}
    \put(-407,29){\small $\displaystyle \, \Tuv$}
    \put(-380,-15){\small $b_0$}
    \put(-245,153){\small $\displaystyle \, \frac{T}{\lcWL}$}
    \put(-235,-15){\small $b_0$}
    \put(-95,153){\small $\displaystyle \, \frac{T}{\lcEE}$}
    \put(-85,-15){\small $b_0$}
	\caption{\small Characteristic periods of a D0-brane released from the boundary (thick) and from the minimum in $A(r)$ (thin) as a function of  $b_0$, in units of the mass of the lightest spin-2 excitation (left), the screening separation of two probe charges (middle) and the critical width for the entanglement entropy of a strip (right).
    } 
    \label{fig:period_gap}
	\end{center}
\end{figure}

Let us now relate our findings to previous results in the literature. First, we believe we can confidently state that \textit{oscillations in the complexity growth of locally excited states are a signature of the smooth end of the dual geometry rather than of confinement.} Refs.~\cite{Fatemiabhari:2025usn,Fatemiabhari:2026goj} found that, in every confining background studied, the probe is trapped between the ultraviolet cutoff and a smooth infrared end of the space, causing $\dot{\CK}(t)$ to oscillate rather than grow monotonically. Ref.~\cite{Fatemiabhari:2026goj} further reported that the frequency of these oscillations is set by the confinement scale. This phenomenon was therefore proposed as a possible ``universal signature of confinement''. In Section~\ref{sec:Krylov} we reproduce the same oscillatory behavior (see, for instance, Figs.~\ref{fig:trajectories}, \ref{fig:oscillations_close_to_CFT}, \ref{fig:trajectoriesD0}, and~\ref{fig:oscillations_close_to_CFT_D0}) throughout the \emph{entire} screened window $0<b_0<1$, including arbitrarily close to $b_0\to0$, where color charges are maximally screened. Together with the caveat already noted in Ref.~\cite{Fatemiabhari:2026goj}---namely, that a finite-dimensional Hilbert space can exhibit recurrences even in the absence of confinement---this indicates that oscillations in complexity alone are not a sufficient diagnostic of confinement. Nevertheless, although the oscillations themselves do not appear to probe confinement, the distinctive nature of the ground state at $b_0=1$ leaves a clear imprint on complexity measures and merits further investigation, as discussed above.

Second, we would also like to comment on the mechanism responsible for the bounce at the end of the space. A closer look at Refs.~\cite{Fatemiabhari:2025usn,Fatemiabhari:2026goj} reveals that the oscillations found there are not perfectly smooth. In fact, the proper momentum jumps by a finite amount at each bounce, which was identified in Ref.~\cite{Fatemiabhari:2026goj} as a coordinate artifact rather than a genuine discontinuity near the end-of-space cap. The subspace spanned by the proper radial coordinate and the shrinking circle is locally a flat plane, and a zero-angular-momentum probe reflects off its origin exactly as a folded Cartesian trajectory would, with its momentum changing sign while preserving its magnitude.\footnote{The situation is similar for a massive probe particle falling radially in the eleven-dimensional uplifted solutions; see Appendix~\ref{app11d}.} This is in contrast to the radial motion of a D0-brane, which, due to its coupling to the dilaton, experiences an effective potential well in the bulk that prevents it from reaching~$\rs$. From the perspective of the eleven-dimensional uplift, the D0-brane is simply an excitation carrying momentum along the M-theory circle, which explains its coupling to the dilaton (see Appendix~\ref{appendix-D0-from-11D}). In this sense, the bounce of the D0-brane is smooth because it is ultimately described by a fundamental object propagating in a regular eleven-dimensional background.

Finally, let us discuss how the scales that emerged in complexity measures ---namely, the periods of the oscillations--- are related to other scales originating from the smooth cap of the geometry. In Fig.~\ref{fig:period_gap} we undertake this comparison. In the left panel, we compare the periods to the mass of the lightest spin-2 state, computed in Ref.~\cite{Elander:2018gte}. From Figs.~4 and 6 of that reference,  in contrast to spin-0 excitations, the spectrum of spin-2 particles is rather featureless. Hence, the lightest state provides a characteristic scale of the discrete spectrum. We see that the two scales seem to be related (up to a factor of order $10$). This connects with the idea that complexity contains \textit{spectroscopic information}~\cite{Fatemiabhari:2025usn,Fatemiabhari:2026goj,Jiang:2025wpj}, showing that there is a qualitative correlation between the periods and the glueball spectrum. Note that this is indeed realized here, despite the absence of confinement.

In addition, we compare the periods to the Wilson--Maldacena screening length $\ell_c^{\rm WL}$ from Ref.~\cite{Faedo:2017fbv} in Fig.~\ref{fig:period_gap}~(middle), and to the critical width for the entanglement entropy of a strip with its complement $\ell_c^{\rm EE}$ in Fig.~\ref{fig:period_gap}~(right), which defines the characteristic scale beyond which the entanglement entropy saturates, reflecting the finite range of quantum correlations, as computed in Ref.~\cite{Jokela:2020wgs}. The agreement is, nevertheless, only qualitative. We conclude that all these scales are related by an approximately constant function of $b_0$.

\subsection*{Outlook}

Our work can be extended in several directions. One natural avenue is to consider other probes~\cite{Nastase:2026lhz,Chatzis:2026ekd}---such as different types of branes or objects carrying momentum in the internal space---in the same geometries and investigate which of them are sensitive to the rich infrared structure of the dual field theories.

It would also be interesting to assess the generality of our results, particularly the appearance of a new maximum in the oscillation period within the walking regime. To this end, one could study complexity in models exhibiting fixed-point annihilation~\cite{Kaplan:2009kr}, such as the Potts model~\cite{Gorbenko:2018dtm,Gorbenko:2018ncu,Cardy:1980wm,Nauenberg:1980nv}. A holographic realization of the same mechanism leading to walking was introduced in Ref.~\cite{Faedo:2019nxw}, where an analogous analysis could be performed.

Finally, the $\Be$ family provides a rich class of regular solutions to eleven-dimensional supergravity in which to investigate other notions of complexity, for instance in connection with tameness and swampland conjectures~\cite{Grimm:2021vpn,Grimm:2024elq,Grimm:2025lip,Grimm:2026haa}. Doing so would require extending these ideas to settings in which the scalar fields parametrizing the moduli space acquire a potential.

\section*{Acknowledgments}

We thank Elena Cáceres, Antón Faedo and David Prieto for useful discussions. CN is supported by  STFC’s grants UKRI4243, ST/Y509644- 1, ST/X000648/1 and ST/T000813/1. JFP is supported by the ‘Atracción de Talento’ program of the Comunidad de Madrid under grant 2020-T1/TIC-20495, by the grants CEX2020-001007-S and PID2021-123017NB-I00 funded by MCIN/AEI/10.13039/501100011033, and by ERDF, EU.

\appendix

\section{Details of the geometry}
\label{app:geometry}

In this Appendix we gather the details concerning the $\Be$ geometries. Their main building block is the squashed complex plane $\CP^3$, and in the UV the metric becomes that of a stack of $N$~coincident D2-branes at the tip of a cone over it. The complex projective space is reproduced as the coset manifold $\rm{Sp}(2)/\rm{U}(2)$, with a two-sphere $\rm{S}^2$ fibered over a four-sphere $\rm{S}^4$.

We use the convenient choice of coordinates of Refs.~\cite{Conde:2011sw,Jokela:2012dw}. Let $\omega^i$ be a set of left-invariant one-forms on a three-sphere. The metric of the unit four-sphere reads
\begin{equation}\label{eq:metricS4}
\dd\Omega_4^2\,=\,\frac{4}{\left(1+\xi^2\right)^2}\left[\dd \xi^2+\frac{\xi^2}{4}\omega^i\omega^i\right]\ ,
\end{equation}
with $\xi$ a non-compact coordinate. Denoting by $\theta$ and $\varphi$ two angles which parametrize the two-sphere, the non-trivial fibration is realized by the vielbeins
\begin{eqnarray}\label{eq:expressionEs}
E^1&=&\dd \theta+\frac{\xi^2}{1+\xi^2}\left(\sin\varphi\,\omega^1-\cos\varphi\,\omega^2\right) \nonumber \\
E^2&=&\sin\theta\left(\dd\varphi-\frac{\xi^2}{1+\xi^2}\omega^3\right)+\frac{\xi^2}{1+\xi^2}\cos\theta\left(\cos\varphi\,\omega^1+\sin\varphi\,\omega^2\right).
\end{eqnarray}
Note that our original ansatz \eqref{10Dansatz} was written in terms of Eqs.~\eqref{eq:metricS4} and~\eqref{eq:expressionEs}.

The set of left-invariant forms on the coset are better expressed in terms of a rotated version of the previous set of coordinates,
\begin{eqnarray}
\mathcal{S}^1&=&\frac{\xi}{1+\xi^2}\left[\sin\varphi\,\omega^1-\cos\varphi\,\omega^2\right] \nonumber\\
\mathcal{S}^2&=&\frac{\xi}{1+\xi^2}\left[\sin\theta\,\omega^3-\cos\theta\left(\cos\varphi\,\omega^1+\sin\varphi\,\omega^2\right)\right] \nonumber\\
\mathcal{S}^3&=&\frac{\xi}{1+\xi^2}\left[\cos\theta\,\omega^3+\sin\theta\left(\cos\varphi\,\omega^1+\sin\varphi\,\omega^2\right)\right] \nonumber\\
\mathcal{S}^4&=&\frac{2}{1+\xi^2}\,\dd\xi \ .
\end{eqnarray}
These satisfy that 
\begin{eqnarray}
\mathcal{S}^1\wedge\mathcal{S}^2\wedge\mathcal{S}^3\wedge\mathcal{S}^4 = \omega_4\,,
\end{eqnarray}
is the volume form in the four-sphere, despite the $\mathcal{S}^i$'s dependence on the angles on the two-sphere. The left-invariant two-forms are now
\begin{equation}\label{eq:X2J2}
 X_2\,=\,E^1\wedge E^2  \, , \qquad  J_2\,=\,\mathcal{S}^1\wedge\mathcal{S}^2+\mathcal{S}^3\wedge\mathcal{S}^4\ ,
\end{equation}
and the three-forms read
\begin{eqnarray}\label{eq:X3J3}
X_3&=&E^1\wedge\left(\mathcal{S}^1\wedge\mathcal{S}^3-\mathcal{S}^2\wedge\mathcal{S}^4\right)-E^2\wedge\left(\mathcal{S}^1\wedge\mathcal{S}^4+\mathcal{S}^2\wedge\mathcal{S}^3\right)\,,
\nonumber\\
J_3&=&-E^1\wedge\left(\mathcal{S}^1\wedge\mathcal{S}^4+\mathcal{S}^2\wedge\mathcal{S}^3\right)-E^2\wedge\left(\mathcal{S}^1\wedge\mathcal{S}^3-\mathcal{S}^2\wedge\mathcal{S}^4\right)\ .
\end{eqnarray}
These satisfy
\begin{equation}
\dd X_2\,=\,\dd J_2\,=\,X_3 \, , \qquad  \dd J_3\,=\,2\left(X_2\wedge J_2+J_2\wedge J_2\right)\ .
\end{equation}
In addition, combining these forms we can construct higher forms which will also be left-invariant. We have the two four-forms 
\begin{equation}
    X_2\wedge J_2\qquad \text{and}\qquad J_2\wedge J_2
\end{equation}
together with the volume form on $\CP^3$ 
\begin{equation}\label{eq:omega6}
    \omega_6 = - (E^1 \wedge E^2)\wedge (\mathcal{S}^1\wedge\mathcal{S}^2\wedge\mathcal{S}^3\wedge\mathcal{S}^4)
\end{equation}
The geometry does not admit left-invariant one- or five-forms. With these elements we can construct the ansatz for the type IIA fluxes given in Eq.~\eqref{eqfluxesansatzFT}.

\section{Eleven-dimensional geometries}\label{app11d}

\subsection{Uplift to eleven dimensions}\label{app:uplift}

The uplift of the ten-dimensional metrics can be performed with the usual ansatz
\begin{equation}
    \dd s_{11}^2 = e^{-\frac{2}{3}\Phi}\dd s_{\rm st}^2  +  e^{\frac{4}{3}\Phi}\ell_p^2\left(\dd\psi + C_1\right)^2\,,
\end{equation}
with $\dd s_{\rm st}^2$ the string frame metric given in Eq.~\eqref{10Dansatz}, $\ell_p$ the eleven-dimensional Planck constant, $\psi$ a coordinate along the M-theory circle and $C_1$ the Ramond–Ramond one-form potential of type IIA,
\begin{equation}
    C_1  = -(\cos\theta \dd\varphi-\xi \mathcal{S}^3)\,,
\end{equation}
see Appendix~\ref{app:geometry}.
We consider here only the horizonless geometries, with $\mathsf{b}  = 1$.

In terms of the vielbein
\begin{equation}
    E^3 = \dd\psi -\cos\theta~ d\phi +\xi ~\mathcal{S}^3
\end{equation}
the eleven-dimensional metric can be written in the M2-brane form 
\begin{equation}
    \dd s_{11}^2 = H^{-2/3}\left(-\dd t^2 + \dd x_1^2 + \dd x_ 2^2\right)  +  H^{1/3}\dd s_8^2\,,
\end{equation}
with 
\begin{equation}
    \dd s_8^2 = e^{-\Lambda} \left[ \mathrm{d}r^2 + e^{2f} \, \mathrm{d}\Omega_4 + e^{2g} \left[ \left(E^1\right)^2 + \left(E^2\right)^2 \right] \right] + e^{\Lambda} Q_k^2 \left(E^3\right)^2\,,
\end{equation}
and warp factor
\begin{equation}
    H = h e^\Lambda\,,
\end{equation}
which is perfectly regular at the end of the space (see Eq.~\eqref{eq:parameters_IR_B8}).

\subsection{Krylov spread complexity from a free-falling particle in eleven dimensions}
\label{appendix-mtheory-particle}

\begin{figure}[t]
	\begin{center}
    \includegraphics[width=0.46\textwidth]{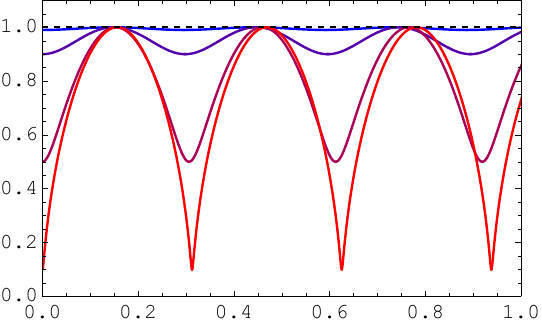} 
   \put(-225,70){$\displaystyle \frac{\rs}{r}$}
    \put(-110,-15){$t \, \Lt$}
    \hfill
    \includegraphics[width=0.46\textwidth]{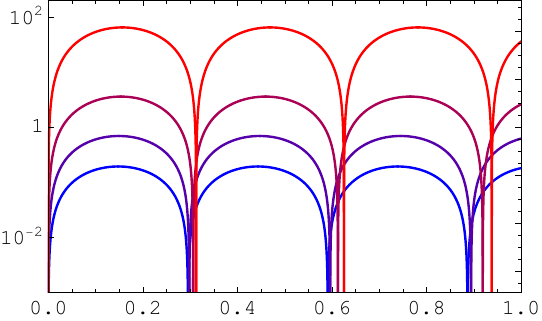} 
   \put(-220,70){$\displaystyle \frac{|\dot\CK|}{m}$}
    \put(-110,-15){$t \, \Lt$}
	\caption{\small 
    Radial trajectories of a massive particle (left) and corresponding Krylov spread complexity growth (right) as a function of time for the theory with $b_0\simeq0.4390$. The different colors stand for different initial positions $r_i$: from blue to red, $\rs/r_i = 0.99$, $0.9$, $0.5$, $0.10$. On the left panel, $\rs/r = 0$ in the vertical axis corresponds to the UV, while $\rs/r = 1$ is the IR, indicated with a horizontal dashed black line.}\label{fig:trajectories-11}
	\end{center}
\end{figure}
It is instructive to analyze the radial motion of a massive particle in the uplifted geometries, a task pursued in this appendix. The world-line of such  particle reads
\begin{equation}
\dd s_{\text{\tiny p, 11D}}^2= e^{-{2\Phi}/{3}}h^{-{1}/{2}}~ \left[-1+ {h}\dot{r}^2 \right] \dd t^2,
\end{equation}
which leads to the action
\begin{equation}
 S =- m \int \dd t \,\sqrt{e^{-{2\Phi}/{3}}h^{-{1}/{2}}\,\left( 1-{h\, \dot{r}^2}\right)}\,.
  \label{action-KS-D0ap}   
\end{equation}
Again, we can use the formulae from Section~\ref{sec:generalformulae} identifying
\begin{equation}
A =  e^{-{2\Phi}/{3}}h^{-{1}/{2}}\,,\qquad B = {h}\,.
\label{eq:definitions11D}
\end{equation}
In this case,
\begin{equation}\label{eq:A_B_particlein11}
    A\sim (r-\rs)^{0}\,,\quad B\sim (r-\rs)^{-1}\,.
\end{equation}
In particular, $A(r)$ is a monotonic function and its minimum value is exactly at $r=\rs$.
Again, by looking at Eq.~\eqref{eq:rdot_Krylov}, we conclude that in this case the particle will always reach the point $r=\rs$, from where it will bounce back. In the IR, the complexity growth reaches a finite (maximum) value due to the finiteness of $A(\rs)$. More precisely, from Eq.~\eqref{eq:CK_growth} we obtain the bound
\begin{equation}
|\dot{\CK}| \leq m\sqrt{\frac{A(\rUV)}{A(r_s)} -1}.
\end{equation}

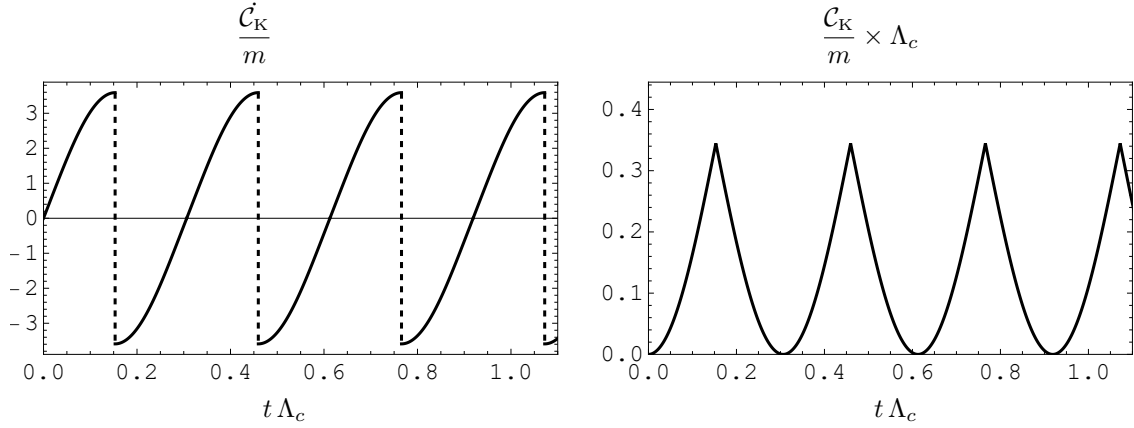
\begin{figure}[t]
    \noindent
    \includegraphics[width=1\textwidth]{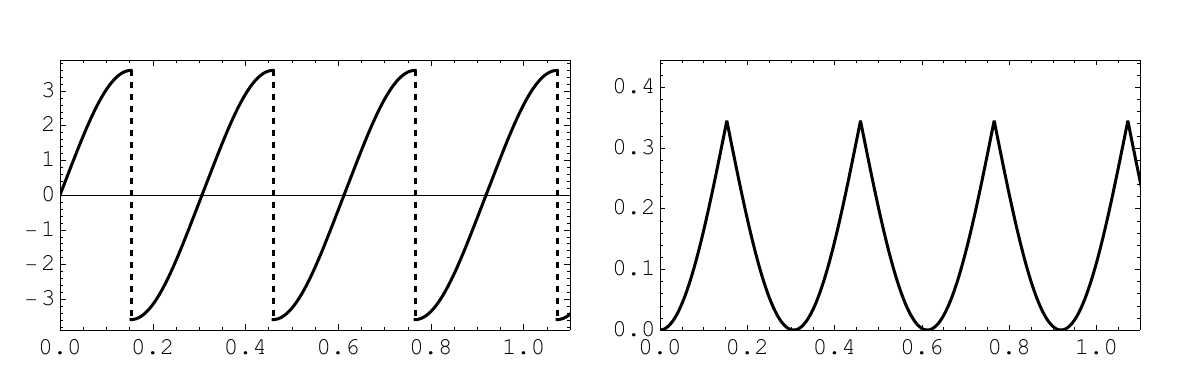} 
    \put(-120,0){\small $t \, \Lt$}
    \put(-350,0){\small $t \, \Lt$}
    \put(-140,140){\small $\displaystyle \frac{\CK}{m}\times \Lt$}
    \put(-360,140){\small $\displaystyle \frac{\dot\CK}{m}$}
	\caption{\small 
    Complexity growth (left) and complexity (right) for a massive particle released from $r=2\rs$ in the eleven-dimensional geometries. Here $b_0=0.4390$. The discontinuity in $|\dot\CK|$ indicated by the dashed line arises, in this case, from the fact that when the particle reaches the IR region it undergoes a change in one of the angular coordinates of the internal manifold. See also related discussion in Refs.~\cite{Fatemiabhari:2025usn,Fatemiabhari:2026goj}.}\label{fig:complexity_and_derivative_11D}
\end{figure}
Similarly to the cases analyzed in the main text, the period of the oscillations is
\begin{equation}\label{eq:time_11Dparticle}
T=2 \int_{\rs}^{r_i}\frac{\dd r \sqrt{B(r)}}{\sqrt{1- \frac{A(r)}{A(r_i)}}}.    
\end{equation}
and the period of a particle released from the boundary is also given by $\Tuv$, Eq.~\eqref{eq:Tmax_particle}. In addition, we can provide an expression for the period of a particle released from close to the IR as follows. The integrand in Eq.~\eqref{eq:time_11Dparticle} diverges at $r=r_i$ (where the denominator vanishes) and at $r=\rs$ (where the numerator blows up). For this reason we find it useful to approximate
\newcommand{\ra}{r_a}
\begin{equation}\label{eq:approx_Ts}
    B(r)^{-1}\left(1-\frac{A(r)}{A(r_i)}\right)\simeq B(\ra)^{-1}\left(1-\frac{A(\ra)}{A(r_i)}\right)\frac{(r_i-r)(r-r_s)}{(\ra-r_i)^2}\,,
\end{equation}
with $\ra = (r_i + \rs)/2$, where both sides of Eq.~\eqref{eq:approx_Ts} coincide. The polynomial on the right hand side captures well the square of the inverse of the integrand when $r_i$ is close to $\rs$. In this regime
\begin{equation}
T\simeq2\frac{(r_i-\ra)\sqrt{B(\ra)}}{\left(\displaystyle1-\frac{A(\ra)} {A(r_i)}\right)^{1/2}}\int_{\rs}^{r_i}\frac{\dd r }{\sqrt{(r_i-r)(r-\rs)}} = 2\pi\, \frac{(r_i-r_a)\sqrt{B(r_a)}}{\left(\displaystyle1-\frac{A(r_a)} {A(r_i)}\right)^{1/2}}.
\end{equation}
Finally, we can take the limit in which $r_i$ approaches $r_s$, resulting in  an expression for the period of a particle oscillation very close to the IR in terms of the parameters from the IR expansions from Eq.~\eqref{eq:parameters_IR_B8},
\begin{equation}\label{eq:TIR_11D}
    \Tir =6\pi\sqrt{\frac{2}{7}} \,  \frac{  f_s^4 h_s}{u_s^2}\, \frac{q_c}{|Q_k|^2}\,.
\end{equation}
Here recall that $\uS=|Q_k|/\rs$.

\begin{figure}[t]
	\begin{center}
    \includegraphics[width=1.05\textwidth]{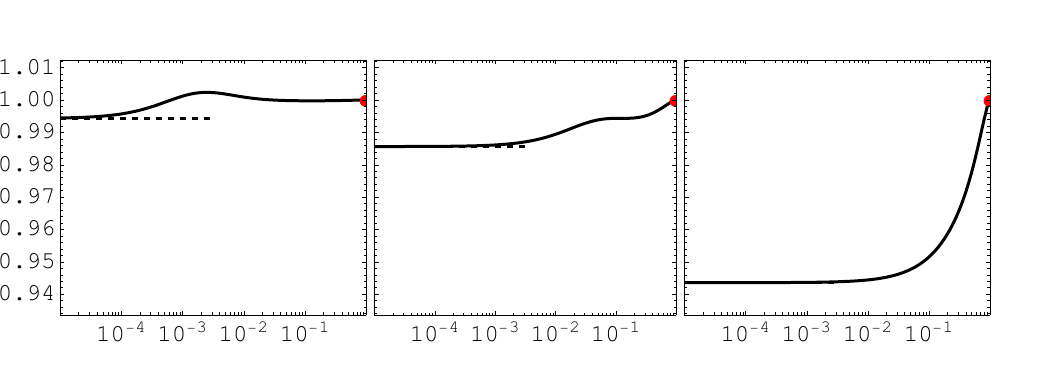} 
   \put(-505,120){\small $\displaystyle \frac{T}{\Tuv}$}
    \put(-125,0){\small $ 1-\rs/\rUV$}
    \put(-260,0){\small $ 1-\rs/\rUV$}
    \put(-400,0){\small $ 1-\rs/\rUV$}
    \put(-120,150){\small $ b_0 = 0.4390$}
    \put(-260,150){\small $ b_0 = 0.1915$}
    \put(-405,150){\small $ b_0 = 0.0605$}
	\caption{\small 
    Periods for a massive particle falling in 11D. The red dot stands for the value $\Tuv$, computed using Eq.~\eqref{eq:Tmax_particle}. The dashed line corresponds to the value of the period close to the IR predicted by Eq.~\eqref{eq:TIR_11D}.
    }\label{fig:period11D}
	\end{center}
\end{figure}

\begin{figure}[t]
    \noindent
    \includegraphics[width=1.1\textwidth]{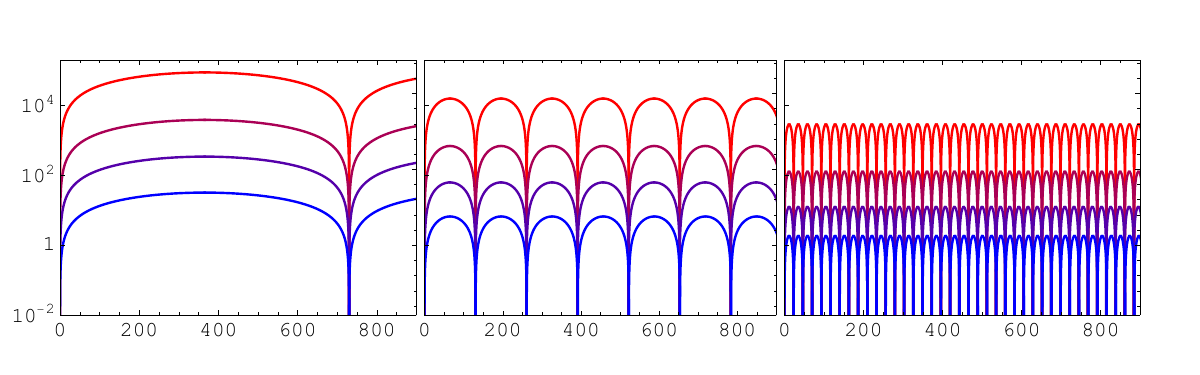} 
    \put(-515,75){$\displaystyle \frac{|\dot\CK|}{m}$}
    \put(-110,0){$t \, \Lt$}
    \put(-260,0){$t \, \Lt$}
    \put(-410,0){$t \, \Lt$}
    \put(-130,135){$b_0=0.1076$}
    \put(-280,135){$b_0=0.0605$}
    \put(-430,135){$b_0=0.0340$}
	\caption{\small 
    Complexity growth obtained from the computation of the massive particle in eleven dimensions, for the different values of $b_0$ indicated on the plots, as a function of time. From blue to red, the value of the initial position is given by $\rs/r_i = 0.99$, $0.9$, $0.5$, $0.10$. The maximum period of the oscillations grows as the conformal theory is approached. The cusps indicate the times at which  turning points of the trajectory occur. }\label{fig:oscillations_close_to_CFT_11D}
\end{figure}
\begin{figure}[t]
	\begin{center}
    \includegraphics[width=0.55\textwidth]{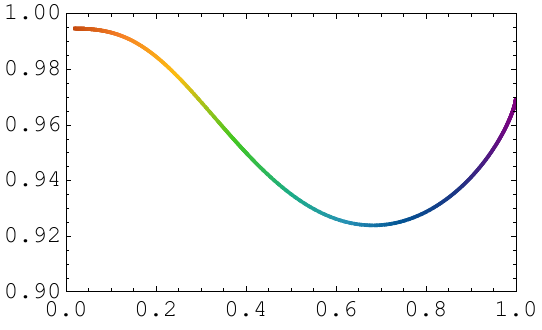} 
   \put(-280,140){$\displaystyle \, \frac{\Tir}{\Tuv}$}
    \put(-120,-15){$b_0$}
	\caption{\small Ratio of the periods of a particle released from the boundary and arbitrarily close to the IR in the eleven-dimensional metric.
    } 
    \label{fig:period_gap_11D}
	\end{center}
\end{figure}
In Fig.~\ref{fig:period11D} we show the values of the periods of the trajectories as a function of the position from which the particle is released, for three different values of $b_0$. Comparing with Fig.~\ref{fig:periodD0}, we conclude that this observable behaves qualitatively similarly for a massive particle in eleven dimensions and for a D0-brane: for sufficiently large values of $b_0$, the period decreases monotonically towards the IR, whereas for smaller values an intermediate maximum appears and eventually becomes the global maximum. 

Finally, let us briefly comment on the behavior in the walking region. As in the previous cases, both scales diverge as the conformal point---see Fig.~\ref{fig:oscillations_close_to_CFT_11D}. Yet, if we compute the ratio between them, we see that they are of the same order for all $b_0$, as shown in Fig.~\ref{fig:period_gap_11D}. 

We conclude that an in-falling massive particle in eleven dimensions yields a qualitatively similar picture to that of an in-falling D0-brane in type IIA, with the caveat that the complexity growth computed for the latter vanishes twice in each period (at the turning points), whereas for the former it vanishes only once, when the particle returns to its initial position. Similarly, most of the conclusions drawn for a massive particle falling in ten dimensions also hold in this case, except for the divergence of the complexity growth, which is absent in M-theory.

\subsection{D0-brane as an excitation of the M-theory circle}\label{appendix-D0-from-11D}

In this appendix we study the action of a {\it massless} particle in eleven dimensions, with momentum along the (eleventh) $\psi$-direction. As we show, this reproduces the action for the  D0 brane in the main body of the paper.  The momentum around the M-theory circle provides the tension of the type IIA D0 excitation.

The compactification of M-theory on a coordinate circle $\psi\sim\psi+4\pi|k|^{-1}$ gives the standard type IIA reduction ansatz
\begin{equation}
ds_{11}^2 \;=\; e^{-2\Phi/3}\,g_{\mu\nu}(x)\,\dd x^\mu \dd x^\nu \;+\; e^{4\Phi/3}\ell_p^2\big(\dd\psi+C_\mu(x)\,\dd x^\mu\big)^2 , \qquad \mu,\nu=0,\dots,9,
\end{equation}
with $g_{\mu\nu}$ the type IIA string-frame metric, $\Phi$ the dilaton, and $C_\mu$ the R-R one-form. 
A massless particle (a KK-wave) in eleven dimensions,  with world-line $X^M(\lambda)=(x^\mu(\lambda),\psi(\lambda))$ and  einbein $e(\lambda)$, has Lagrangian
\begin{equation}
\mathcal L_{11} \;=\; \frac1e\,G_{MN}\dot X^M\dot X^N \;=\; \frac1e\Big[\,e^{-2\Phi/3}g_{\mu\nu}\dot x^\mu\dot x^\nu \;+\; \ell_p^2e^{4\Phi/3}\big(\dot\psi+C_\mu\dot x^\mu\big)^2\,\Big], \qquad \dot{A}\equiv \frac{dA}{d\lambda}.
\end{equation}
Since translations in $\psi$ are isometries, its conjugate momentum is conserved
\begin{equation}
P_\psi \;\equiv\; \frac{\partial\mathcal L_{11}}{\partial\dot\psi} \;=\; \frac{2}{e}\ell_p^2\,e^{4\Phi/3}\big(\dot\psi+C_\mu\dot x^\mu\big) \qquad\Longrightarrow\qquad \dot\psi \;=\; \frac {e}{2\ell_p^2}\,e^{-4\Phi/3}P_\psi \;-\; C_\mu\dot x^\mu .
\end{equation}
The Kaluza--Klein momentum along $\psi$, dubbed $P_\psi$, plays the role of the D0-brane charge. To work at a fixed value of this conserved quantity, we write the Routhian via a Legendre transform, $R\equiv P_\psi\dot\psi-\mathcal L_{11}$ (in other words, we eliminate $\dot\psi$ in favour of $P_\psi$). The Routhian plays the role  of \emph{minus} the effective Lagrangian for the remaining coordinates $x^\mu$. In units where the Planck length $\ell_P=1$, we find,
\begin{equation}
R(e) \;=\; \frac e4\,e^{-4\Phi/3}P_\psi^2 \;+\; \frac1e\,e^{-2\Phi/3}\big(-g_{\mu\nu}\dot x^\mu\dot x^\nu\big) \;-\; P_\psi\,C_\mu\dot x^\mu .
\end{equation}
The einbein is non-dynamical, so it is fixed algebraically by its own equation of motion, $\partial R/\partial e=0$. In more detail,
\begin{equation}
e \;=\; \frac{2\,e^{\Phi/3}}{P_\psi}\sqrt{-g_{\mu\nu}\dot x^\mu\dot x^\nu}\ .
\end{equation}
Substituting back, the two terms quadratic in $e$ are equal and combine to,
\begin{equation}
R\big|_{\rm on\text{-}shell} \;=\; P_\psi\, e^{-\Phi}\sqrt{-g_{\mu\nu}\dot x^\mu\dot x^\nu} \;-\; P_\psi\, C_\mu\dot x^\mu .
\end{equation}
We now fix the world-line reparametrisation invariance in static gauge, $\lambda=x^0\equiv t$ (so $\dot x^0=1$), and identify the D0-brane tension with the KK momentum, $\TD\equiv P_\psi$ (in the standard normalization $\TD=1/(g_s\ell_s)=1/R_{11}$, this is the usual statement that the $n^{\rm th}$ KK mode, $P_\psi=n/R_{11}$, is $n$ coincident D0-branes). Since $\mathcal L_{D0}=-R|_{\rm on\text{-}shell}$,
\begin{equation}
\boxed{\ \mathcal L_{D0} \;=\; \underbrace{-\,\TD\,e^{-\Phi}\sqrt{-g_{\mu\nu}\dot x^\mu\dot x^\nu}\,}_{\text{Born--Infeld}} \;+\; \underbrace{\TD\,C_\mu\dot x^\mu}_{\text{Wess--Zumino}}\ .}
\end{equation}
This is the standard bosonic D0-brane action. We consider now the radial in-fall of Section \ref{sectionD0-IIA}.
For a purely radial trajectory, $x^\mu=(t,r)$ (all other coordinates fixed), in a background of the form used throughout the $B_8$-family,
\begin{equation}
g_{tt}=-h(r)^{-1/2}, \qquad g_{rr}=h(r)^{1/2}, \qquad \Phi=\Phi(r), \qquad C_1=-\cos\theta \dd\varphi +\xi {\cal S}^3\equiv 0,
\end{equation}
we have
\begin{equation}
\mathcal L_{D0} = -\,\TD\,e^{-\Phi}\sqrt{h^{-1/2}(1-h~\dot r^2)}\,,
\end{equation}
which is the action we used in Eq.~(\ref{action-KS-D0}). In this way we have shown why the result of Section \ref{sectionD0-IIA} is robust. When calculating in the non-singular M-theory context, we find the same result as when calculating in type IIA.

\begin{figure}[t]
    \begin{center}
    \includegraphics[width=0.7\textwidth]{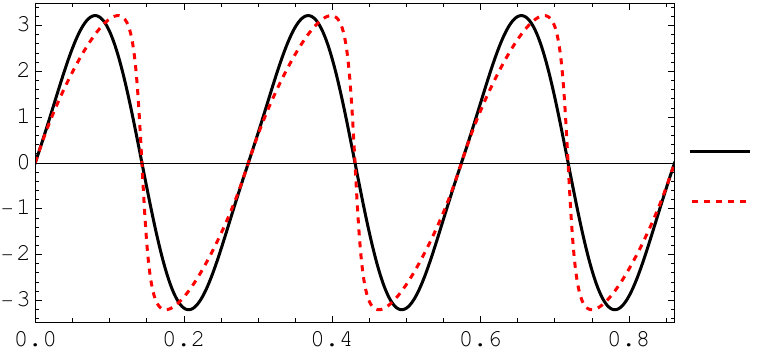}
    \put(-345,85){$\displaystyle \frac{\dot\CK}{\TD}$}
    \put(-175,-12){$t\,\Lt$}
    \put(-3,81){\small \textit{10D (D0-frame)}}
    \put(-3,59){\small \textit{11D}}
    \caption{\small Complexity growth for a D0-brane released from $r_i=2\rs$ in the $b_0=0.4390$ background, computed with the proper momentum of Eq.~\eqref{complexityBBgeneric} (solid) and with the proper radial momentum
    of the eleven-dimensional geometry (dashed). The eleven-dimensional curve is $e^{-2\Phi/3}$ times the ten-dimensional one, rescaled by a constant so that both have the same maximum.}
    \label{fig:D0_vs_11D}
    \end{center}
\end{figure}
Note that the 10D proper momentum of the D0-brane does not coincide with the 11D radial proper momentum of the KK-wave. Indeed,
\begin{equation}
P_{\bar y}=\frac{P_r}{\sqrt{\hat g_{rr}}}=\frac{P_r}{e^{-\Phi}h^{1/4}}\,,\qquad
P^{(11)}_{\hat r}=\frac{P_r}{\sqrt{G^{(11)}_{rr}}}=\frac{P_r}{e^{-\Phi/3}h^{1/4}}\,,
\end{equation}
and therefore
\begin{equation}
    P^{(11)}_{\hat r}=e^{-2\Phi/3}\,P_{\bar y}\,.
\end{equation}
The factor $e^{2\Phi/3}$ is (up to constants) the local radius of the M-theory circle. Because the D0-brane stays in $r_f\le r\le r_i$ along its trajectory and never reaches the tip, this factor is smooth, positive and bounded. As a consequence, both $P_{\bar y}$ and $P_r$
\begin{itemize}
\item vanish at the same instants (the two turning points $r_i$ and $r_f$) and have the same sign at all times;
\item have exactly the same period, which is a property of the trajectory alone; all the periods $\Tuv$, $\Tmin$, $T(r_i)$ discussed in the paper are therefore frame independent;
\item are both bounded and continuous.
\end{itemize}
If we compare them as candidate definitions for $\dot\CK$, we conclude that they differ only in the profile of $\dot\CK$ within each period, as illustrated in Fig.~\ref{fig:D0_vs_11D}. Nevertheless, we note that the eleven-dimensional object is \textit{massless}, so the massive-particle prescription from Refs.\cite{Caputa:2024sux,He:2024pox,Fan:2024iop} cannot be applied to it directly and the D0-frame quantity is the natural ten-dimensional realization of complexity growth.

\section{Confining limit ($b_0 = 1$)}
\label{app:Bconf}

In this appendix we give a few details of the limiting $b_0=1$ case, for which $Q_k = 0$, and the way it is approached. It is useful to work with the variable, defined through 
\begin{equation}\label{eq:change_r_to_xi}
    \dd\xi = -\frac{\xi ^2 \sqrt{1-\xi ^4}}{\rz}\dd r\,.
\end{equation}
In this coordinate, the ground-state geometry has $\xi\in(0,1)$ and 
\begin{equation}
    e^{2f} = \frac{\rho _0^2}{2 \xi ^2}\,,\qquad e^{2g} = \frac{\left(1-\xi ^4\right) \rho _0^2}{4 \xi ^2}\,,\qquad  e^\Lambda = \mathsf{b} =  1\,.
\end{equation}
Then the fluxes are given by
\begin{equation}\label{eq:fluxes_conf}
    \begin{aligned}
   b_J &=\frac{2 q_c}{3 \rho _0}\left[\frac{ \xi\sqrt{1-\xi ^4}  +\left(\xi ^4-3\right) \mathcal{X}(\xi )}{1-\xi ^4}\right]\,,\\[3mm]
   b_X &=-\frac{2 q_c}{3 \rho _0}\left[\xi\sqrt{1-\xi ^4} +\left(\xi ^4-3\right) \mathcal{X}(\xi )\right]\,,\\[3mm]
   a_J &= \frac{q_c}{6} + \frac{2q_c}{3} \cdot \frac{\xi ^3\mathcal{X}(\xi )}{\sqrt{1-\xi ^4}}\,,
    \end{aligned}
\end{equation}
where we have defined the function
\begin{equation}
    \mathcal{X}(\xi) = K(-1)-F\left(\left.\sin ^{-1}(\xi )\right|-1\right)\,,
\end{equation}
with $F(\phi|m)$ the elliptic integral of the first kind and $K(m)$ the complete elliptic integral of the first kind.

Finally, to find the warp factor we need to perform the numerical integration of 

\begin{equation}
    h = \frac{128 q_c^2}{9\rho_0^6}\int_0^\xi \dd\sigma \left(
    \frac{\left(2 \sigma ^4-3\right) \sigma ^5}{\left(\sigma ^4-1\right)^2}
    +\frac{\left(4 \sigma ^8-9 \sigma ^4+9\right) \sigma ^4
   \mathcal{X}(\sigma )}{\left(1-\sigma ^4\right)^{5/2}}
    -\frac{2
   \left(\sigma ^4-3\right) \sigma ^{11} \mathcal{X}(\sigma )^2}{\left(\sigma
   ^4-1\right)^3}
   \right)\,.
\end{equation}
Note that this is finite in the IR, where
$h(\xi = 1) = 128 q_c^2 \hir^{\text{\tiny conf}}/(9\rho_0^6)$ with $\hir^{\text{\tiny conf}} \simeq 0.85545132956$. In particular, this metric is regular in type~IIA supergravity.

In contrast to the ground state, geometries describing finite temperature deconfined phases are only known numerically. These are constructed by connecting the UV expansion
\begin{equation}\label{eq:UV_expansions_conf}
\begin{aligned}
e^{2f}
&=
\frac{\rz^2}{2\xi^2}
\left[
1+2 f_4 \xi ^4+2 f_5 \xi ^5
+\cdots
\right],
\qquad
e^{2g}
=
\frac{\rz^2}{4\xi^2}
\left[
1+\left(-4 f_4-1\right) \xi ^4+2 f_5 \xi ^5
+\cdots
\right],
\\[4pt]
e^{{\Lambda}}
&=
1+2 f_5 \xi ^5+\cdots,
\qquad
\mathsf{b}
=
1+\mathsf{b}_5 \xi^5+\cdots,
\qquad
h
=
-\frac{128 q_c^2 }{15 \rho _0^6}\, \bconf\xi ^5
+\cdots .\\[4pt]
b_X
&=
\frac{2q_c}{3\rz}
\left(-\bconf-4 \xi +\frac{b_4 \xi ^4}{2}+\cdots
\right),
\qquad
b_J
=
\frac{2q_c}{3\rz}
\left(\bconf+4 \xi +b_4 \xi ^4+\cdots
\right),
\\[4pt]
a_J
&=
\frac{q_c}{6}
\left(
1-2 b_4 \xi ^3+\left(-24 f_4-4\right) \xi ^4+\cdots
\right);
\end{aligned}
\end{equation}
with the expansion about the black brane horizon
    \begin{equation}\label{eqBH_expansions_conf}
\begin{array}{lll}
     e^{2f} = \rz^2 \,\fH+\cdots\,, \quad
     & e^{2g} = \rz^2 \,\gH+\cdots \,, \quad
     &e^{\Lambda} =  \lH + \cdots \,, 
     \\[2mm]
     \mathsf{b} = \bH (\xi-\xih)+\cdots \,,\quad 
      &\displaystyle h 
=
\frac{128 q_c^2 }{15 \rho _0^6} \hH +\cdots \,, \\
\end{array}
    \end{equation}
where the fluxes are also constant. These two boundary conditions are connected by means of a \textit{shooting} procedure. 

\begin{figure}[t]
	\begin{center}
    \textit{Massive particle in ten dimensions.}\vspace{2mm}
    
    \includegraphics[width=0.46\textwidth]{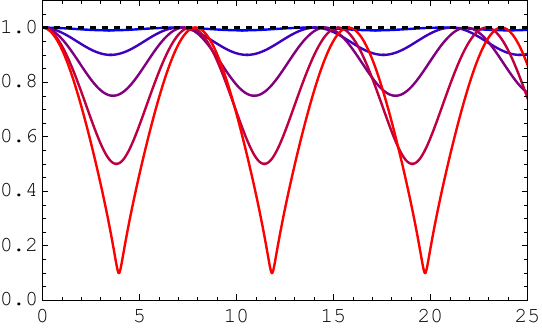} 
   \put(-225,70){$\displaystyle \xi$}
    \put(-120,-15){$t \, \rho_0^2/q_c$}
    \hfill
    \includegraphics[width=0.46\textwidth]{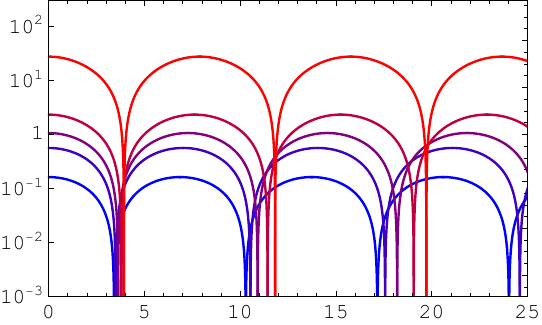} 
   \put(-225,70){$\displaystyle \frac{\dot\CK}{m}$}
    \put(-120,-15){$t \, \rho_0^2/q_c$}
    \vspace{7mm}

    \textit{D0-brane.}\vspace{2mm}
    
    \includegraphics[width=0.46\textwidth]{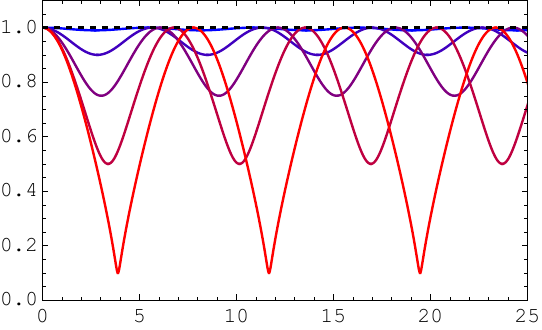} 
   \put(-225,70){$\displaystyle \xi$}
    \put(-120,-15){$t \, \rho_0^2/q_c$}
    \hfill
    \includegraphics[width=0.46\textwidth]{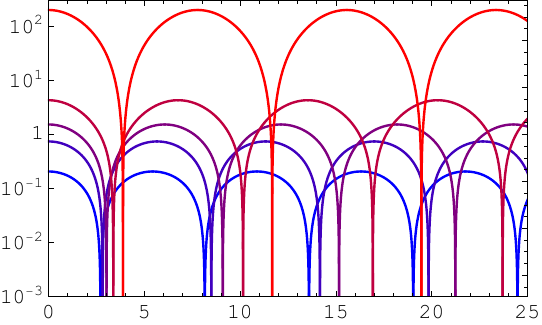} 
   \put(-225,70){$\displaystyle \frac{\dot\CK}{\TD}$}
    \put(-120,-15){$t \, \rho_0^2/q_c$}
    \vspace{7mm}
    
    \textit{Massive particle in eleven dimensions.}\vspace{2mm}
    
    \includegraphics[width=0.46\textwidth]{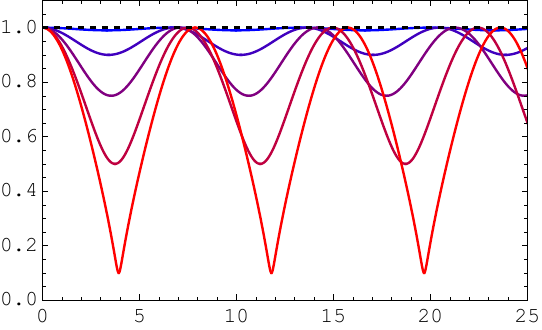} 
   \put(-225,70){$\displaystyle \xi$}
    \put(-120,-15){$t \, \rho_0^2/q_c$}
    \hfill
    \includegraphics[width=0.46\textwidth]{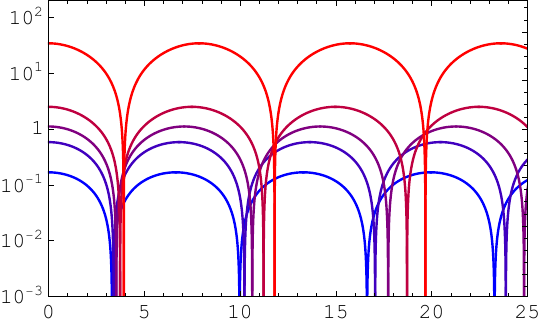} 
   \put(-225,70){$\displaystyle \frac{\dot\CK}{m}$}
    \put(-120,-15){$t \, \rho_0^2/q_c$}
	\caption{\small 
    Trajectories (left) and Krylov spread complexity growth (right) as a function of time for the confining theory $\Bconf$, for the cases indicated in the title of the different plots. The colors stand for various initial positions $\xi_i$: from blue to red, $\xi_i = 0.99$, $0.9$, $0.75$, $0.5$, $0.10$. Note that the divergence for the in-falling massive particle in ten dimensions is absent.}\label{fig:trajectories_B8conf}
	\end{center}
\end{figure}
For these black brane solutions to correspond to the finite temperature states of the ground state just presented, it is necessary that the metrics approach the boundary with the same leading order coefficients. Demanding that the fluxes in Eq.~\eqref{eq:UV_expansions_conf} attain the same value at the boundary as in Eq.~\eqref{eq:fluxes_conf}, we obtain the condition
\begin{equation}
    \bconf = -3 K(-1) = -\frac{3 \Gamma \left(1/4\right)^2}{4 \sqrt{2 \pi }} \simeq -3.9331\,.
\end{equation}
With this it is possible to understand how to take the limit $b_0\to1$ of the $\Be$ family so that these geometries are recovered. Recall that $\Bconf$ has $|Q_k| = 0$. Demanding that the leading order of the warp factors in the original coordinate coincide in this limit\footnote{Close to the boundary, $\displaystyle \xi = \frac{\rz}{r}+\cdots$, which can be seen by integrating Eq.~\eqref{eq:change_r_to_xi}.}
\begin{equation}
    \lim_{b_0\to1} \frac{64 \left(1-b_0^2\right) q_c^2}{15 |Q_k| r^5} +\cdots= \frac{128 K(-1) q_c^2}{5\rz r^5} +\cdots\,
\end{equation}
we conclude that
\begin{equation}\label{eq:rescaling_Qk}
    \rz = \lim_{b_0\to1}\frac{6 K(-1)}{1-b_0^2}|Q_k|\,,
\end{equation}
which means that vanishes as $Q_k\simeq(1-b_0^2)$ in this limit. With this rescaling we saw, for example, that the growth of complexity computed at finite temperature from wormhole solutions in the $\Bconf$ theory is recovered as the limit of $Q_k\to0$ of the $\Be$ family in Fig.~\ref{fig:complexity_growth_confinement}.

Let us now discuss Krylov spread complexity measures for this theory. In this case, the geometries are regular solutions of type IIA, with a warp factor and a dilaton that become constant in the IR. This feature introduces a drastic qualitative difference from the $b_0\in(0,1)$ cases. Indeed, as can be seen in Fig.~\ref{fig:trajectories_B8conf}, the absence of a singularity in the ten-dimensional description removes the divergence that we found in the complexity growth computed from a particle falling in ten dimensions. In addition, the motion of the D0-brane does not stop in the IR, but bounces back, as happens for ordinary particles. As a consequence, the complexity growth no longer vanishes in the IR, in contrast to the case with $Q_k\neq0$, where $\dot\CK$ vanishes twice in every period. From its eleven-dimensional description, the absence of an ``IR barrier'' for these D0-branes is a consequence of the trivial fibration of the M-theory circle.
\begin{figure}[t]
	\begin{center}
    \includegraphics[width=.6\textwidth]{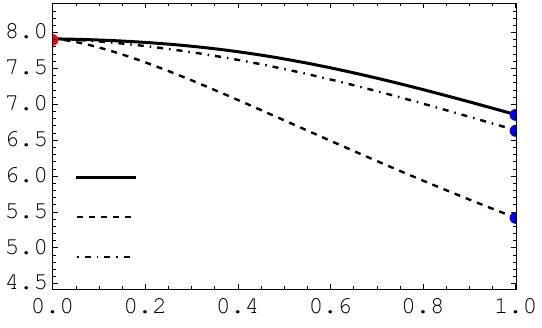} 
   \put(-505,120){\small$\displaystyle  \frac{T}{\Tuv}$}
    \put(-135,-15){ $ \xi_i $}
    \put(-310,145){$\displaystyle \,  T \, \frac{{ \rz^2}}{q_c}$}
    \put(-200,72){\small \textit{particle in 10D}}
    \put(-200,52){\small \textit{D0-brane}}
    \put(-200,32){\small \textit{particle in 11D}}
	\caption{\small 
    Period of the oscillations of different in-falling objects as a function of the position from which they are released in $\Bconf$. The three different curves correspond to the three probes indicated in the plot. The red dot stands for the $\Tuv$, given by Eq.~\eqref{eq:Tmax_particle}. The blue dots correspond to the values in Eq.~\eqref{eq:approx_periods_11D}.
    }\label{fig:period_B8conf}
	\end{center}
\end{figure}

Finally, it is also instructive to compute the value of the periods for these different objects as a function of the point from which they are released, as depicted in Fig.~\ref{fig:period_B8conf}. In this case, the periods monotonically decrease as the object is released closer to the IR, as we found for a massive particle in eleven dimensions in theories with CS interactions.
The maximum value of the period is again given by Eq.~\eqref{eq:Tmax_particle} and is in this case 
\begin{equation}\label{eq:Tuv_conf}
\Tuv^{(Q_k=0)} \simeq 7.9068477704 \frac{q_c}{\rz^2}\,.
\end{equation}
This coincides with the limit $b_0\to1$ of the values of $\Tuv$ that we obtained for $Q_k\neq0$, once the
rescaling~\eqref{eq:rescaling_Qk} is performed. Indeed, constructing backgrounds up to $1-b_0\simeq 3\times10^{-5}$
and extrapolating we find
\begin{equation}\label{eq:Tuv_limit}
    \lim_{b_0\to1}\Tuv^{(Q_k\neq0)} = 7.90685(5)\, \frac{q_c}{\rz^2}\,,
\end{equation}
in agreement with Eq.~\eqref{eq:Tuv_conf}. This is expected: $\Tuv$ is an integral over the whole geometry, and the region near the tip in which the $Q_k\neq0$ and $Q_k=0$ solutions differ by order-one amounts shrinks to zero size as $b_0\to1$, while $\sqrt h$ remains integrable there. Note that the comparison only relies on the identification of
the UV scale, Eq.~\eqref{eq:rescaling_Qk}, which matches the leading asymptotic D2-brane behaviour of both geometries; any matching that agrees with it at leading order differs by terms vanishing as $b_0\to1$, and gives the same limit.

\begin{table}
\centering
\begin{tabular}{lccc}
\toprule
 & ~~10D particle~~ & ~~D0-brane ($\Tmin$)~~ & ~~11D particle~~\\
\midrule
$\lim_{b_0\to1}$ of the ratio to $\Tuv$~~ & $0$ & $0.791$ & $0.969$\\
ratio to $\Tuv$ in $\Bconf$ & $0.8667$ & $0.6852$ & $0.8391$\\
\toprule
\end{tabular}
\caption{Comparison of the different IR periods in the $b_0\to1$ limit.}\label{table}
\end{table}
In contrast, the periods of probes released close to the IR are different: they are sensitive to the near-tip region and their confining limit is not uniform. Using a procedure similar to the one that led to Eqs.~\eqref{eq:Tmin_D0} and~\eqref{eq:TIR_11D}, we find that in this case the periods are bounded from below by
\begin{equation}\label{eq:approx_periods_11D}
\begin{aligned}
    \Tir &=\frac{32\pi}{3} \sqrt{\frac{2}{35}}  \, \hir^{\text{\tiny conf}}\simeq 6.853 \frac{q_c}{\rz^2}\,,\qquad \text{\textit{(particle in 10D),}}\\[2mm]
    \Tir &=\frac{16 \pi  }{3 \sqrt{7}}\, \hir^{\text{\tiny conf}}\simeq 5.417 \frac{q_c}{\rz^2}\,,\qquad \qquad \text{\textit{(D0-brane),}}\\[2mm]
    \Tir &= 8\pi  \sqrt{\frac{2}{21}} \,  \hir^{\text{\tiny conf}}\simeq 6.635\frac{q_c}{\rz^2}\,,\qquad \text{\textit{(particle in 11D).}}\\ 
\end{aligned}
\end{equation}
The key differences are:
\begin{itemize}
\item{  for the in-falling particle in ten dimensions the period does not vanish in the IR.}
\item{  for the D0-brane the minimum period is also provided by the case in which it is released close to the IR, rather than by an intermediate value of the radial coordinate.}
\end{itemize}
As summarised in Table~\ref{table}, the ratios of these periods to $\Tuv$  do not coincide with the $b_0\to1$ limits of the corresponding ratios of the theories with $Q_k\neq0$. For the massive particle in ten dimensions the period vanishes when it is released close to the tip, see Fig.~\ref{fig:period}. For the D0-brane, the minimum period tends to $\Tmin/\Tuv\to0.791$ (Fig.~\ref{fig:period_walking_D0}), and for the massive particle in eleven dimensions $\Tir/\Tuv\to0.969$ (Fig.~\ref{fig:period_gap_11D}). In other words, the limits
$b_0\to1$ and $r_i\to\rs$ do not commute.

\bibliographystyle{JHEP}
\bibliography{main.bib}

\providecommand{\href}[2]{#2}\begingroup\raggedright\begin{thebibliography}{100}

\bibitem{Cvetic:2001bw}
M.~Cvetic, G.~W. Gibbons, J.~T. Liu, H.~Lu and C.~N. Pope, \emph{{A New fractional D2-brane, G(2) holonomy and T duality}}, \href{https://doi.org/10.1088/0264-9381/19/20/310}{\emph{Class. Quant. Grav.} {\bfseries 19} (2002) 5163} [\href{https://arxiv.org/abs/hep-th/0106162}{{\ttfamily hep-th/0106162}}].

\bibitem{Herzog:2002ss}
C.~P. Herzog, \emph{{String tensions and three-dimensional confining gauge theories}}, \href{https://doi.org/10.1103/PhysRevD.66.065009}{\emph{Phys. Rev. D} {\bfseries 66} (2002) 065009} [\href{https://arxiv.org/abs/hep-th/0205064}{{\ttfamily hep-th/0205064}}].

\bibitem{Cvetic:2001pga}
M.~Cvetic, G.~W. Gibbons, H.~Lu and C.~N. Pope, \emph{{New complete noncompact spin(7) manifolds}}, \href{https://doi.org/10.1016/S0550-3213(01)00559-4}{\emph{Nucl. Phys. B} {\bfseries 620} (2002) 29} [\href{https://arxiv.org/abs/hep-th/0103155}{{\ttfamily hep-th/0103155}}].

\bibitem{Faedo:2017fbv}
A.~F. Faedo, D.~Mateos, D.~Pravos and J.~G. Subils, \emph{{Mass Gap without Confinement}}, \href{https://doi.org/10.1007/JHEP06(2017)153}{\emph{JHEP} {\bfseries 06} (2017) 153} [\href{https://arxiv.org/abs/1702.05988}{{\ttfamily 1702.05988}}].

\bibitem{GomezSubils:2021dzb}
J.~G. Subils, \emph{{Non-perturbative Aspects of Quantum Field Theories from Holography}}, Ph.D. thesis, Universitat de Barcelona, 2021.
\newblock \href{https://arxiv.org/abs/2107.01954}{{\ttfamily 2107.01954}}.

\bibitem{Elander:2018gte}
D.~Elander, A.~F. Faedo, D.~Mateos, D.~Pravos and J.~G. Subils, \emph{{Mass spectrum of gapped, non-confining theories with multi-scale dynamics}}, \href{https://doi.org/10.1007/JHEP05(2019)175}{\emph{JHEP} {\bfseries 05} (2019) 175} [\href{https://arxiv.org/abs/1810.04656}{{\ttfamily 1810.04656}}].

\bibitem{Faedo:2025nqc}
A.~F. Faedo, C.~Hoyos and J.~G. Subils, \emph{{Spontaneous breaking of baryon symmetry in strongly coupled three-dimensional theories}}, \href{https://doi.org/10.1016/j.physletb.2025.139894}{\emph{Phys. Lett. B} {\bfseries 870} (2025) 139894} [\href{https://arxiv.org/abs/2507.15998}{{\ttfamily 2507.15998}}].

\bibitem{Gubser:2004qj}
S.~S. Gubser, C.~P. Herzog and I.~R. Klebanov, \emph{{Symmetry breaking and axionic strings in the warped deformed conifold}}, \href{https://doi.org/10.1088/1126-6708/2004/09/036}{\emph{JHEP} {\bfseries 09} (2004) 036} [\href{https://arxiv.org/abs/hep-th/0405282}{{\ttfamily hep-th/0405282}}].

\bibitem{Butti:2004pk}
A.~Butti, M.~Grana, R.~Minasian, M.~Petrini and A.~Zaffaroni, \emph{{The Baryonic branch of Klebanov-Strassler solution: A supersymmetric family of SU(3) structure backgrounds}}, \href{https://doi.org/10.1088/1126-6708/2005/03/069}{\emph{JHEP} {\bfseries 03} (2005) 069} [\href{https://arxiv.org/abs/hep-th/0412187}{{\ttfamily hep-th/0412187}}].

\bibitem{Dymarsky:2005xt}
A.~Dymarsky, I.~R. Klebanov and N.~Seiberg, \emph{{On the moduli space of the cascading SU(M+p) x SU(p) gauge theory}}, \href{https://doi.org/10.1088/1126-6708/2006/01/155}{\emph{JHEP} {\bfseries 01} (2006) 155} [\href{https://arxiv.org/abs/hep-th/0511254}{{\ttfamily hep-th/0511254}}].

\bibitem{Elander:2020rgv}
D.~Elander, A.~F. Faedo, D.~Mateos and J.~G. Subils, \emph{{Phase transitions in a three-dimensional analogue of Klebanov-Strassler}}, \href{https://doi.org/10.1007/JHEP06(2020)131}{\emph{JHEP} {\bfseries 06} (2020) 131} [\href{https://arxiv.org/abs/2002.08279}{{\ttfamily 2002.08279}}].

\bibitem{Jokela:2020wgs}
N.~Jokela and J.~G. Subils, \emph{{Is entanglement a probe of confinement?}}, \href{https://doi.org/10.1007/JHEP02(2021)147}{\emph{JHEP} {\bfseries 02} (2021) 147} [\href{https://arxiv.org/abs/2010.09392}{{\ttfamily 2010.09392}}].

\bibitem{Jokela:2023lvr}
N.~Jokela, H.~Ruotsalainen and J.~G. Subils, \emph{{Limitations of entanglement entropy in detecting thermal phase transitions}}, \href{https://doi.org/10.1007/JHEP01(2024)186}{\emph{JHEP} {\bfseries 01} (2024) 186} [\href{https://arxiv.org/abs/2310.11205}{{\ttfamily 2310.11205}}].

\bibitem{Maldacena:2001kr}
J.~M. Maldacena, \emph{{Eternal black holes in anti-de Sitter}}, \href{https://doi.org/10.1088/1126-6708/2003/04/021}{\emph{JHEP} {\bfseries 04} (2003) 021} [\href{https://arxiv.org/abs/hep-th/0106112}{{\ttfamily hep-th/0106112}}].

\bibitem{Susskind:2014rva}
L.~Susskind, \emph{{Computational Complexity and Black Hole Horizons}}, \href{https://doi.org/10.1002/prop.201500092}{\emph{Fortsch. Phys.} {\bfseries 64} (2016) 24} [\href{https://arxiv.org/abs/1403.5695}{{\ttfamily 1403.5695}}].

\bibitem{Susskind:2014moa}
L.~Susskind, \emph{{Entanglement is not enough}}, \href{https://doi.org/10.1002/prop.201500095}{\emph{Fortsch. Phys.} {\bfseries 64} (2016) 49} [\href{https://arxiv.org/abs/1411.0690}{{\ttfamily 1411.0690}}].

\bibitem{Nielsen:2005mkt}
M.~A. Nielsen, \emph{{A geometric approach to quantum circuit lower bounds}}, \href{https://doi.org/10.26421/QIC6.3-2}{\emph{Quant. Inf. Comput.} {\bfseries 6} (2006) 213} [\href{https://arxiv.org/abs/quant-ph/0502070}{{\ttfamily quant-ph/0502070}}].

\bibitem{Nielsen:2006cea}
M.~A. Nielsen, M.~R. Dowling, M.~Gu and A.~C. Doherty, \emph{{Quantum Computation as Geometry}}, \href{https://doi.org/10.1126/science.1121541}{\emph{Science} {\bfseries 311} (2006) 1133} [\href{https://arxiv.org/abs/quant-ph/0603161}{{\ttfamily quant-ph/0603161}}].

\bibitem{Jefferson:2017sdb}
R.~Jefferson and R.~C. Myers, \emph{{Circuit complexity in quantum field theory}}, \href{https://doi.org/10.1007/JHEP10(2017)107}{\emph{JHEP} {\bfseries 10} (2017) 107} [\href{https://arxiv.org/abs/1707.08570}{{\ttfamily 1707.08570}}].

\bibitem{Chapman:2017rqy}
S.~Chapman, M.~P. Heller, H.~Marrochio and F.~Pastawski, \emph{{Toward a Definition of Complexity for Quantum Field Theory States}}, \href{https://doi.org/10.1103/PhysRevLett.120.121602}{\emph{Phys. Rev. Lett.} {\bfseries 120} (2018) 121602} [\href{https://arxiv.org/abs/1707.08582}{{\ttfamily 1707.08582}}].

\bibitem{Stanford:2014jda}
D.~Stanford and L.~Susskind, \emph{{Complexity and Shock Wave Geometries}}, \href{https://doi.org/10.1103/PhysRevD.90.126007}{\emph{Phys. Rev. D} {\bfseries 90} (2014) 126007} [\href{https://arxiv.org/abs/1406.2678}{{\ttfamily 1406.2678}}].

\bibitem{Pedraza:2021mkh}
J.~F. Pedraza, A.~Russo, A.~Svesko and Z.~Weller-Davies, \emph{{Lorentzian Threads as Gatelines and Holographic Complexity}}, \href{https://doi.org/10.1103/PhysRevLett.127.271602}{\emph{Phys. Rev. Lett.} {\bfseries 127} (2021) 271602} [\href{https://arxiv.org/abs/2105.12735}{{\ttfamily 2105.12735}}].

\bibitem{Pedraza:2021fgp}
J.~F. Pedraza, A.~Russo, A.~Svesko and Z.~Weller-Davies, \emph{{Sewing spacetime with Lorentzian threads: complexity and the emergence of time in quantum gravity}}, \href{https://doi.org/10.1007/JHEP02(2022)093}{\emph{JHEP} {\bfseries 02} (2022) 093} [\href{https://arxiv.org/abs/2106.12585}{{\ttfamily 2106.12585}}].

\bibitem{Caceres:2025ypk}
E.~C{\'a}ceres, R.~Carrasco and J.~F. Pedraza, \emph{{Lorentzian threads and nonlocal computation in holography}}, \href{https://doi.org/10.1103/bk1l-3ffb}{\emph{Phys. Rev. D} {\bfseries 113} (2026) 106024} [\href{https://arxiv.org/abs/2512.07963}{{\ttfamily 2512.07963}}].

\bibitem{Pedraza:2022dqi}
J.~F. Pedraza, A.~Russo, A.~Svesko and Z.~Weller-Davies, \emph{{Computing spacetime}}, \href{https://doi.org/10.1142/S021827182242010X}{\emph{Int. J. Mod. Phys. D} {\bfseries 31} (2022) 2242010} [\href{https://arxiv.org/abs/2205.05705}{{\ttfamily 2205.05705}}].

\bibitem{Carrasco:2023fcj}
R.~Carrasco, J.~F. Pedraza, A.~Svesko and Z.~Weller-Davies, \emph{{Gravitation from optimized computation: Einstein and beyond}}, \href{https://doi.org/10.1007/JHEP09(2023)167}{\emph{JHEP} {\bfseries 09} (2023) 167} [\href{https://arxiv.org/abs/2306.08503}{{\ttfamily 2306.08503}}].

\bibitem{Brown:2015bva}
A.~R. Brown, D.~A. Roberts, L.~Susskind, B.~Swingle and Y.~Zhao, \emph{{Holographic Complexity Equals Bulk Action?}}, \href{https://doi.org/10.1103/PhysRevLett.116.191301}{\emph{Phys. Rev. Lett.} {\bfseries 116} (2016) 191301} [\href{https://arxiv.org/abs/1509.07876}{{\ttfamily 1509.07876}}].

\bibitem{Brown:2015lvg}
A.~R. Brown, D.~A. Roberts, L.~Susskind, B.~Swingle and Y.~Zhao, \emph{{Complexity, action, and black holes}}, \href{https://doi.org/10.1103/PhysRevD.93.086006}{\emph{Phys. Rev. D} {\bfseries 93} (2016) 086006} [\href{https://arxiv.org/abs/1512.04993}{{\ttfamily 1512.04993}}].

\bibitem{Belin:2021bga}
A.~Belin, R.~C. Myers, S.-M. Ruan, G.~S{\'a}rosi and A.~J. Speranza, \emph{{Does Complexity Equal Anything?}}, \href{https://doi.org/10.1103/PhysRevLett.128.081602}{\emph{Phys. Rev. Lett.} {\bfseries 128} (2022) 081602} [\href{https://arxiv.org/abs/2111.02429}{{\ttfamily 2111.02429}}].

\bibitem{Myers:2024vve}
R.~C. Myers and S.-M. Ruan, \emph{{Complexity Equals (Almost) Anything}},  in \emph{{Gravity, Strings and Fields}: {A Conference in Honour of Gordon Semenoff}}, 3, 2024, \href{https://arxiv.org/abs/2403.17475}{{\ttfamily 2403.17475}}, \href{https://doi.org/10.1007/978-3-031-91266-5_11}{DOI}.

\bibitem{Caceres:2025myu}
E.~C{\'a}ceres, R.~Carrasco, V.~Patil, J.~F. Pedraza and A.~Svesko, \emph{{The landscape of complexity measures in 2D gravity}}, \href{https://doi.org/10.1007/JHEP10(2025)218}{\emph{JHEP} {\bfseries 10} (2025) 218} [\href{https://arxiv.org/abs/2503.20943}{{\ttfamily 2503.20943}}].

\bibitem{Carmi:2016wjl}
D.~Carmi, R.~C. Myers and P.~Rath, \emph{{Comments on Holographic Complexity}}, \href{https://doi.org/10.1007/JHEP03(2017)118}{\emph{JHEP} {\bfseries 03} (2017) 118} [\href{https://arxiv.org/abs/1612.00433}{{\ttfamily 1612.00433}}].

\bibitem{Carmi:2017jqz}
D.~Carmi, S.~Chapman, H.~Marrochio, R.~C. Myers and S.~Sugishita, \emph{{On the Time Dependence of Holographic Complexity}}, \href{https://doi.org/10.1007/JHEP11(2017)188}{\emph{JHEP} {\bfseries 11} (2017) 188} [\href{https://arxiv.org/abs/1709.10184}{{\ttfamily 1709.10184}}].

\bibitem{Chapman:2016hwi}
S.~Chapman, H.~Marrochio and R.~C. Myers, \emph{{Complexity of Formation in Holography}}, \href{https://doi.org/10.1007/JHEP01(2017)062}{\emph{JHEP} {\bfseries 01} (2017) 062} [\href{https://arxiv.org/abs/1610.08063}{{\ttfamily 1610.08063}}].

\bibitem{Reynolds:2016rvl}
A.~Reynolds and S.~F. Ross, \emph{{Divergences in Holographic Complexity}}, \href{https://doi.org/10.1088/1361-6382/aa6925}{\emph{Class. Quant. Grav.} {\bfseries 34} (2017) 105004} [\href{https://arxiv.org/abs/1612.05439}{{\ttfamily 1612.05439}}].

\bibitem{Chapman:2018dem}
S.~Chapman, H.~Marrochio and R.~C. Myers, \emph{{Holographic complexity in Vaidya spacetimes. Part I}}, \href{https://doi.org/10.1007/JHEP06(2018)046}{\emph{JHEP} {\bfseries 06} (2018) 046} [\href{https://arxiv.org/abs/1804.07410}{{\ttfamily 1804.07410}}].

\bibitem{Jorstad:2023kmq}
E.~J{\o}rstad, R.~C. Myers and S.-M. Ruan, \emph{{Complexity=Anything: Singularity Probes}}, \href{https://doi.org/10.1007/JHEP07(2023)223}{\emph{JHEP} {\bfseries 07} (2023) 223} [\href{https://arxiv.org/abs/2304.05453}{{\ttfamily 2304.05453}}].

\bibitem{Arean:2024pzo}
D.~Are{\'a}n, H.-S. Jeong, J.~F. Pedraza and L.-C. Qu, \emph{{Kasner interiors from analytic hairy black holes}}, \href{https://doi.org/10.1007/JHEP11(2024)138}{\emph{JHEP} {\bfseries 11} (2024) 138} [\href{https://arxiv.org/abs/2407.18430}{{\ttfamily 2407.18430}}].

\bibitem{Nozaki:2013wia}
M.~Nozaki, T.~Numasawa and T.~Takayanagi, \emph{{Holographic Local Quenches and Entanglement Density}}, \href{https://doi.org/10.1007/JHEP05(2013)080}{\emph{JHEP} {\bfseries 05} (2013) 080} [\href{https://arxiv.org/abs/1302.5703}{{\ttfamily 1302.5703}}].

\bibitem{Caputa:2014eta}
P.~Caputa, J.~Sim{\'o}n, A.~{\v{S}}tikonas and T.~Takayanagi, \emph{{Quantum Entanglement of Localized Excited States at Finite Temperature}}, \href{https://doi.org/10.1007/JHEP01(2015)102}{\emph{JHEP} {\bfseries 01} (2015) 102} [\href{https://arxiv.org/abs/1410.2287}{{\ttfamily 1410.2287}}].

\bibitem{Agon:2020fqs}
C.~A. Ag{\'o}n, S.~F. Lokhande and J.~F. Pedraza, \emph{{Local quenches, bulk entanglement entropy and a unitary Page curve}}, \href{https://doi.org/10.1007/JHEP08(2020)152}{\emph{JHEP} {\bfseries 08} (2020) 152} [\href{https://arxiv.org/abs/2004.15010}{{\ttfamily 2004.15010}}].

\bibitem{Balasubramanian:2022tpr}
V.~Balasubramanian, P.~Caputa, J.~M. Magan and Q.~Wu, \emph{{Quantum chaos and the complexity of spread of states}}, \href{https://doi.org/10.1103/PhysRevD.106.046007}{\emph{Phys. Rev. D} {\bfseries 106} (2022) 046007} [\href{https://arxiv.org/abs/2202.06957}{{\ttfamily 2202.06957}}].

\bibitem{Parker:2018yvk}
D.~E. Parker, X.~Cao, A.~Avdoshkin, T.~Scaffidi and E.~Altman, \emph{{A Universal Operator Growth Hypothesis}}, \href{https://doi.org/10.1103/PhysRevX.9.041017}{\emph{Phys. Rev. X} {\bfseries 9} (2019) 041017} [\href{https://arxiv.org/abs/1812.08657}{{\ttfamily 1812.08657}}].

\bibitem{Chattopadhyay:2023spread}
A.~Chattopadhyay, A.~Mitra and H.~J.~R. van Zyl, \emph{{Spread complexity as classical dilaton solutions}}, \href{https://doi.org/10.1103/PhysRevD.108.025013}{\emph{Phys. Rev. D} {\bfseries 108} (2023) 025013} [\href{https://arxiv.org/abs/2302.10489}{{\ttfamily 2302.10489}}].

\bibitem{Lv:2023jbv}
C.~Lv, R.~Zhang and Q.~Zhou, \emph{{Building Krylov complexity from circuit complexity}}, \href{https://doi.org/10.1103/PhysRevResearch.6.L042001}{\emph{Phys. Rev. Res.} {\bfseries 6} (2024) L042001} [\href{https://arxiv.org/abs/2303.07343}{{\ttfamily 2303.07343}}].

\bibitem{Craps:2023ivc}
B.~Craps, O.~Evnin and G.~Pascuzzi, \emph{{A Relation between Krylov and Nielsen Complexity}}, \href{https://doi.org/10.1103/PhysRevLett.132.160402}{\emph{Phys. Rev. Lett.} {\bfseries 132} (2024) 160402} [\href{https://arxiv.org/abs/2311.18401}{{\ttfamily 2311.18401}}].

\bibitem{Beetar:2025spread}
C.~Beetar, E.~L. Graef, J.~Murugan, H.~Nastase and H.~J.~R. van Zyl, \emph{{A Quantum Computational Perspective on Spread Complexity}}, \href{https://doi.org/10.1016/j.physletb.2026.140570}{\emph{Phys. Lett. B} {\bfseries 878} (2026) 140570} [\href{https://arxiv.org/abs/2506.07257}{{\ttfamily 2506.07257}}].

\bibitem{Craps:2025explicit}
B.~Craps, G.~Pascuzzi, J.~F. Pedraza, L.-C. Qu and S.-M. Ruan, \emph{{Explicit connections between Krylov and Nielsen complexity}}, \href{https://doi.org/10.1103/spd2-l836}{\emph{Phys. Rev. D} {\bfseries 114} (2026) 046024} [\href{https://arxiv.org/abs/2511.15799}{{\ttfamily 2511.15799}}].

\bibitem{Rabinovici:2023yex}
E.~Rabinovici, A.~S{\'a}nchez-Garrido, R.~Shir and J.~Sonner, \emph{{A bulk manifestation of Krylov complexity}}, \href{https://doi.org/10.1007/JHEP08(2023)213}{\emph{JHEP} {\bfseries 08} (2023) 213} [\href{https://arxiv.org/abs/2305.04355}{{\ttfamily 2305.04355}}].

\bibitem{Heller:2024ldz}
M.~P. Heller, J.~Papalini and T.~Schuhmann, \emph{{Krylov Spread Complexity as Holographic Complexity beyond Jackiw-Teitelboim Gravity}}, \href{https://doi.org/10.1103/spcr-jgm6}{\emph{Phys. Rev. Lett.} {\bfseries 135} (2025) 151602} [\href{https://arxiv.org/abs/2412.17785}{{\ttfamily 2412.17785}}].

\bibitem{Balasubramanian:2024lqk}
V.~Balasubramanian, J.~M. Magan, P.~Nandi and Q.~Wu, \emph{{Spread complexity and the saturation of wormhole size}}, \href{https://doi.org/10.1103/vpyr-b3fb}{\emph{Phys. Rev. D} {\bfseries 113} (2026) 046004} [\href{https://arxiv.org/abs/2412.02038}{{\ttfamily 2412.02038}}].

\bibitem{Fu:2025kkh}
Y.~Fu, H.-S. Jeong, K.-Y. Kim and J.~F. Pedraza, \emph{{Toward Krylov-based holography in double-scaled SYK}}, \href{https://doi.org/10.1007/JHEP05(2026)056}{\emph{JHEP} {\bfseries 05} (2026) 056} [\href{https://arxiv.org/abs/2510.22658}{{\ttfamily 2510.22658}}].

\bibitem{Erdmenger:2023wjg}
J.~Erdmenger, S.-K. Jian and Z.-Y. Xian, \emph{{Universal chaotic dynamics from Krylov space}}, \href{https://doi.org/10.1007/JHEP08(2023)176}{\emph{JHEP} {\bfseries 08} (2023) 176} [\href{https://arxiv.org/abs/2303.12151}{{\ttfamily 2303.12151}}].

\bibitem{Balasubramanian:2023kwd}
V.~Balasubramanian, J.~M. Magan and Q.~Wu, \emph{{Quantum chaos, integrability, and late times in the Krylov basis}}, \href{https://doi.org/10.1103/PhysRevE.111.014218}{\emph{Phys. Rev. E} {\bfseries 111} (2025) 014218} [\href{https://arxiv.org/abs/2312.03848}{{\ttfamily 2312.03848}}].

\bibitem{Caputa:2024vrn}
P.~Caputa, H.-S. Jeong, S.~Liu, J.~F. Pedraza and L.-C. Qu, \emph{{Krylov complexity of density matrix operators}}, \href{https://doi.org/10.1007/JHEP05(2024)337}{\emph{JHEP} {\bfseries 05} (2024) 337} [\href{https://arxiv.org/abs/2402.09522}{{\ttfamily 2402.09522}}].

\bibitem{Camargo:2024deu}
H.~A. Camargo, K.-B. Huh, V.~Jahnke, H.-S. Jeong, K.-Y. Kim and M.~Nishida, \emph{{Spread and spectral complexity in quantum spin chains: from integrability to chaos}}, \href{https://doi.org/10.1007/JHEP08(2024)241}{\emph{JHEP} {\bfseries 08} (2024) 241} [\href{https://arxiv.org/abs/2405.11254}{{\ttfamily 2405.11254}}].

\bibitem{Baggioli:2024wbz}
M.~Baggioli, K.-B. Huh, H.-S. Jeong, K.-Y. Kim and J.~F. Pedraza, \emph{{Krylov complexity as an order parameter for quantum chaotic-integrable transitions}}, \href{https://doi.org/10.1103/PhysRevResearch.7.023028}{\emph{Phys. Rev. Res.} {\bfseries 7} (2025) 023028} [\href{https://arxiv.org/abs/2407.17054}{{\ttfamily 2407.17054}}].

\bibitem{Huh:2024ytz}
K.-B. Huh, H.-S. Jeong, L.~A. Pando~Zayas and J.~F. Pedraza, \emph{{Krylov complexity in mixed phase space}}, \href{https://doi.org/10.1103/gmy7-dn7l}{\emph{Phys. Rev. D} {\bfseries 111} (2025) L121902} [\href{https://arxiv.org/abs/2412.04963}{{\ttfamily 2412.04963}}].

\bibitem{Baggioli:2025knt}
M.~Baggioli, K.-B. Huh, H.-S. Jeong, X.~Jiang, K.-Y. Kim and J.~F. Pedraza, \emph{{Quantum Chaos Diagnostics for non-Hermitian Systems from Bi-Lanczos Krylov Dynamics}},  \href{https://arxiv.org/abs/2508.13956}{{\ttfamily 2508.13956}}.

\bibitem{Erdmenger:2026iga}
J.~Erdmenger, K.-B. Huh, H.-S. Jeong and J.~F. Pedraza, \emph{{Unfolded Krylov complexity: universal chaotic dynamics without false positives}},  \href{https://arxiv.org/abs/2609.02228}{{\ttfamily 2609.02228}}.

\bibitem{Nandy:2024evd}
P.~Nandy, A.~S. Matsoukas-Roubeas, P.~Mart{\'\i}nez-Azcona, A.~Dymarsky and A.~del Campo, \emph{{Quantum dynamics in Krylov space: Methods and applications}}, \href{https://doi.org/10.1016/j.physrep.2025.05.001}{\emph{Phys. Rept.} {\bfseries 1125-1128} (2025) 1} [\href{https://arxiv.org/abs/2405.09628}{{\ttfamily 2405.09628}}].

\bibitem{Baiguera:2025dkc}
S.~Baiguera, V.~Balasubramanian, P.~Caputa, S.~Chapman, J.~Haferkamp, M.~P. Heller et~al., \emph{{Quantum complexity in gravity, quantum field theory, and quantum information science}}, \href{https://doi.org/10.1016/j.physrep.2025.11.001}{\emph{Phys. Rept.} {\bfseries 1159} (2026) 1} [\href{https://arxiv.org/abs/2503.10753}{{\ttfamily 2503.10753}}].

\bibitem{Rabinovici:2025otw}
E.~Rabinovici, A.~S{\'a}nchez-Garrido, R.~Shir and J.~Sonner, \emph{{Krylov Complexity}},  \href{https://arxiv.org/abs/2507.06286}{{\ttfamily 2507.06286}}.

\bibitem{Jeong:2026toappear}
H.-S. Jeong and J.~F. Pedraza, \emph{{Quantum Chaos and Spread of States in Krylov Subspace: A Topical Review}},  \href{https://arxiv.org/abs/2608.26825}{{\ttfamily 2608.26825}}.

\bibitem{Caputa:2024sux}
P.~Caputa, B.~Chen, R.~W. McDonald, J.~Sim{\'o}n and B.~Strittmatter, \emph{{Spread complexity rate as proper momentum}}, \href{https://doi.org/10.1103/7zs8-9zpg}{\emph{Phys. Rev. D} {\bfseries 113} (2026) L041901} [\href{https://arxiv.org/abs/2410.23334}{{\ttfamily 2410.23334}}].

\bibitem{He:2024pox}
P.-Z. He, \emph{{Revisit the relationship between spread complexity rate and radial momentum}},  \href{https://arxiv.org/abs/2411.19172}{{\ttfamily 2411.19172}}.

\bibitem{Fan:2024iop}
Z.-Y. Fan, \emph{{Momentum-Krylov complexity correspondence}},  \href{https://arxiv.org/abs/2411.04492}{{\ttfamily 2411.04492}}.

\bibitem{Li:2025fqz}
Z.~Li and J.~Tian, \emph{{The holography of spread complexity. A story of observers}}, \href{https://doi.org/10.1007/JHEP07(2026)187}{\emph{JHEP} {\bfseries 07} (2026) 187} [\href{https://arxiv.org/abs/2506.13481}{{\ttfamily 2506.13481}}].

\bibitem{Fatemiabhari:2025usn}
A.~Fatemiabhari, H.~Nastase, C.~Nunez and D.~Roychowdhury, \emph{{Holographic Krylov complexity in confining gauge theories}}, \href{https://doi.org/10.1007/JHEP09(2026)155}{\emph{JHEP} {\bfseries 09} (2026) 155} [\href{https://arxiv.org/abs/2511.22717}{{\ttfamily 2511.22717}}].

\bibitem{Fatemiabhari:2025poq}
A.~Fatemiabhari, H.~Nastase, C.~Nunez and D.~Roychowdhury, \emph{{Holographic Krylov complexity for conformal quiver gauge theories}}, \href{https://doi.org/10.1016/j.nuclphysb.2026.117402}{\emph{Nucl. Phys. B} {\bfseries 1025} (2026) 117402} [\href{https://arxiv.org/abs/2512.14812}{{\ttfamily 2512.14812}}].

\bibitem{Fatemiabhari:2026goj}
A.~Fatemiabhari and C.~Nunez, \emph{{Krylov complexity, confinement and universality}}, \href{https://doi.org/10.1007/JHEP07(2026)035}{\emph{JHEP} {\bfseries 07} (2026) 035} [\href{https://arxiv.org/abs/2602.17757}{{\ttfamily 2602.17757}}].

\bibitem{Fatemiabhari:2026rob}
A.~Fatemiabhari, C.~Nunez and R.~T. Santamaria, \emph{{Complexity and Operator Growth in Holographic 6d SCFTs}}, \href{https://doi.org/10.1016/j.nuclphysb.2026.117496}{\emph{Nucl. Phys. B} {\bfseries 1028} (2026) 117496} [\href{https://arxiv.org/abs/2603.10106}{{\ttfamily 2603.10106}}].

\bibitem{Nastase:2026lhz}
H.~Nastase, C.~Nunez and D.~Roychowdhury, \emph{{Holographic Krylov complexity for charged, composite, and extended probes}}, \href{https://doi.org/10.1103/3lxz-17p9}{\emph{Phys. Rev. D} {\bfseries 114} (2026) 086001} [\href{https://arxiv.org/abs/2604.07432}{{\ttfamily 2604.07432}}].

\bibitem{Chatzis:2026ekd}
D.~Chatzis, M.~Hammond, C.~Nunez, A.~V. Ramallo and R.~T. Santamaria, \emph{{Holographic Spread Complexity from Branes and Strings}},  \href{https://arxiv.org/abs/2607.00074}{{\ttfamily 2607.00074}}.

\bibitem{Nunez:2026kwr}
C.~Nunez and D.~Roychowdhury, \emph{{Krylov Complexity and $c$-function along RG Flows}},  \href{https://arxiv.org/abs/2608.02715}{{\ttfamily 2608.02715}}.

\bibitem{Roychowdhury:2026eds}
D.~Roychowdhury, \emph{{Krylov complexity for $\eta $ deformed superstring backgrounds}}, \href{https://doi.org/10.1140/epjc/s10052-026-16265-z}{\emph{Eur. Phys. J. C} {\bfseries 86} (2026) 1005} [\href{https://arxiv.org/abs/2601.06555}{{\ttfamily 2601.06555}}].

\bibitem{Zoakos:2026obl}
D.~Zoakos, \emph{{Holographic Krylov complexity in the Coulomb branch of {\ensuremath{\mathscr{N}}} = 4 SYM}}, \href{https://doi.org/10.1007/JHEP06(2026)066}{\emph{JHEP} {\bfseries 06} (2026) 066} [\href{https://arxiv.org/abs/2603.15435}{{\ttfamily 2603.15435}}].

\bibitem{Roychowdhury:2026sgg}
D.~Roychowdhury, \emph{{Krylov complexity for Lin-Maldacena geometries and their holographic duals}}, \href{https://doi.org/10.1007/JHEP05(2026)197}{\emph{JHEP} {\bfseries 05} (2026) 197} [\href{https://arxiv.org/abs/2604.16977}{{\ttfamily 2604.16977}}].

\bibitem{Roychowdhury:2026vzq}
D.~Roychowdhury, \emph{{Krylov state complexity for BMN matrix model}},  \href{https://arxiv.org/abs/2605.10786}{{\ttfamily 2605.10786}}.

\bibitem{Roychowdhury:2026igc}
D.~Roychowdhury, \emph{{Krylov Complexity for Plane Wave Matrix Model}},  \href{https://arxiv.org/abs/2605.26055}{{\ttfamily 2605.26055}}.

\bibitem{Alfinito:2026vah}
E.~Alfinito and M.~Beccaria, \emph{{Krylov correlators in $ \mathfrak{sl}\left(2,\mathbb{R}\right) $ models: exact results and holographic complexity}}, \href{https://doi.org/10.1007/JHEP08(2026)144}{\emph{JHEP} {\bfseries 08} (2026) 144} [\href{https://arxiv.org/abs/2605.17550}{{\ttfamily 2605.17550}}].

\bibitem{BitaghsirFadafan:2026lek}
K.~Bitaghsir~Fadafan and M.~R. Mohammadi~Mozaffar, \emph{{Holographic Krylov complexity with Lifshitz scaling and hyperscaling violation}}, \href{https://doi.org/10.1007/JHEP09(2026)278}{\emph{JHEP} {\bfseries 09} (2026) 278} [\href{https://arxiv.org/abs/2606.31724}{{\ttfamily 2606.31724}}].

\bibitem{Graef:2026pzv}
E.~L. Graef, J.~Murugan, H.~Nastase and H.~J.~R. Van~Zyl, \emph{{On the Universality of Probe Complexity in $\mathcal{N}=4$ SYM}},  \href{https://arxiv.org/abs/2606.21662}{{\ttfamily 2606.21662}}.

\bibitem{Roychowdhury:2026mpd}
D.~Roychowdhury, \emph{{Krylov complexity and spectral density of BMN matrix model}},  \href{https://arxiv.org/abs/2607.24632}{{\ttfamily 2607.24632}}.

\bibitem{Qu:2025lgo}
L.-C. Qu, \emph{{Lanczos meets orthogonal polynomials}}, \href{https://doi.org/10.1007/JHEP05(2026)225}{\emph{JHEP} {\bfseries 05} (2026) 225} [\href{https://arxiv.org/abs/2512.15857}{{\ttfamily 2512.15857}}].

\bibitem{Alfinito:2026yex}
E.~Alfinito and M.~Beccaria, \emph{{Krylov complexity in the Veneziano-Wosiek model}}, \href{https://doi.org/10.1088/1751-8121/ae99a1}{\emph{J. Phys. A} {\bfseries 59} (2026) 345402} [\href{https://arxiv.org/abs/2603.16291}{{\ttfamily 2603.16291}}].

\bibitem{Baume:2026jyt}
F.~Baume, A.~{\c{C}}avu{\c{s}}o{\u{g}}lu, V.~Chakrabhavi and J.~J. Heckman, \emph{{Controlled Chaos in 4D SCFTs}},  \href{https://arxiv.org/abs/2606.23785}{{\ttfamily 2606.23785}}.

\bibitem{Muck:2026top}
W.~M{\"u}ck, \emph{{Krylov complexity has it all}}, \href{https://doi.org/10.1103/3g6c-jx8t}{\emph{Phys. Rev. D} {\bfseries 114} (2026) L061702} [\href{https://arxiv.org/abs/2605.28681}{{\ttfamily 2605.28681}}].

\bibitem{Li:2026pdh}
Z.~Li and J.~Tian, \emph{{Comments on holographic spread complexity}},  \href{https://arxiv.org/abs/2607.18024}{{\ttfamily 2607.18024}}.

\bibitem{Pedraza:2026zji}
J.~F. Pedraza and L.-C. Qu, \emph{{Recursion Coefficients and Krylov Dynamics in Polynomial Random Matrix Models}},  \href{https://arxiv.org/abs/2608.10072}{{\ttfamily 2608.10072}}.

\bibitem{Qu:2026dmv}
L.-C. Qu, \emph{{A geodesic distance interpretation of Lanczos coefficients}},  \href{https://arxiv.org/abs/2608.16982}{{\ttfamily 2608.16982}}.

\bibitem{Bhattacharya:2026ybq}
A.~Bhattacharya, M.~Flory, M.~P. Heller, E.~Rizza and T.~Schuhmann, \emph{{The quadratic growth of Krylov spread complexity in the BTZ black hole}},  \href{https://arxiv.org/abs/2608.09922}{{\ttfamily 2608.09922}}.

\bibitem{Bhattacharyya:2026zpu}
A.~Bhattacharyya, S.~Pal and J.~F. Pedraza, \emph{{Krylov complexity and the growth of the black hole interior in 3D gravity}},  \href{https://arxiv.org/abs/2608.19373}{{\ttfamily 2608.19373}}.

\bibitem{Jokela:2021knd}
N.~Jokela, J.~Kastikainen, E.~Kiritsis and F.~Nitti, \emph{{Flavored ABJM theory on the sphere and holographic F-functions}}, \href{https://doi.org/10.1007/JHEP03(2022)091}{\emph{JHEP} {\bfseries 03} (2022) 091} [\href{https://arxiv.org/abs/2112.08715}{{\ttfamily 2112.08715}}].

\bibitem{Aharony:2008ug}
O.~Aharony, O.~Bergman, D.~L. Jafferis and J.~Maldacena, \emph{{N=6 superconformal Chern-Simons-matter theories, M2-branes and their gravity duals}}, \href{https://doi.org/10.1088/1126-6708/2008/10/091}{\emph{JHEP} {\bfseries 10} (2008) 091} [\href{https://arxiv.org/abs/0806.1218}{{\ttfamily 0806.1218}}].

\bibitem{Loewy:2002hu}
A.~Loewy and Y.~Oz, \emph{{Branes in special holonomy backgrounds}}, \href{https://doi.org/10.1016/S0370-2693(02)01878-6}{\emph{Phys. Lett. B} {\bfseries 537} (2002) 147} [\href{https://arxiv.org/abs/hep-th/0203092}{{\ttfamily hep-th/0203092}}].

\bibitem{Aharony:2008gk}
O.~Aharony, O.~Bergman and D.~L. Jafferis, \emph{{Fractional M2-branes}}, \href{https://doi.org/10.1088/1126-6708/2008/11/043}{\emph{JHEP} {\bfseries 11} (2008) 043} [\href{https://arxiv.org/abs/0807.4924}{{\ttfamily 0807.4924}}].

\bibitem{Hashimoto:2010bq}
A.~Hashimoto, S.~Hirano and P.~Ouyang, \emph{{Branes and fluxes in special holonomy manifolds and cascading field theories}}, \href{https://doi.org/10.1007/JHEP06(2011)101}{\emph{JHEP} {\bfseries 06} (2011) 101} [\href{https://arxiv.org/abs/1004.0903}{{\ttfamily 1004.0903}}].

\bibitem{Aharony:2009fc}
O.~Aharony, A.~Hashimoto, S.~Hirano and P.~Ouyang, \emph{{D-brane Charges in Gravitational Duals of 2+1 Dimensional Gauge Theories and Duality Cascades}}, \href{https://doi.org/10.1007/JHEP01(2010)072}{\emph{JHEP} {\bfseries 01} (2010) 072} [\href{https://arxiv.org/abs/0906.2390}{{\ttfamily 0906.2390}}].

\bibitem{Benini:2007gx}
F.~Benini, F.~Canoura, S.~Cremonesi, C.~Nunez and A.~V. Ramallo, \emph{{Backreacting flavors in the Klebanov-Strassler background}}, \href{https://doi.org/10.1088/1126-6708/2007/09/109}{\emph{JHEP} {\bfseries 09} (2007) 109} [\href{https://arxiv.org/abs/0706.1238}{{\ttfamily 0706.1238}}].

\bibitem{Aramini:2025twg}
F.~Aramini, R.~Argurio, M.~Bertolini, E.~Garc{\'\i}a-Valdecasas and P.~Moroni, \emph{{Gravity, finite duality cascades and confinement}}, \href{https://doi.org/10.1007/JHEP09(2025)054}{\emph{JHEP} {\bfseries 09} (2025) 054} [\href{https://arxiv.org/abs/2506.18988}{{\ttfamily 2506.18988}}].

\bibitem{Aramini:2026stj}
F.~Aramini, R.~Argurio, M.~Bertolini, P.~Moroni and V.~Tatitscheff, \emph{{Confinement in a finite duality cascade}}, \href{https://doi.org/10.1007/JHEP07(2026)081}{\emph{JHEP} {\bfseries 07} (2026) 081} [\href{https://arxiv.org/abs/2604.18702}{{\ttfamily 2604.18702}}].

\bibitem{Maldacena:1998im}
J.~M. Maldacena, \emph{{Wilson loops in large N field theories}}, \href{https://doi.org/10.1103/PhysRevLett.80.4859}{\emph{Phys. Rev. Lett.} {\bfseries 80} (1998) 4859} [\href{https://arxiv.org/abs/hep-th/9803002}{{\ttfamily hep-th/9803002}}].

\bibitem{Sonnenschein:1999if}
J.~Sonnenschein, \emph{{What does the string / gauge correspondence teach us about Wilson loops?}},  in \emph{{Advanced School on Supersymmetry in the Theories of Fields, Strings and Branes}}, pp.~219--269, 7, 1999, \href{https://arxiv.org/abs/hep-th/0003032}{{\ttfamily hep-th/0003032}}.

\bibitem{Nunez_2010}
C.~Nunez, M.~Piai and A.~Rago, \emph{{Wilson Loops in string duals of Walking and Flavored Systems}}, \href{https://doi.org/10.1103/PhysRevD.81.086001}{\emph{Phys. Rev. D} {\bfseries 81} (2010) 086001} [\href{https://arxiv.org/abs/0909.0748}{{\ttfamily 0909.0748}}].

\bibitem{Dunne:1998qy}
G.~V. Dunne, \emph{{Aspects of Chern-Simons theory}},  in \emph{{Les Houches Summer School in Theoretical Physics, Session 69: Topological Aspects of Low-dimensional Systems}}, 7, 1998, \href{https://arxiv.org/abs/hep-th/9902115}{{\ttfamily hep-th/9902115}}.

\bibitem{Karabali:1999ef}
D.~Karabali, C.-j. Kim and V.~P. Nair, \emph{{Gauge invariant variables and the Yang-Mills-Chern-Simons theory}}, \href{https://doi.org/10.1016/S0550-3213(99)00701-4}{\emph{Nucl. Phys. B} {\bfseries 566} (2000) 331} [\href{https://arxiv.org/abs/hep-th/9907078}{{\ttfamily hep-th/9907078}}].

\bibitem{Ooguri:2008dk}
H.~Ooguri and C.-S. Park, \emph{{Superconformal Chern-Simons Theories and the Squashed Seven Sphere}}, \href{https://doi.org/10.1088/1126-6708/2008/11/082}{\emph{JHEP} {\bfseries 11} (2008) 082} [\href{https://arxiv.org/abs/0808.0500}{{\ttfamily 0808.0500}}].

\bibitem{Susskind:2018tei}
L.~Susskind, \emph{{Why do Things Fall?}},  \href{https://arxiv.org/abs/1802.01198}{{\ttfamily 1802.01198}}.

\bibitem{Susskind:2019ddc}
L.~Susskind, \emph{{Complexity and Newton's Laws}}, \href{https://doi.org/10.3389/fphy.2020.00262}{\emph{Front. in Phys.} {\bfseries 8} (2020) 262} [\href{https://arxiv.org/abs/1904.12819}{{\ttfamily 1904.12819}}].

\bibitem{Ageev:2018msv}
D.~S. Ageev and I.~Y. Aref'eva, \emph{{When things stop falling, chaos is suppressed}}, \href{https://doi.org/10.1007/JHEP01(2019)100}{\emph{JHEP} {\bfseries 01} (2019) 100} [\href{https://arxiv.org/abs/1806.05574}{{\ttfamily 1806.05574}}].

\bibitem{Jiang:2025wpj}
X.~Jiang, J.~C. Halimeh and N.~S. Srivatsa, \emph{{Krylov Complexity Meets Confinement}}, \href{https://doi.org/10.1103/1gsg-zb8h}{\emph{Phys. Rev. D} (2025) } [\href{https://arxiv.org/abs/2511.03783}{{\ttfamily 2511.03783}}].

\bibitem{Kormos:2016osj}
M.~Kormos, M.~Collura, G.~Tak{\'a}cs and P.~Calabrese, \emph{{Real-time confinement following a quantum quench to a non-integrable model}}, \href{https://doi.org/10.1038/nphys3934}{\emph{Nature Phys.} {\bfseries 13} (2016) 246}.

\bibitem{Anabalon:2021tua}
A.~Anabalon and S.~F. Ross, \emph{{Supersymmetric solitons and a degeneracy of solutions in AdS/CFT}}, \href{https://doi.org/10.1007/JHEP07(2021)015}{\emph{JHEP} {\bfseries 07} (2021) 015} [\href{https://arxiv.org/abs/2104.14572}{{\ttfamily 2104.14572}}].

\bibitem{Chatzis:2024kdu}
D.~Chatzis, A.~Fatemiabhari, C.~Nunez and P.~Weck, \emph{{SCFT deformations via uplifted solitons}}, \href{https://doi.org/10.1016/j.nuclphysb.2024.116659}{\emph{Nucl. Phys. B} {\bfseries 1006} (2024) 116659} [\href{https://arxiv.org/abs/2406.01685}{{\ttfamily 2406.01685}}].

\bibitem{Chatzis:2024top}
D.~Chatzis, A.~Fatemiabhari, C.~Nunez and P.~Weck, \emph{{Conformal to confining SQFTs from holography}}, \href{https://doi.org/10.1007/JHEP08(2024)041}{\emph{JHEP} {\bfseries 08} (2024) 041} [\href{https://arxiv.org/abs/2405.05563}{{\ttfamily 2405.05563}}].

\bibitem{Chatzis:2025hek}
D.~Chatzis, M.~Hammond, G.~Itsios, C.~Nunez and D.~Zoakos, \emph{{Supersymmetric AdS solitons, Coulomb branch flows and twisted compactifications}}, \href{https://doi.org/10.1007/JHEP04(2026)184}{\emph{JHEP} {\bfseries 04} (2026) 184} [\href{https://arxiv.org/abs/2511.18128}{{\ttfamily 2511.18128}}].

\bibitem{Anabalon:2024che}
A.~Anabal{\'o}n, H.~Nastase and M.~Oyarzo, \emph{{Supersymmetric AdS solitons and the interconnection of different vacua of $ \mathcal{N} $ = 4 Super Yang-Mills}}, \href{https://doi.org/10.1007/JHEP05(2024)217}{\emph{JHEP} {\bfseries 05} (2024) 217} [\href{https://arxiv.org/abs/2402.18482}{{\ttfamily 2402.18482}}].

\bibitem{Fatemiabhari:2024aua}
A.~Fatemiabhari and C.~Nunez, \emph{{From conformal to confining field theories using holography}}, \href{https://doi.org/10.1007/JHEP03(2024)160}{\emph{JHEP} {\bfseries 03} (2024) 160} [\href{https://arxiv.org/abs/2401.04158}{{\ttfamily 2401.04158}}].

\bibitem{Anabalon:2026yxk}
A.~Anabal{\'o}n, H.~Nastase, C.~Nunez, M.~Oyarzo and R.~Stuardo, \emph{{Moduli space of $ \mathcal{N} $ = 4 super Yang-Mills from AdS/CFT}}, \href{https://doi.org/10.1007/JHEP05(2026)251}{\emph{JHEP} {\bfseries 05} (2026) 251} [\href{https://arxiv.org/abs/2603.18141}{{\ttfamily 2603.18141}}].

\bibitem{Nunez:2023nnl}
C.~Nunez, M.~Oyarzo and R.~Stuardo, \emph{{Confinement in (1 + 1) dimensions: a holographic perspective from I-branes}}, \href{https://doi.org/10.1007/JHEP09(2023)201}{\emph{JHEP} {\bfseries 09} (2023) 201} [\href{https://arxiv.org/abs/2307.04783}{{\ttfamily 2307.04783}}].

\bibitem{Nunez:2023xgl}
C.~Nunez, M.~Oyarzo and R.~Stuardo, \emph{{Confinement and D5-branes}}, \href{https://doi.org/10.1007/JHEP03(2024)080}{\emph{JHEP} {\bfseries 03} (2024) 080} [\href{https://arxiv.org/abs/2311.17998}{{\ttfamily 2311.17998}}].

\bibitem{Witten:1998zw}
E.~Witten, \emph{{Anti-de Sitter space, thermal phase transition, and confinement in gauge theories}}, \href{https://doi.org/10.4310/ATMP.1998.v2.n3.a3}{\emph{Adv. Theor. Math. Phys.} {\bfseries 2} (1998) 505} [\href{https://arxiv.org/abs/hep-th/9803131}{{\ttfamily hep-th/9803131}}].

\bibitem{Klebanov:2000hb}
I.~R. Klebanov and M.~J. Strassler, \emph{{Supergravity and a confining gauge theory: Duality cascades and chi SB resolution of naked singularities}}, \href{https://doi.org/10.1088/1126-6708/2000/08/052}{\emph{JHEP} {\bfseries 08} (2000) 052} [\href{https://arxiv.org/abs/hep-th/0007191}{{\ttfamily hep-th/0007191}}].

\bibitem{Maldacena:2000yy}
J.~M. Maldacena and C.~Nunez, \emph{{Towards the large N limit of pure N=1 superYang-Mills}}, \href{https://doi.org/10.1103/PhysRevLett.86.588}{\emph{Phys. Rev. Lett.} {\bfseries 86} (2001) 588} [\href{https://arxiv.org/abs/hep-th/0008001}{{\ttfamily hep-th/0008001}}].

\bibitem{Leichenauer:2013kaa}
S.~Leichenauer and V.~Rosenhaus, \emph{{AdS black holes, the bulk-boundary dictionary, and smearing functions}}, \href{https://doi.org/10.1103/PhysRevD.88.026003}{\emph{Phys. Rev. D} {\bfseries 88} (2013) 026003} [\href{https://arxiv.org/abs/1304.6821}{{\ttfamily 1304.6821}}].

\bibitem{Aguilar-Gutierrez:2025kmw}
S.~E. Aguilar-Gutierrez, H.~A. Camargo, V.~Jahnke, K.-Y. Kim and M.~Nishida, \emph{{Krylov operator complexity in holographic CFTs: Smeared boundary reconstruction and the dual proper radial momentum}}, \href{https://doi.org/10.1103/6bgg-vglp}{\emph{Phys. Rev. D} {\bfseries 112} (2025) 126014} [\href{https://arxiv.org/abs/2506.03273}{{\ttfamily 2506.03273}}].

\bibitem{Kaplan:2009kr}
D.~B. Kaplan, J.-W. Lee, D.~T. Son and M.~A. Stephanov, \emph{{Conformality Lost}}, \href{https://doi.org/10.1103/PhysRevD.80.125005}{\emph{Phys. Rev. D} {\bfseries 80} (2009) 125005} [\href{https://arxiv.org/abs/0905.4752}{{\ttfamily 0905.4752}}].

\bibitem{Gorbenko:2018dtm}
V.~Gorbenko, S.~Rychkov and B.~Zan, \emph{{Walking, Weak first-order transitions, and Complex CFTs II. Two-dimensional Potts model at $Q>4$}}, \href{https://doi.org/10.21468/SciPostPhys.5.5.050}{\emph{SciPost Phys.} {\bfseries 5} (2018) 050} [\href{https://arxiv.org/abs/1808.04380}{{\ttfamily 1808.04380}}].

\bibitem{Gorbenko:2018ncu}
V.~Gorbenko, S.~Rychkov and B.~Zan, \emph{{Walking, Weak first-order transitions, and Complex CFTs}}, \href{https://doi.org/10.1007/JHEP10(2018)108}{\emph{JHEP} {\bfseries 10} (2018) 108} [\href{https://arxiv.org/abs/1807.11512}{{\ttfamily 1807.11512}}].

\bibitem{Cardy:1980wm}
J.~L. Cardy, M.~Nauenberg and D.~J. Scalapino, \emph{{Scaling Theory of the Potts Model Multicritical Point}}, \href{https://doi.org/10.1103/PhysRevB.22.2560}{\emph{Phys. Rev. B} {\bfseries 22} (1980) 2560}.

\bibitem{Nauenberg:1980nv}
M.~Nauenberg and D.~J. Scalapino, \emph{{Singularities and Scaling Functions at the Potts Model Multicritical Point}}, \href{https://doi.org/10.1103/PhysRevLett.44.837}{\emph{Phys. Rev. Lett.} {\bfseries 44} (1980) 837}.

\bibitem{Faedo:2019nxw}
A.~F. Faedo, C.~Hoyos, D.~Mateos and J.~G. Subils, \emph{{Holographic Complex Conformal Field Theories}}, \href{https://doi.org/10.1103/PhysRevLett.124.161601}{\emph{Phys. Rev. Lett.} {\bfseries 124} (2020) 161601} [\href{https://arxiv.org/abs/1909.04008}{{\ttfamily 1909.04008}}].

\bibitem{Grimm:2021vpn}
T.~W. Grimm, \emph{{Taming the landscape of effective theories}}, \href{https://doi.org/10.1007/JHEP11(2022)003}{\emph{JHEP} {\bfseries 11} (2022) 003} [\href{https://arxiv.org/abs/2112.08383}{{\ttfamily 2112.08383}}].

\bibitem{Grimm:2024elq}
T.~W. Grimm and M.~van Vliet, \emph{{On the complexity of quantum field theory}}, \href{https://doi.org/10.1007/JHEP06(2025)215}{\emph{JHEP} {\bfseries 06} (2025) 215} [\href{https://arxiv.org/abs/2410.23338}{{\ttfamily 2410.23338}}].

\bibitem{Grimm:2025lip}
T.~W. Grimm, D.~Prieto and M.~van Vliet, \emph{{Tame embeddings, volume growth, and complexity of moduli spaces}}, \href{https://doi.org/10.1103/d51c-j1s9}{\emph{Phys. Rev. D} {\bfseries 112} (2025) 106015} [\href{https://arxiv.org/abs/2503.15601}{{\ttfamily 2503.15601}}].

\bibitem{Grimm:2026haa}
T.~W. Grimm, D.~Prieto and M.~van Vliet, \emph{{Tame complexity of effective field theories in the quantum gravity landscape}}, \href{https://doi.org/10.1007/JHEP06(2026)192}{\emph{JHEP} {\bfseries 06} (2026) 192} [\href{https://arxiv.org/abs/2601.18863}{{\ttfamily 2601.18863}}].

\bibitem{Conde:2011sw}
E.~Conde and A.~V. Ramallo, \emph{{On the gravity dual of Chern-Simons-matter theories with unquenched flavor}}, \href{https://doi.org/10.1007/JHEP07(2011)099}{\emph{JHEP} {\bfseries 07} (2011) 099} [\href{https://arxiv.org/abs/1105.6045}{{\ttfamily 1105.6045}}].

\bibitem{Jokela:2012dw}
N.~Jokela, J.~Mas, A.~V. Ramallo and D.~Zoakos, \emph{{Thermodynamics of the brane in Chern-Simons matter theories with flavor}}, \href{https://doi.org/10.1007/JHEP02(2013)144}{\emph{JHEP} {\bfseries 02} (2013) 144} [\href{https://arxiv.org/abs/1211.0630}{{\ttfamily 1211.0630}}].

\end{thebibliography}\endgroup
\end{document}